\documentclass[twoside,12pt]{article}
\usepackage{graphicx}

\usepackage{color}
\usepackage{float}
\graphicspath{ {./figures/},  {./Esym_clus/}, {./Gianluca_figures/}, {./doubleratio_comd/}}
\def\Journal#1#2#3#4{{#1} {#2} (#4) #3 }
\def\PTP{{\em Progress of Theoretical Physics}}
\def\ADNDT{{\em Atomic Data and Nuclear Data Tables}}
\def\JPCS{{\em Journal of Physics: Conference Series}}
\def\EPJA{{\em Eur. Phys. J. A}}
\def\PHY{{\em Physics}}
\def\JPG{{\em J. Phys. G: Nucl. Part. Phys.}}
\def\RPP{{\em Rep. Prog. Phys.}}
\def\JCP{{\em Journal of Computational Physics}}
\def\ANN{{\em Ann. Rev. Nucl. Part. Sci.}}
\def\ANNAST{{\em Ann. Rev. Astron. Astrophys.}}
\def\AP{{\em Ann. Phys}}
\def\APJ{{\em Astrophysical Journal}}
\def\APJL{{\em The Astrophysical Journal Letters}}
\def\APJS{{\em Astrophys. J. Suppl. Ser.}}
\def\EJP{{\em Eur. J. Phys.}}
\def\LANC{{\em Lettere Al Nuovo Cimento}}
\def\NCA{{\em Nuovo Cimento} A}
\def\NC{{\em La Rivista del Nuovo Cimento}}

\def\NP{{\em Nucl. Phys}}
\def\NPA{{\em Nucl. Phys.} A}

\def\JPAM{{\em J. Phys. A: Math. Gen.}}
\def\PRO{{\em Prog. Theor. Phys.}}
\def\PROG{{\em Progress in Particle and Nuclear Physics}}

\def\PLA{{\em Phys. Lett.} A}
\def\PLB{{\em Phys. Lett.} B}

\def\PRL{\em Phys. Rev. Lett.}
\def\PREV{\em Phys. Rev.}
\def\PREP{\em Phys. Rep.}
\def\PRA{{\em Phys. Rev.} A}
\def\PRD{{\em Phys. Rev.} D}
\def\PRC{{\em Phys. Rev.} C}

\def\SCI{{\em Science}}
\def\NAT{{\em Nature}}
\def\ZP{{\em Z. Phys.}}

\def\ZPA{{\em Z. Phys.} A}

\def\RMP{{\em Rev. Mod. Phys.}}
\def\CHEM{{\em J. Chem. Phys.}}
\def\INT{{\em Int. J. Mod. Phys.} E}

\newcommand{\be}{\begin{equation}}
\newcommand{\ee}{\end{equation}}
\newcommand{\bea}{\begin{eqnarray}}
\newcommand{\eea}{\end{eqnarray}}

\begin{document}

\title{ \vspace{1cm} The many facets of the (non relativistic) Nuclear Equation of State}
\author{G.\ Giuliani,$^1$, H.\ Zheng,$^{1,2}$  and A.\ Bonasera,$^{1,3}$ \\ 
\\
$^1$Cyclotron Institute, Texas A\&M University, College Station, TX 77843, USA\\
$^2$Physics Department, Texas A\&M University, College Station, TX 77843, USA\\
$^3$Laboratori Nazionali del Sud, INFN, via Santa Sofia, 62, 95123 Catania, Italy}

\maketitle

\begin{abstract} 
A nucleus is a quantum many body system made of strongly interacting Fermions, protons and neutrons (nucleons). This produces a rich Nuclear Equation of State
whose knowledge is crucial to our understanding of the composition and evolution of celestial objects. The nuclear equation of state displays many different features; first neutrons and protons might be treated as identical particles or nucleons, but when the differences between protons and neutrons are spelled out, we can have
 completely different scenarios, just by changing slightly their interactions. At zero temperature and for neutron rich matter, a quantum liquid gas phase transition at low densities or a quark-gluon plasma at high densities might occur. Furthermore, the large binding energy of the $\alpha$ particle, a Boson, might also open the possibility
 of studying a system made of a mixture of Bosons and Fermions, which adds to the open problems of the nuclear equation of state.
\end{abstract}

%\eject
\tableofcontents
\section{Introduction}
Many aspects of the Nuclear Equation of State (NEOS) have been studied in large detail in the past years. Finite nuclei resemble classical liquid drops, the crucial difference is that the nucleus in its ground state, or zero temperature, does not `solidify' similarly to a drop at low temperatures \cite{landau, r2ad_gr1, ring, preston, povh,  aldobook, khuang}.  This is due to the quantum nature of the nucleus: more specifically its constituents, neutrons ($n$) and protons ($p$), are Fermions. They obey the Pauli principle which forbids two equal Fermions, two protons with the same spin or two neutrons with the same spin (either both up or
both down), to occupy the same quantum state. Thus at zero temperature, two or more Fermions cannot be at rest (a solid) when confined in a finite volume. In intuitive terms, we can express the Pauli principle by saying that a volume $V=\frac{4\pi}{3}R^3$ in coordinate space and $V_p=\frac{4\pi}{3}p_F^3$ in momentum space of size $h^3=(2\pi \hbar)^3$ can at most contain $g=(2s+1)(2\tau+1)$ nucleons, where $\hbar=197.3 MeV\cdot fm$ is the Planck constant, $s$ and $\tau$ are the spin and isospin of the considered Fermion. Thus:
\begin{eqnarray}
\frac{\frac{4\pi}{3}R^3\frac{4\pi}{3}p_F^3}{h^3}=\frac{A}{g}.\label{Pauli}
\end{eqnarray}
Since the density is given by $\rho=\frac{A}{V}$, where $A=N+Z$ is the total number of nucleons (protons+neutrons), we can easily invert equation (\ref{Pauli}) and express the Fermi momentum $p_F$ as function of density \cite{bertschrep88, kimura1}:
\begin{eqnarray}
p_F=(\frac{3\rho}{4\pi g})^{1/3}h.\label{Pf}
\end{eqnarray}
For a nucleus in the ground state $\rho_0=0.16 fm^{-3}$ and $p_F=263 MeV/c$. This means that the nucleons in the nucleus are moving,
even at zero temperature, with a maximum momentum $p_F$ corresponding to a Fermi energy $\epsilon_F=\frac{p_F^2}{2m}=36.8 MeV$. Because of the Fermi energy, the nucleus
or any Fermionic system would expand if there is no confining external potential or interactions among them. Since the total energy of a nucleus in its ground state is about $E \approx -8A MeV$, the average kinetic energy from fermi motion is $\frac{3}{5}\epsilon_F = 22.08 MeV/A$, then the interaction must account for an average $-30 MeV/A$, which is a large value. Because of the relentless motion of the nucleons in the nuclei confined to a finite space due to the nuclear force, we can compare the nucleus to a drop or a liquid. Similarly to a drop, we can compress it and it will oscillate with a typical frequency known as Isoscalar Giant Monopole Resonance (ISGMR) \cite{shlomo2, youngblood1, youngblood2, youngblood3, piekarewicz1, chen1, lipparini1, cao1,  khan1, khan2, colo1, blaizotrep80}:
 \begin{eqnarray}
E_{GMR}=80A^{-1/3}\approx \hbar \sqrt{\frac{K_A}{m\langle r^2 \rangle}} ,\label{GMR}
\end{eqnarray}
where $K_A = K +$surface, Coulomb, symmetry and pairing corrections \cite{blaizotrep80}, $\langle r^2\rangle = \frac{3}{5}R^2$,
$R=r_0A^{1/3}=1.14A^{1/3}fm$ is the average radius of a nucleus of mass $A$, and $K$ is the nuclear compressibility which could be derived from the NEOS if known. From experiments and comparison to theory we know that $K=250\pm25MeV$ which implies that the nucleus is quite `incompressible'. Other nuclear modes are possible where the volume of the nucleus remains constant, such as shape oscillations. A significant example of shape oscillation is the Isoscalar Giant Quadrupole Resonance (ISGQR) mode. Most of these oscillations can be described quantum mechanically as we will discuss later, but also, in some limit, using hydrodynamics \cite{lipparini1, landaufluid, bertschhyro}. An important and maybe crucial feature of nuclei is the fact that its constituents, protons and neutrons, can be described as two different quantum fluids. The fluids might behave as one fluid, such as in the Giant Resonance (GR) cases we briefly discussed before and therefore called Isoscalar GR (ISGR). There are resonances where $n$ and $p$ oscillate against each other and these are
called Isovector GR (IVGR). An important example, related to this last feature, is the Isovector Giant Dipole Resonance (IVGDR). We will discuss this case in some detail
below, section \ref{giantR}, since it gives important informations about the $n-p$ restoring force or potential, thus giving important constraints about the NEOS.

The situations discussed above regard the nucleus near its ground state. However, important phenomena and objects in the universe, such as the Big-Bang (BB) \cite{bigbang1, bigbang2, bigbang3, bigbang4, bradly}, Supernovae explosions (SN) \cite{bradly, suzuki1, supernova, add_nslattimer} or Neutron Stars (NS) \cite{bradly, add_nslattimer, csquark07, nsob1, nsob2, nsob3, nsob4, nsob5, nsob6, nsob7, nsob8, nsob9, nsob10, nsob11, nsob12} require the knowledge of the nuclear interactions in extreme situations, this means we need to pin down the NEOS not only near $\rho_0$ but also at very high or very low densities and/or temperatures. Because of the liquid drop analogy, we expect that if we decrease the density and increase the temperature, the system will become unstable and we will get a ``quantum liquid gas" (QLG) phase transition. Not only because the nucleus is a quantum system, but also because it is made of two strongly interacting fluids,
thus the ``symmetry energy", i.e. the energy of interaction between $n$ and $p$, will be crucial. At very high densities, even at zero temperature, the nucleons will break into their constituents, quarks and gluons, and we get a state of matter called Quark-Gluon Plasma (QGP). Such a state occurred at the very beginning of the BB \cite{bigbang1, bigbang2, bigbang3, bigbang4, bradly}, at very high temperature, and it might occur in the center
part of massive stars including neutron stars \cite{bradly, add_nslattimer, csquark07, nsob1, nsob2, nsob3, nsob4, nsob5, nsob6, nsob7, nsob8, nsob9, nsob10, nsob11, nsob12} as well as heavy ion collisions at relativistic energies \cite{starcoll1, starcoll2, phenixcoll1, phenixcoll2, alicecoll1, alicecoll2}. It is very surprising that slightly changing the symmetry energy we can get at zero temperature either a QLG phase transition or a QGP  by increasing the neutron concentration. In section \ref{eost0}, we will assume {\it for illustration} that at $T=0$, nuclear matter undergoes a second order phase transition at large concentration of neutrons, such as in a neutron star. Assuming that the ground state symmetry energy $E_{sym}=S(\rho_0)=a_a(\infty)=32 MeV$ (following the literature we use different symbols for the symmetry energy and we hope it should not create confusion) and imposing the relevant conditions on pressure and compressibility \cite{landau, khuang, pathria}, we will get two solutions for the critical density: one solution indicates a QLG and the other a QGP! 

It is clear that because of this extreme sensitivity of the NEOS to the symmetry energy, a large effort, both experimental and theoretical, must be implemented \cite{tsang1, tsang2}. It is naive to think that we can constrain the NEOS through astrophysical observations alone \cite{nsob4, r1ad_astr1, r1ad_astr2}, since celestial objects are so complex and observations rare and difficult sometimes to interpret. The NEOS
must also be constrained by laboratory experiments in such a way that our understanding of the universe can steadily improve. New laboratories producing exotic nuclei, either neutron rich or poor, are being built or in operation and this will have a large impact not only in our studies of the NEOS \cite{exotic00} but also in practical applications such as medicine. 

Past studies have been rather effective in constraining the isoscalar part of the NEOS and this will be our starting point in this paper. We learned a lot from those studies about the NEOS of finite systems. We have some ideas on how to deal with the Coulomb field which is present in nuclei and it is a very important ingredient even though might bother
sometimes. Adding to the wealth of informations coming from nuclear physics studies, is the fact that nucleons form very stable systems which are Bosons, for instance $\alpha$ particles \cite{sternbraid, hua2}. In some situations it seems that the nucleus can be thought as formed of $\alpha$ particles, the classical example of the decay of $^{12}C$ into 3$\alpha$, or the $\alpha$ decay of radioactive nuclei. This might suggest that in some situations nuclei can form a Bose-Einstein condensate (BEC) as proposed by many authors \cite{ohkubo1, ohkubo2, kokalova1, tohsaki1, ropke1, sogo1, funaki1, funaki2, raduta1, schuck1}.
 If these conjectures will be experimentally confirmed, we will have the smallest BEC made at most of $10\sim 20$ Bosons. As we will briefly discuss later, the Coulomb repulsion might help in forming a condensate but it hinders the possibility of having large BEC made of $\alpha$ particles (or deuterons) \cite{hua6, smith1}. Thus the NEOS can be discussed not only in terms of the mixture of $n$ and $p$, but also by the mixture of Bosons and Fermions. Clearly, quantum tools must be used to unveil these features: using classical mechanics and some free parameters can  be misleading, BEC or Fermion quenching (FQ) \cite{esteve1, muller1, sanner1, westbrook1} are not classical phenomena!
 
In order to constrain the NEOS, we need to use thermodynamical concepts, therefore we need to create in laboratory equilibrated systems at temperature $T$ and density $\rho$. This is an important task when dealing with finite systems, not impossible, as we have seen already in the past. Since we would like to study the finest details of the NEOS, we need to determine precisely the source size, i.e. the number of $n$ and $p$, which means that we have to detect, event by event, $A_s=N_s+Z_s$ and its excitation energy, which requires the measurement of the kinetic energies of the fragments, their charges and masses with good precision. This can be accomplished both by a suitable choice of the colliding nuclei and beam energy, or better by a careful choice of the equilibrated source, thus eliminating particles emitted before equilibrium is reached. Informations about the neutrons emitted during the process is also crucial and usually hard to have because of experimental difficulties, thus sophisticated models must be implemented. Careful analysis must be able to distinguish between dynamical and equilibrium effects and it is important to stress that dynamical effects give also very important informations about the NEOS, usually through comparison to models. These dynamical effects include observed collective flows, $\pi$, $\gamma$, $K$ and other particle productions which give informations about the time development of the reactions and the sensitivity to different ingredients of the NEOS. In this work we will not discuss about dynamical effects and we will refer to the literature to have more pieces of the NEOS puzzle \cite{bertschrep88, baoanrep08, aldoepja06, aldorep94, aldonc00, aichelinrep91, cassingrep90, csernairep86, dasrep05, onoprog04, stockerrep86, papajcp05, baranrep05, fmd1, fmd7}.
  
Once an isolated and approximately equilibrated hot source is determined, there are different methods proposed in the literature to obtain, $T$, $\rho$, the pressure $P$, entropy $S$ and energy density $\epsilon$. Those methods could be based on classical or quantum assumptions. We know that the nucleus is a quantum system, but, in some conditions, classical approximations could be valid or simply of guidance. At the end of the day the validity of classical approximations must be confirmed by quantum physics. The reason why classical approximation might give a good description of nuclear phenomena is the use of parameters, fitted to experiments, but also to the fact that densities are rather low and temperatures high, i.e. high entropy, where classical approximations are valid. Of course we might claim that we have reached a good understanding of nuclear phenomena only when we can describe them through quantum mechanics.
 
The starting point of the NEOS is the understanding of the nucleus in its ground state. The paradigm of nuclei is the liquid drop model and the Weizs\" acker mass formula which we will briefly discuss next.

\section{ Empirical Mass Formula } \label{eos}
The ground state energy of a nucleus, which is defined as the negative of the binding energy $B(A, Z)$, with mass $A$, charge $Z$ and neutron number $N=A-Z$, can be rather well described by the semi-empirical mass formula first proposed by Weizs\"acker \cite{ring, preston, povh, aldobook, bradly, mf0, mf1, mf3, mf3_1, mf4, mf5, csernai, wong}:
%\begin{equation}
%B(A, Z) = a_vA-a_sA^{2/3}-a_c \frac{Z^2}{A^{1/3}}-a_{sym}\frac{(N-Z)^2}{A}+a_p\delta A^{-3/4}\label{bind} 
%\end{equation}
\begin{eqnarray}
\frac{E}{A} &=& -\frac{B(A, Z)}{A} \nonumber\\
&=& -\frac{a_vA-a_sA^{2/3}-a_c \frac{Z^2}{A^{1/3}}-a_a \frac{(N-Z)^2}{A}+a_p\frac{\delta}{A^{1/2}}}{A}\nonumber\\
&=& -a_v+\frac{a_s}{A^{1/3}}+a_c \frac{Z^2}{A^{4/3}}+a_a(\frac{N-Z}{A})^2- a_p \frac{\delta}{A^{3/2}} \nonumber\\
&=& -a_v+\frac{a_s}{A^{1/3}}+\frac{a_c}{4} A^{2/3}(1-m_\chi)^2+a_am_\chi^2- a_p \frac{\delta}{A^{3/2}}. \label{bind}
\end{eqnarray}
Similarly to a liquid drop, it contains a volume term, $a_vA $ and a surface term $a_sA^{2/3}$. However, nuclei are made of neutrons and charged protons, thus a Coulomb term, $a_c \frac{Z^2}{A^{1/3}}$, a symmetry term, $a_a\frac{(N-Z)^2}{A}$, which takes into account the symmetry of the nucleon-nucleon force, must be added.  As we will discuss later, the proton/neutron 
 concentration might play a role similar to an order parameter which we define as :
 \begin{equation}
m_\chi=\frac{N-Z}{A}.\label{op}
\end{equation}
 Furthermore, quantum effects, such as pairing, 
 are described through the last term, $a_p\frac{\delta}{A^{1/2}}$, where $\delta=\frac{(-1)^N+(-1)^Z}{2}$.  Typical values of the parameters entering equation (\ref{bind}) are:
\begin{equation}
a_v = 15.835 MeV, \quad a_s= 18.33 MeV, \quad a_c = 0.714 MeV, \quad a_a = 23.2 MeV, \quad a_p= 11.2 MeV.
\end{equation}
The Coulomb term can be easily calculated by assuming a uniformly spherical charge distribution of radius $R=r_0A^{1/3}$ where $r_0=1.14fm$. There is an almost universal consensus about the value of the volume term $a_v = 15.5\pm 0.5 MeV$ which corresponds to the value that an infinite system with $N=Z$ nucleons would have if the Coulomb term is neglected. This fictitious state of matter is usually referred as infinite nuclear matter (INM). The pairing term is also reasonably well constrained while the symmetry energy might vary between $a_{a} = 20\sim 40MeV$. The reason why this term is not so well constrained is because corrections might be added to the simple formula equation (\ref{bind}), in order to get a better reproduction of binding energies along the (so far) known periodic table \cite{ring, preston, povh, aldobook, bradly, mf0, mf1, mf3, mf3_1, mf4, mf5, csernai, wong, r2ad_mf1, r2ad_mf2, r2ad_mf3}. The actual value of the symmetry term is of fundamental importance for our understanding, for instance, of neutron stars. In fact, taking the limit of infinite mass and $m_\chi=1$, i.e. a neutron star, gives a binding energy of about $\frac{B(\infty, 0)}{A}\approx16-20(40) MeV$ i.e. an energy per particle which varies between $4$ and $24 MeV$, which means that the neutron star is unbound and a strong gravitational energy is required to keep it confined. This value is of importance for understanding stellar structure and evolution even though it is not the only quantity needed to constrain the neutron star mass and its radius.

For finite nuclei, equation (\ref{bind}), gives the limit of stability as well.  For instance, for superheavy nuclei, assuming $Z=148$, neglecting pairing, we find a maximum value for the binding energy $\frac{B(A, Z)}{A}$ at $A=380$.  Shell corrections are crucial to understand the possibility of existence of superheavy nuclei \cite{she1, she2, she3, she4, she5, she6, she7, foldenshe1, foldenshe2}. We can also make neutron rich/poor nuclei which are the goal of existing and planned facilities worldwide such as FRIB-Michigan State University, FAIR-GSI, Spiral-GANIL,  T-REX-Texas $A\&M$, SPES-Legnaro, RIKEN-Japan, BRIF-China etc. For Z=3, the simple mass formula above gives a binding energy equal to zero for $A\approx 15$. The neutron and proton limiting stability regions define the so called drip-lines and are subject of vigorous theoretical and experimental studies.

\begin{figure}[tb]
\begin{center}
\begin{minipage}[t]{12 cm}
\includegraphics[width=1.0\columnwidth]{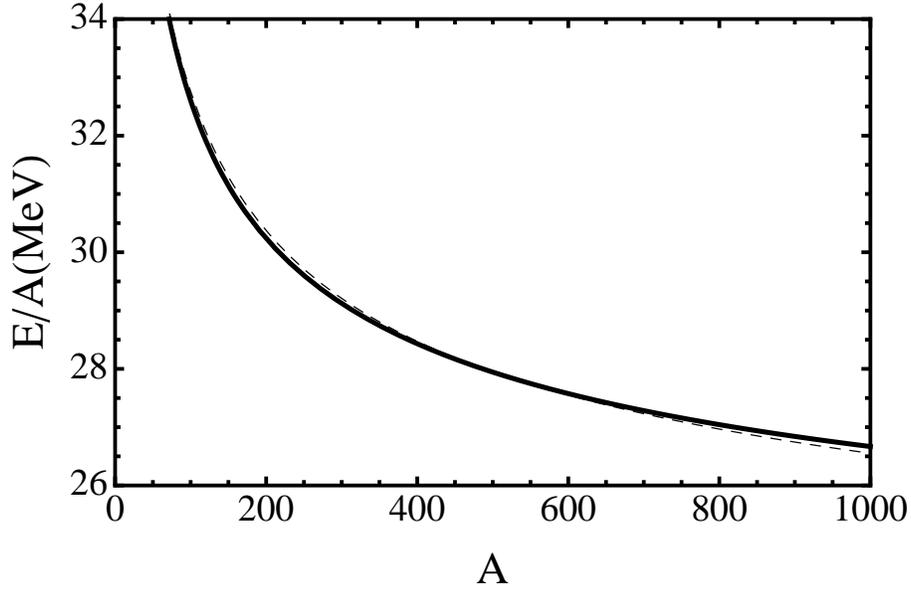}
\end{minipage}
\begin{minipage}[t]{16.5 cm}
\caption{Energy per particle versus the mass number for symmetric nuclear matter within the Fermi gas model with finite size corrections. Dashed line equation (\ref{eafgs}-1)  and long dashed line equation (\ref{eafgs}-2) cannot be distinguished from the exact results, solid line. }\label{sym_fg}
\end{minipage}
\end{center}
\end{figure}

We can understand the origin of some terms entering equation (\ref{bind}) using a simple Fermi gas formula which takes into account finite sizes and neutron-proton asymmetries.  We follow a method first proposed by Hill and Wheeler \cite{hill1} and later extended to excited nuclei as well in ref. \cite{mekjian0}.  Assuming a gas of nucleons contained in a sphere of radius $R$, the number of states with wave numbers between $k$ and $k+dk$ is given by \cite{mekjian0}:
\begin{equation}
dN = V \frac{k^2dk}{2\pi^2} - S\frac{kdk}{8\pi}+L\frac{dk}{8\pi},\label{dnfs}
\end{equation}
where $k=\frac{p}{\hbar}$. $V$, $S$ and $L$ are the volume, surface and circumference of the sphere which are connected through the relations:
\begin{equation}
\frac{S}{V}=\frac{3}{R}; \quad \frac{L}{V}=\frac{3}{2R^2}.
\end{equation}
The density and energy are given respectively by:
\begin{eqnarray}
\rho &=& \frac{A}{V}\nonumber\\
&=& \frac{g}{V} \int dN \nonumber\\
&=&  \frac{g}{V}\int_0^{k_f}(V \frac{k^2dk}{2\pi^2} - S\frac{kdk}{8\pi}+L\frac{dk}{8\pi} )\nonumber\\
&=&g (\frac{k_f^3}{6\pi^2}-\frac{S}{V} \frac{k_f^2}{16\pi}+\frac{L}{V}\frac{k_f}{8\pi}) \nonumber\\
&=&g (\frac{1}{6\pi^2}k_f^3-\frac{3}{16 \pi R} k_f^2+\frac{3}{16 \pi R^2} k_f) ,
\end{eqnarray}
\begin{eqnarray}
E &=& g\int_0^{k_f} \frac{\hbar^2 k^2}{2m}dN \nonumber\\
&=&g \frac{\hbar^2}{2m}\int_0^{k_f} (V \frac{k^4dk}{2\pi^2} - S\frac{k^3dk}{8\pi}+L\frac{k^2dk}{8\pi} )\nonumber\\
&=& g\frac{\hbar^2}{2m} (V\frac{k_f^5}{10\pi^2}-S\frac{k_f^4}{32\pi}+L\frac{k_f^3}{24\pi}) \nonumber\\
&=&g\frac{\hbar^2}{2m} (\frac{2R^3}{15\pi}k_f^5-\frac{R^2}{8} k_f^4+\frac{R}{12}k_f^3). \label{etfs}
\end{eqnarray}
where $g$ is the degeneracy. Thus the energy per particle follows:
\begin{eqnarray}
\frac{E}{A}&=& \frac{g\frac{\hbar^2}{2m} (\frac{2R^3}{15\pi}k_f^5-\frac{R^2}{8} k_f^4+\frac{R}{12}k_f^3) }{\rho V}\nonumber\\
&=&\frac{g\frac{\hbar^2}{2m} (\frac{2R^3}{15\pi}k_f^5-\frac{R^2}{8} k_f^4+\frac{R}{12}k_f^3) }{g (\frac{1}{6\pi^2}k_f^3-\frac{3}{16 \pi R} k_f^2+\frac{3}{16 \pi R^2} k_f)  V}\nonumber\\
&=&\frac{\hbar^2}{2m} \frac{\frac{2R^2}{15\pi}k_f^4-\frac{R}{8} k_f^3+\frac{1}{12}k_f^2}{\frac{2R^2}{9\pi}k_f^2-\frac{R}{4} k_f+\frac{1}{4}} \nonumber\\
&=& \frac{3}{5}\frac{\hbar^2k_f^2}{2m}+\frac{3}{5}\frac{\hbar^2k_f^2}{2m} \frac{\frac{R}{24}k_f-\frac{1}{9}}{\frac{2R^2}{9\pi}k_f^2-\frac{R}{4} k_f+\frac{1}{4}}.
\end{eqnarray}
We know that the nuclear density is $\rho_0 = 0.16 fm^{-3}$, corresponding to $R = 1.14 A^{1/3}$. Thus we can  plot $E/A$ at ground state density as function of the mass number $A$, figure \ref{sym_fg}. The results are fitted with two equations, which take into account explicitly
surface and curvature corrections:
\begin{equation}
(1) \frac{E}{A} = 21.1743+\frac{53.7755}{A^{1/3}}, \quad (2) \frac{E}{A} = 21.9697+\frac{44.7559}{A^{1/3}}+\frac{21.33}{A^{2/3}}. \label{eafgs}
\end{equation}

\begin{figure}[tb]
\begin{center}
\begin{minipage}[t]{12 cm}
\includegraphics[width=1.0\columnwidth]{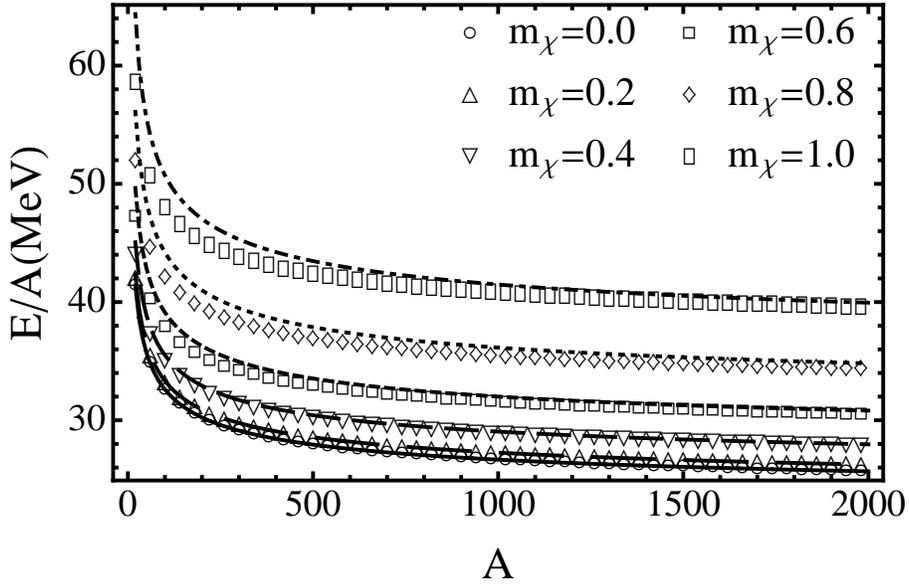}
\end{minipage}
\begin{minipage}[t]{16.5 cm}
\caption{Energy per particle versus the mass number in a Fermi gas model which takes into account finite size effects. Different values of $m_\chi$ are included:  lines are for $\frac{E(A, m_\chi=0)}{A}\times (1+\frac{5}{9}m_\chi^2)$ and symbols are for $\frac{E(A, m_\chi)}{A}$.}\label{sym_fg2}
\end{minipage}
\end{center}
\end{figure}

This simple model shows that for infinite symmetric (uncharged) nuclei the Fermi gas gives a contribution to the total energy per particle of the system of about $\bar\epsilon_f=21.5 MeV$, which is the 
average Fermi energy. To have bound matter, we need a potential energy contribution of about $-37.5 MeV$ for $N=Z$. However, for finite nuclei the Fermi energy increases and this must be
balanced by the nuclear attraction which now might have surface, as in the simple mass formula above, and curvature corrections, which are included in more involved mass formulas \cite{mf6, mf6_1}.  The net result for finite nuclei, with Coulomb and pairing, is an energy per particle of about $-8 MeV$, i.e. half of the value of infinite symmetric nuclear matter (without Coulomb).

We can learn from the Fermi gas also regarding the $N\ne Z$ systems.  Within a Fermi gas approximation, we can separate protons and neutrons:
\begin{equation}
\rho_p = \frac{Z}{V} = \frac{1-m_\chi}{2}\rho_0, \quad \rho_n = \frac{N}{V} = \frac{1+m_\chi}{2}\rho_0,
\end{equation}
and assuming for simplicity that $p$ and $n$ are contained within a sphere of radius R:
\begin{equation}
\rho_p =g_p  (\frac{1}{6\pi^2}k_{fp}^3-\frac{3}{16 \pi R} k_{fp}^2+\frac{3}{16 \pi R^2} k_{fp}), \label{1}
\end{equation}
\begin{equation}
\rho_n =g_n  (\frac{1}{6\pi^2}k_{fn}^3-\frac{3}{16 \pi R} k_{fn}^2+\frac{3}{16 \pi R^2} k_{fn}), \label{2}
\end{equation}
\begin{equation}
E_p= g_p\frac{\hbar^2}{2m} (\frac{2R^3}{15\pi}k_{fp}^5-\frac{R^2}{8} k_{fp}^4+\frac{R}{12}k_{fp}^3), 
\end{equation}
\begin{equation}
E_n= g_n\frac{\hbar^2}{2m} (\frac{2R^3}{15\pi}k_{fn}^5-\frac{R^2}{8} k_{fn}^4+\frac{R}{12}k_{fn}^3). 
\end{equation}
The total energy is given by the sum of the two contributions and gives exactly the value calculated before.
%\begin{equation}
%E = E_p+E_n
%\end{equation}

We can define the symmetry energy as:
\begin{equation}
S(A,m_\chi) = \frac{E}{A}(A, m_\chi)-  \frac{E}{A}(A, 0)\approx \frac{1}{2}a_{sym} m_\chi^2+\frac{1}{4}b_{sym} m_\chi^4+\frac{1}{6}c_{sym} m_\chi^6, \label{symexp}
\end{equation}
where we have expanded the symmetry energy in terms of the order parameter $m_\chi$, equation (\ref{op}). From this we can define the average symmetry energy as \cite{symTypel10}:
\begin{equation}
\bar E_s=\frac{\frac{E}{A}(A, 1)+\frac{E}{A}(A, -1)}{2}-\frac{E}{A}(A, 0). \label{avEs}  
\end{equation}
We will use this definition later on table \ref{aw}. This definition of the average symmetry energy is equivalent to the definition entering in the mass formula, equation (\ref{bind}), $a_a=\frac{1}{2}a_{sym}$ only when higher order terms in equation (\ref{symexp}) can be neglected. If higher order terms are important, then the average symmetry energy is more suitable for neutron rich (or poor) nuclei. Notice that even terms only have been retained to take into account 
the symmetry of nuclear forces under the exchanges of protons with neutrons. Of course such a symmetry is broken by the Coulomb force, equation (\ref{bind}), and by the small mass difference between neutrons and protons.
We can test the approximation above using the Fermi gas model which includes finite size effects.  In figure \ref{sym_fg2} we plot the energy per particle versus $A$ for different values of $m_\chi$.

 As we see from the figure, the lowest order approximation is rather good for very large nuclei, while small systems require the inclusion of higher order terms in the expansion, equation (\ref{symexp}). The corrections we obtained in the naive Fermi gas model will also be valid when interactions are included. In particular the symmetry energy might contain another term arising from antisymmetrization, 
 as proposed by Wigner \cite{wigner}:
\begin{equation}
E_{Wig}=W |m_\chi|, \label {wig}
\end{equation}
where $W$ is another fitting parameter. Other corrections due to Pauli blocking entering the Wigner term are possible as discussed for instance in \cite{ring, preston}.

\begin{figure}[tb]
\begin{center}
\begin{minipage}[t]{12 cm}
\includegraphics[width=1.0\columnwidth]{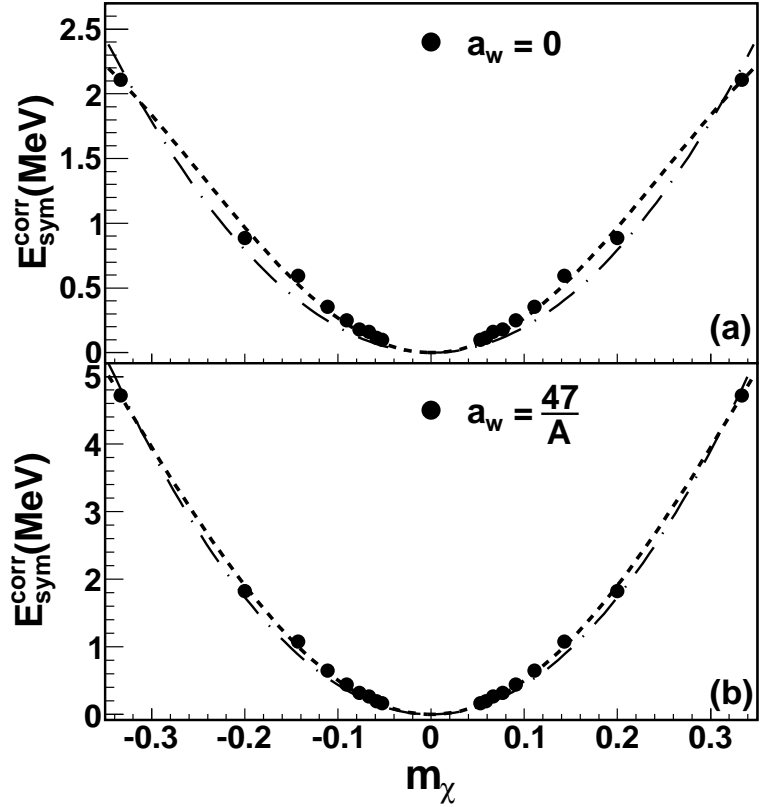}
\end{minipage}
\begin{minipage}[t]{16.5 cm}
\caption{ Symmetry energy per particle versus $n$, $p$ concentration corrected for pairing and Coulomb contributions. 
The two panels are for different Wigner corrections to the symmetry energy: $a_w=0,  \frac{47}{A}$. 
The dash-dotted line refers to  the O($m_\chi^2$) fit and the dashed line to the  O($m_\chi^6$) fit.}\label{symO6}
\end{minipage}
\end{center}
\end{figure}  

In ref. \cite{aldobook}, a different way to extract the symmetry energy is discussed. We can define:
\begin{eqnarray}
\frac{\delta E}{A}&=&\frac{1}{A}(\frac{E_{2p}+E_{2n}}{2}-E_d) \nonumber\\
&=&-\frac{1}{A}(\frac{B_{2p}+B_{2n}}{2}-B_d) \nonumber\\
&=& -(\frac{1}{2}\frac{B_{2p}+B_{2n}}{A}-\frac{B_d}{A}),
\end{eqnarray}
where $E_{2p,n}$, $E_d$ are the ground state energies of a nucleus having $2p (2n)$ or a deuteron $(d)$, outside a closed shell, compare to equation (\ref{symexp}). This equation would give exactly the symmetry energy per particle if the Coulomb and pairing 
contributions are neglected. More precisely, substituting the mass formula equation (\ref{bind}) into the above equation, we obtain:
\begin{equation}
\frac{\delta E}{A}=(a_a+S_c)m_{\chi}^2-2\frac{a_p}{A^{3/2}}+O(m_\chi^4),
\end{equation}
where $S_c=\frac{a_c}{4}A^{2/3}$ and $a_p = 11.2 MeV$. Notice that it is rather difficult, if not impossible, to avoid either Coulomb or pairing contributions from differences of binding energies
of nuclei. We can subtract the Coulomb, Wigner and pairing contributions:
\begin{equation}
\frac{\delta E}{A}-S_c m_{\chi}^2+2\frac{a_p}{A^{3/2}}+a_w |m_\chi|= a_a m_{\chi}^2+O(m_\chi^4)=E_{sym}^{corr}(m_\chi), \label{esymcorr}
\end{equation}
where $a_w=\frac{W}{A}$. The corrected symmetry energy obtained from experimental binding energy data of several nuclei is plotted in figure \ref{symO6}, together with different parametrization, equations (\ref{bind}, \ref{symexp}, \ref{wig}).
Clearly the higher order approximation gives a better reproduction of the data, with and without the inclusion of the Wigner term.  However, the range of $m_\chi$ from data is rather limited, thus the constraints to the higher order terms in the symmetry energy are rather poor. We fixed the parameter $c_{sym}$ in such a way to give $\bar E_s=32 MeV$, which is the average of the values obtained at the $O(m_\chi^2)$ with and without Wigner term. This summarizes the difficulty in obtaining the symmetry energy: the range of ground state neutron to proton concentrations is too small, not enough to find a unique value and more `exotic' species are needed. The value of the Wigner term is taken from ref. \cite{wignerterm}. We notice that the
inclusion of the Wigner term gives a good reproduction of the data already at $O(m_{\chi}^2)$, thus we can conclude that higher order term can mimic rather well the Wigner term as well. The Wigner term gives also a contribution to the average symmetry energy $\propto \frac{1}{A}$, which goes to zero for infinite nuclear matter. 
%\begin{equation}
%\frac{E_{sym}}{A}(m) = \frac{1}{2}a_{sym} m^2+\frac{1}{4} b_{sym} m^4 +\frac{1}{6} c_{sym} m^6+O(m^8)-a_w|m|
%\end{equation}
%and 
%\begin{equation}
%a_{sym}=m1, \quad b_{sym}=m2, \quad c_{sym}=m3, \quad a_w=m4
%\end{equation}

The values obtained from the fits are reported in table \ref{aw}. The large variation of the fitting parameters depends on the approximation used which contributes to the uncertainty on the symmetry energy. Notice that $a_a=\frac{1}{2}a_{sym}$. 

\begin{table}
\begin{center}
\begin{minipage}[t]{16.5 cm}
\caption{Fitting values of the symmetry energy obtained from experimental data and equation (\ref{esymcorr}).  The fitting function is $\frac{1}{2}a_{sym} m_\chi^2+\frac{1}{4} b_{sym} m_\chi^4 +\frac{1}{6} c_{sym} m_\chi^6$.}
\label{aw}
\end{minipage}
\begin{tabular}{|c|c|c|c|c|}
\hline
&&&\\[-2mm]
    $a_w$ (MeV)          &   $a_{sym}$ (MeV)  &   $b_{sym}$ (MeV)   &    $c_{sym}$ (MeV) & $\bar E_s$ (MeV)        \\[2mm]
 \hline
 $0 $  &    39.679   &       0               &        0     & 19.84     \\
 \hline
$0 $  &   55.241   &      -355.26       &        559.167   &  32  \\
 \hline
  $\frac{47}{A} $  &    86.679   &       0       &        0    &  43.34     \\
 \hline
$\frac{47}{A} $  &   102.07   &       -340.07       &    395.895 & 32     \\
 \hline
\end{tabular}
\end{center}
\end{table}

\subsection{\it Isobaric Analog States} \label{IAS}
An elegant and powerful method to extract the symmetry energy
coefficients has been proposed by P. Danielewicz and collaborators. It is based on the systematic study of Isobaric Analog
States (IAS) \cite{Dan1, Dan2, Dan3}. Such a method relies on the calculation of
the excitation energies between nuclei belonging to the same IAS chain. The
starting point is the introduction of the isospin operator $\hat{T}=
\sum_{i=1}^{A} \hat{t}_{i}$, where $\hat{t}_{i}$ is the
isospin operator for the single nucleon,
in the symmetry term of the Weizs\"{a}cker formula (see equation (\ref{bind})). On this
basis the third component
$T_{z}=\frac{N-Z}{2}$ of the $\hat{T}$ operator can be easily related to
the $m_{\chi}$ order parameter:
\begin{equation} 
m_{\chi}=\frac{2T_{z}}{A}.
\end{equation}
As a consequence the symmetry term $E_{1}$ will assume the following form:
\begin{equation}
E_{1}=\frac{4a_{a}(A)}{A}T_{z}^{2}. \label{iassym}
\end{equation}
In equation (\ref{iassym}) a functional dependence  of the symmetry energy coefficient $a_{a}$,
which already appears in equation (\ref{bind}), on the mass number $A$ is considered. The
$E_{1}$ energy is invariant
for reflections in the isospin space; furthermore, under the hypothesis of
charge invariance all the
components of the isospin operator can be treated in the same way as $T_{z}$ and
equation (\ref{iassym}) can be generalized as follows:
\begin{equation}
E_{1}=\frac{4a_{a}(A)}{A}\hat{T}^{2}=\frac{4a_{a}(A)}{A}T(T+1),
\end{equation}
where $T(T+1)$ is the eigenvalue of the $\hat{T}^{2}$ operator. For
a given nuclear ground state (gs) the $T$ number corresponds to the lowest
possible value of $T= | T_{z} | =\frac{1}{2} | N-Z |$. 

\begin{figure}[tb]
\begin{center}
\begin{minipage}[t]{12 cm}
\includegraphics[width=1.0\columnwidth]{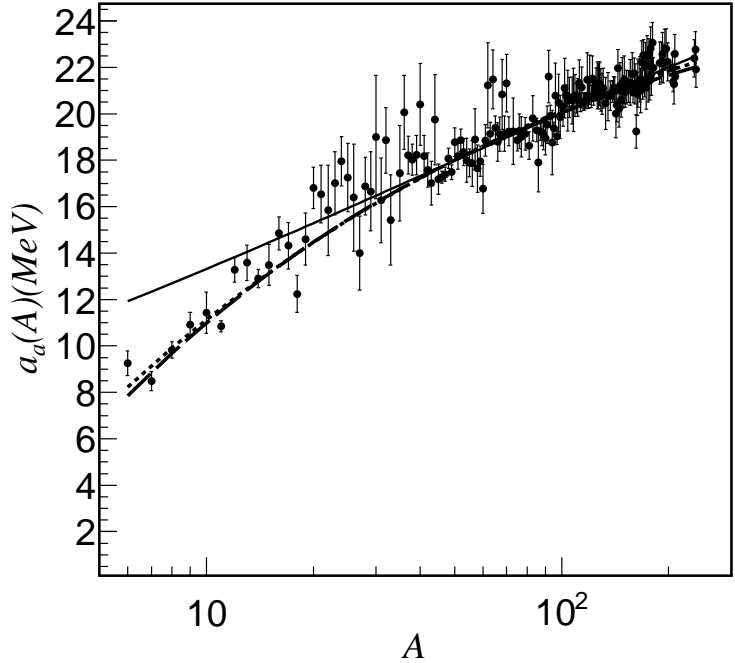}
\end{minipage}
\begin{minipage}[t]{16.5 cm}
\caption{ $a_{a}(A)$ symmetry coefficients extracted from the $E^{*}_{IAS}$
excitation energies. The solid line represents the fit using equation (\ref{iasaafdan}) while the
long dashed and short dashed lines using equation (\ref{iasaafgia1}) and equation (\ref{iasaafgia2}) respectively.} \label{ias}
\end{minipage}
\end{center}
\end{figure}

The definition of the
excitation energy to the IAS $E^{*}_{IAS}$ goes as follows: the lowest energy
state for a given $T> | T_{z} |$ of a nucleus could be the IAS of the gs of a
neighboring isobar nucleus, and the excitation energy is:
\begin{equation}
E^{*}_{IAS}=E_{IAS}-E_{gs}=\frac{4a_{a}(A)}{A} \Delta \hat{T}^{2}_{\bot}+
\Delta E_{mic}. \label{iasestar}
\end{equation}
$\Delta \hat{T}^{2}_{\bot}$ represents the variation in the isospin transverse
squared between the IAS and the gs:
\begin{equation}
\Delta \hat{T}^{2}_{\bot}= \Delta \hat{T}^{2} = T(T+1)- | T_{z} | (| T_{z} |
+1) \equiv \hat{T}^{2}_{\bot} - | T_{z} |.
\end{equation}
$| T_{z} |$ accounts for the Wigner term contribution (see equation (\ref{wig})),
the correction $\Delta
E_{mic}$ is related to pairing and shell effects as proposed by Koura
{\it et al.} \cite{Koura}. The $E^{*}_{IAS}$ energy gives the
possibility to investigate the variations of the symmetry energy coefficient
$a_{a}(A)$ on a ``moving nucleus-by-nucleus'' basis. By using the $E^{*}_{IAS}$
values
provided by the data compilations \cite{Dan1, Table}, the symmetry energy
coefficient is obtained reversing equation (\ref{iasestar}):
\begin{equation}
a_{a}(A)=\frac{A}{4}\frac{(E^{*}_{IAS}- \Delta
E_{mic})}{\Delta (T(T+1))} , \label{iasaa}
\end{equation}
where $\Delta (T(T+1)) = \Delta \hat{T}^{2}$. Figure \ref{ias} shows the behavior as a
function of the mass number of the $a_{a}(A)$ coefficients, calculated with
equation (\ref{iasaa}). One issue of the exposed method is related to the fluctuations of the
$a_{a}(A)$ values coming from the subtractions of the single ground state
energy to the energies of different excited states. To address this issue a
dedicated fitting procedure has been developed, the $a_{a}(A)$
coefficients resulting of such procedure are represented by the full circles in
figure \ref{ias}.

The figure shows an increase of the symmetry coefficient with the increase of
the mass number $A$ up to a value of $A \approx 40$, for the heavier nuclei
instead an almost constant value of approximately $22 MeV$ is reached.
The solid line in the figure represents a fit performed by
assuming volume and surface contributions to the symmetry energy coefficient
\cite{Dan1}:
\begin{equation}
\frac{1}{a_{a}(A)}=\frac{1}{a_{a}^{V}}+\frac{A^{-\frac{1}{3}}}{a_{a}^{S}},\label{iasaafdan}
\end{equation}
where $a_{a}^{V}=35.51 MeV$ and $a_{a}^{S}=9.89 MeV$. By looking at the figure
a certain volume-surface competition can be noticed while moving from low to
high $A$ values, eg. for $A \approx 10$, $a_{a} \approx 10 MeV$ with
$a_{a}^{S}=9.89 MeV$ \cite{Dan3}. However, the $a_{a}(A)$ values corresponding to the mass
numbers in the $5 \leq A \leq 30$ interval are better reproduced by the
following formulas, suggested by the Fermi gas model corrected for surface discussed in section \ref{eos}:
\begin{equation}
a_{a}(A)=27.893-36.427A^{-\frac{1}{3}}, \label{iasaafgia1}
\end{equation}
\begin{equation}
a_{a}(A)=28.761-41.857A^{-\frac{1}{3}}+8.2744A^{-\frac{2}{3}}. \label{iasaafgia2}
\end{equation}

The global behavior of $a_{a}(A)$ as a function of $A$ is better reproduced by
equation (\ref{iasaafgia1}) especially in the lower mass region (long-dashed line).
A further curvature contribution is introduced in equation (\ref{iasaafgia2}) (short-dashed line).
In contrast to the results of equation (\ref{iasaafdan}), the fits given in equations (\ref{iasaafgia1}, \ref{iasaafgia2}), provide a bulk contribution to
the symmetry coefficient coming from the volume term $a_{a}^{V}$ of $28\sim 29
MeV$. In following works, Danielewicz and collaborators have used these results together with Hartree-Fock (HF) calculations to constrain the density dependence of the symmetry energy \cite{Dan1, Dan2, Dan3, r2ad_ias1}.

\section{ The Nuclear Equation of State at Zero Temperature } \label{eost0}
\subsection{\it Momentum independent NEOS}\label{mideos}
The results discussed in the previous sections refer to properties of nuclei in their ground state or at small excitation energies. In all cases the density is the ground state one, $\rho_0=0.16 fm^{-3}$, plus surface effects. In nuclear astrophysics it is necessary to know nuclear properties not only at different densities, but at different temperatures as well. We have seen already the density dependence of the energy of a nucleus in the simple Fermi gas model. On similar grounds we need to introduce the density dependence, the momentum dependence of the interactions among nucleons and we need to distinguish between protons and neutrons. A simple approximation to the nuclear interaction was proposed by Skyrme \cite{skyrme} and it is widely used in the literature \cite{shlomo2, stone2, stone3, shlomo0, shlomo1,  skyrme1_1, skyrme1_2, skyrme1_3, skyrme1_4, skyrme1_5, skyrme1_6, skyrme1_7, skyrme1_8, skyrme1_9, skyrme1_10, skyrme1_11, skyrme1_12, skyrme1_13,  skyrme1_14, skyrme1_15, skyrme1_16, skyrme2,  skyrme3,  skyrme4,   skyrme5,  skyrme6, skyrme8,  skyrme9,  skyrme10,  skyrme11,  skyrme12,  skyrme13,  skyrme14,  skyrme15,  skyrme16, ad_skyrme1, ad_skyrme2, ad_skyrme3, ad_skyrme4, ad_skyrme5, ad_skyrme6, ad_skyrme7, ad_skyrme8, ad_skyrme9, ad_skyrme10, ad_skyrme11, ad_skyrme12, ad_skyrme13, ad_skyrme14, ad_skyrme15}. Hundreds of interactions have been proposed but stringent experimental quantities are also available which should, in principle reduce this huge inflation. The knowledge of the kinetic and potential energies of nucleons leads to the Nuclear Equation of State (NEOS). In heavy ion collisions, highly excited systems might be formed and, in some conditions, a temperature and a density might be recovered from experimental observations or models. In this way it is possible to investigate the NEOS at finite temperature.  First we will discuss the NEOS at zero temperature and we will assume that the interaction among nucleons is local. 

It is possible to write the energy per particle as:
 \begin{equation}
\frac{E}{A}(\rho, m_{\chi}) = (1+\frac{5}{9}m_{\chi}^2) \bar \epsilon_f \tilde \rho^{2/3} + (1+c_1m_{\chi}^2)\frac{A_1}{2}\tilde \rho + (1+c_2 m_{\chi}^2) \frac{B_1}{1+\sigma} \tilde \rho^\sigma,\label{ck225}
\end{equation}
where $\tilde \rho=\frac{\rho}{\rho_0}$, $\bar \epsilon_f$ is the average fermi energy, $A_1$, $B_1$, $\sigma$, $c_1$ and $c_2$ are the parameters to be determined in order to reproduce some properties of INM. The assumed form for the energy per particle in equation (\ref{ck225}) is for guidance only and many different forms can be found in the literature \cite{shlomo2, hua2, baoanrep08, stone2, stone3, shlomo0, shlomo1, stone1, hua1, usmani1}.  It is a simple expansion to second order in $m_\chi$, and higher order terms might be added once more constraints to the NEOS are determined.

This equation refers to an hypothetical infinite nuclear system with $N$ neutrons and $Z$ protons without Coulomb interaction. In order to fix the
parameters entering equation (\ref{ck225}), we impose some constraints coming from observations. In particular for symmetric nuclear matter we require that:
\begin{equation}
\left\{\begin{array}{l} 
\frac{E}{A}\vert_{\rho=\rho_0}= -15 MeV,\\
P\vert_{\rho=\rho_0}=\rho^2 \frac{\partial(\frac{E}{A})}{\partial \rho}\vert_{\rho=\rho_0}=0,\\
K\vert_{\rho=\rho_0}=9\frac{\partial P}{\partial \rho}\vert_{\rho=\rho_0}=225 MeV.
\end{array}
\right.\label{ck225_eq2}
\end{equation}
Where the pressure must be zero for a system in the ground state and the compressibility is fixed by the ISGMR \cite{shlomo2,  youngblood1, youngblood2, youngblood3, piekarewicz1,  chen1, lipparini1, cao1, khan1, khan2, colo1, blaizotrep80}. There is a general consensus that $K=250\pm25 MeV$,
here we will assume $K=225 MeV$ which is the same value obtained in a simple Fermi gas \cite{aldoeos}. The latter condition implies  that interactions give no contribution to the compressibility at ground state density. Solving equations (\ref{ck225_eq2}) gives:
$A_1=-210 MeV$, $B_1=157.5 MeV$ and $\sigma = \frac{4}{3}$. We remind that the repulsive higher order density dependence is needed in order to get nuclear saturation. Once the interaction is known, it is easy to calculate the forces acting on a particle from the gradient respect to ${\bf r}$ of the mean field \cite{bertschrep88, baoanrep08, aldorep94, aichelinrep91, cassingrep90}.

The value of the ground state energy is obtained from the mass formula and precisely from the volume term. In order to fix  the parameters for the asymmetric NEOS we need to know the value of the symmetry energy that,
as we have discussed above, is somehow constrained between $20 MeV$ and $40MeV$.  
The definition of symmetry energy $S(\rho)$  to order $m_\chi^2$ is
\begin{eqnarray}
\frac{E}{A}(\rho, m_\chi)-\frac{E}{A}(\rho, 0)&=& (\frac{5}{9} \bar \epsilon_f \tilde \rho^{2/3} +c_1\frac{A_1}{2}\tilde \rho + c_2 \frac{B_1}{1+\sigma} \tilde \rho^\sigma)m_\chi^2\nonumber\\
&=&S(\rho)m_\chi^2. \
\end{eqnarray}
Therefore
\begin{equation}
S(\rho) = \frac{5}{9} \bar \epsilon_f \tilde \rho^{2/3} +c_1\frac{A_1}{2}\tilde \rho + c_2 \frac{B_1}{1+\sigma} \tilde \rho^\sigma.\label{s}
\end{equation}
Similarly to the pressure and compressibility defined above, we can define the following quantities:
\begin{eqnarray}
L(\rho)&=&3\rho_0 \frac{\partial S(\rho)}{\partial \rho} \nonumber\\
&=& 3[\frac{10}{27} \bar \epsilon_f \tilde \rho^{-1/3} + c_1\frac{A_1}{2} + c_2 \frac{B_1\sigma}{1+\sigma} \tilde \rho^{\sigma-1}],\label{l}
\end{eqnarray}
\begin{eqnarray}
K_{sym}(\rho) &=& 9\rho_0^2 \frac{\partial^2 S(\rho)}{\partial \rho^2}\nonumber\\
&=& 9[-\frac{10}{81} \bar \epsilon_f \tilde \rho^{-4/3} + c_2 \frac{B_1\sigma(\sigma -1)}{1+\sigma} \tilde \rho^{\sigma-2}]\label{ksym}.
\end{eqnarray}
The definitions above help in understanding the sensitivity of different observables to each one of them. For instance, we have seen that the ISGMR is sensitive to the compressibility, on similar grounds we might expect
that the IVGDR is sensitive to $K_{sym}$. Furthermore, they might be useful when comparing different forms of proposed nuclear interactions.
However, we can only constraint equation (\ref{s})  from properties of finite nuclei.  To have a better grasp of the symmetry energy we need more constraints
to fix the values of $c_1,c_2$.   

It is instructive to calculate the values of  $L$ and $K_{sym}$ at ground state density. Simple calculations give:
\begin{equation}
L(\rho_0)=L=3[\frac{10}{27} \bar \epsilon_f  + c_1\frac{A_1}{2} + c_2 \frac{B_1\sigma}{1+\sigma}], \label{l0}
\end{equation}
\begin{equation}
K_{sym}(\rho_0)=K_{sym}=9[-\frac{10}{81} \bar \epsilon_f + c_2 \frac{B_1\sigma(\sigma -1)}{1+\sigma}].\label{ksym0}
\end{equation}
Substituting into the symmetry energy equation (\ref{s}):
\begin{equation}
S(\rho_0)=a_a=\frac{5}{27}\bar\epsilon_f-\frac{10}{81}\frac{\bar\epsilon_f}{\sigma}+\frac{L}{3}-\frac{K_{sym}}{9\sigma}.\label{s0}
\end{equation}

\begin{figure}[tb]
\begin{center}
\begin{minipage}[t]{16 cm}
 \begin{tabular}{cc}
 \includegraphics[width=0.5\columnwidth]{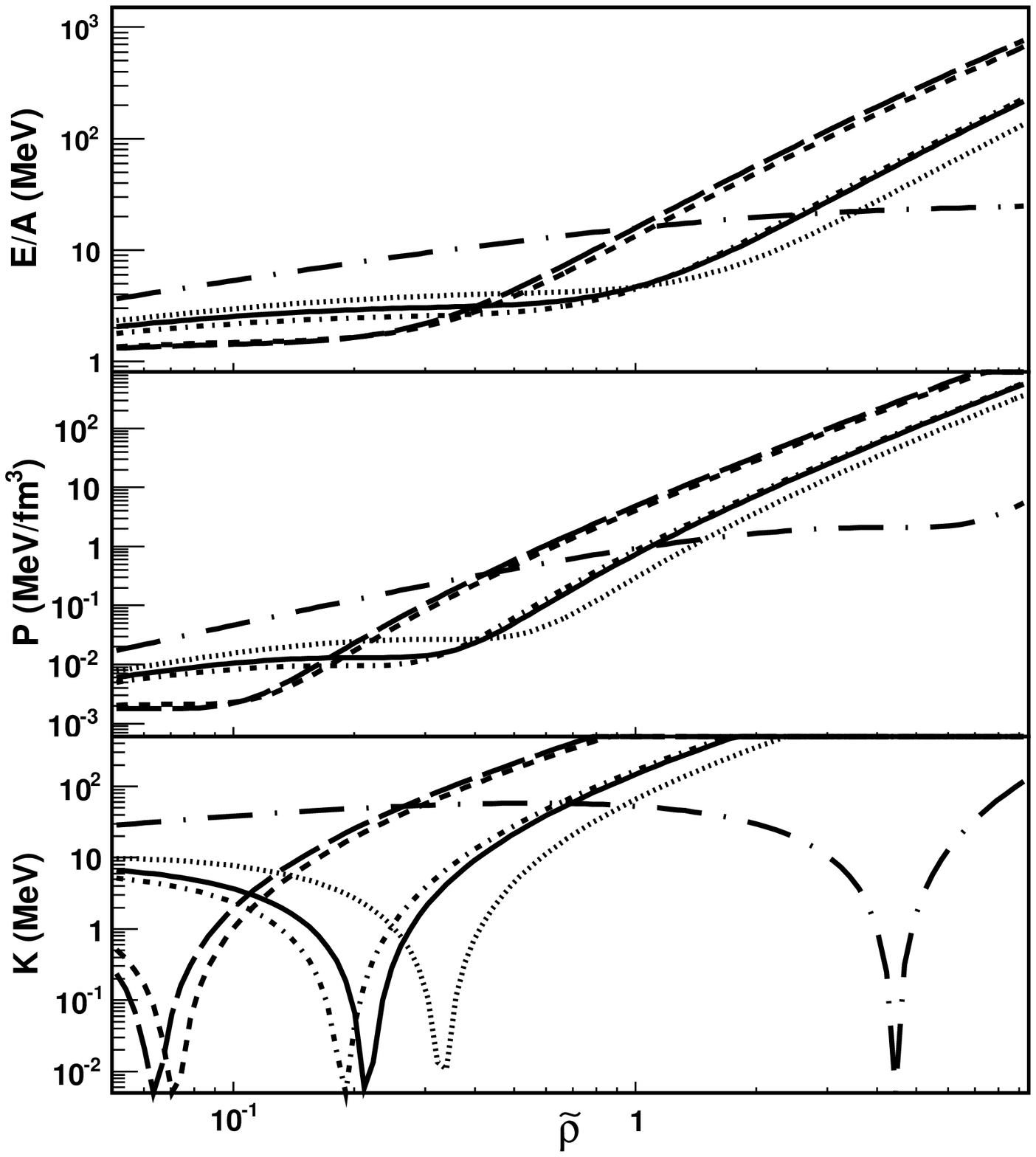}\includegraphics[width=0.5\columnwidth]{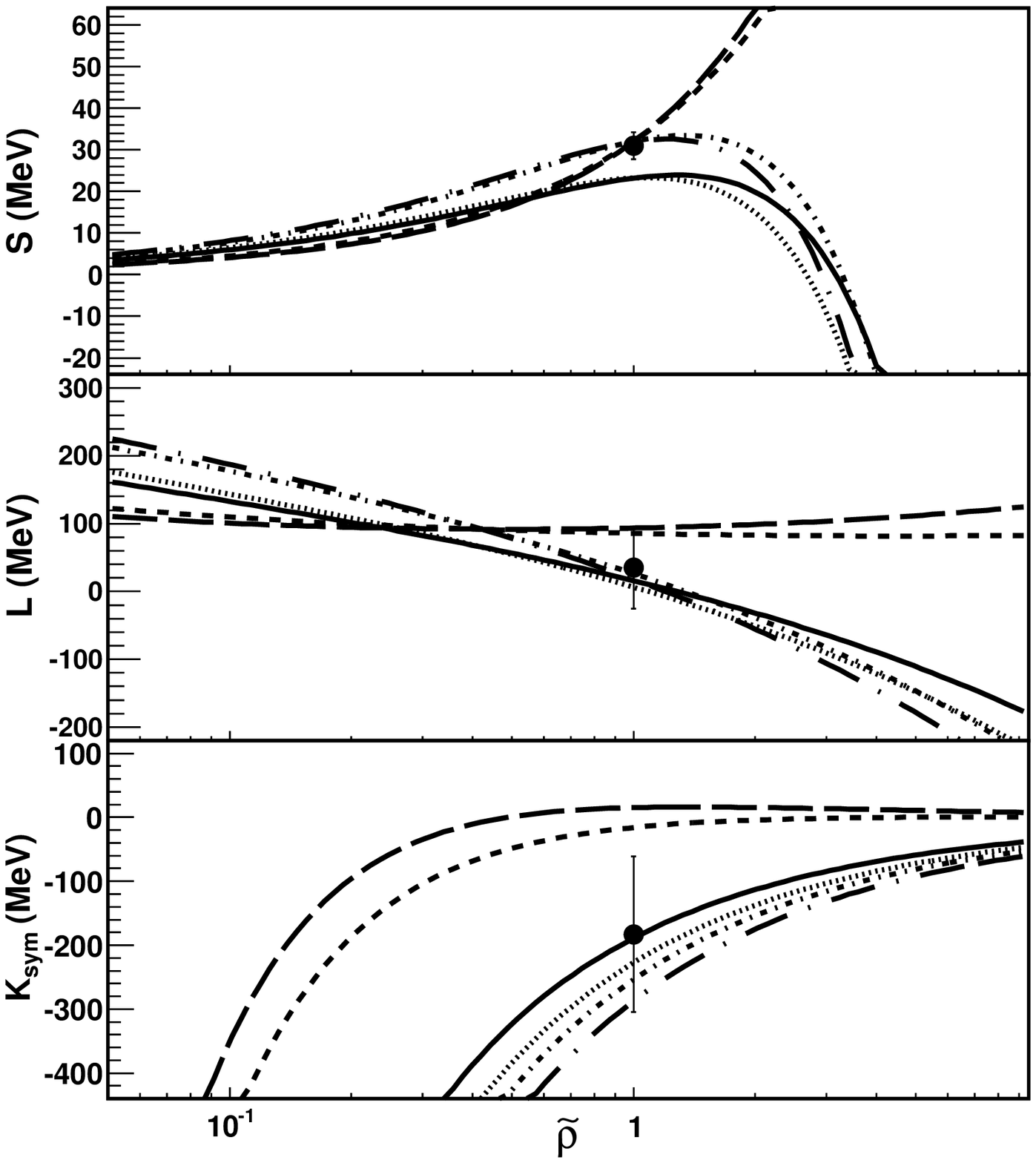}
 \end{tabular}
\end{minipage}
\begin{minipage}[t]{16.5 cm}
\caption{ The $E/A$, Pressure, $K$, $S$, $L$ and $K_{sym}$, for cases (1-6) in table \ref{c1c2}. The solid line refers to case (1), dotted line refers to case (2), the dash-dotted line refers to case (3), the dashed line refers to case (4), the long dashed line refers to case (5) and the long dash-dotted line refers to case (6).}\label{slk}
\end{minipage}
\end{center}
\end{figure}

The latter equation links the values of $L$ and $K_{sym}$ to the symmetry energy value and to $\sigma$. Recall that the value of $\sigma$ is connected to the nuclear compressibility $K$ and it is  greater than 1 in order to get nuclear saturation.
For $\sigma=2$ we have $K=380 MeV$. From equation (\ref{s0}) we can estimate $K_{sym}=-159 MeV$, for $S(\rho_0)=32 MeV$, $L=50 MeV$, $\sigma=\frac{4}{3}$; and $K_{sym}=-226 MeV$ if $\sigma=2$, which shows the sensitivity of $K_{sym}$ to the compressibility. Thus it is difficult to find physical quantities which depend on one ingredient rather than another one. Of course equations (\ref{l0}, \ref{ksym0}, \ref{s0}) refer to the particular NEOS we are using and these relations will change for different choices such as including momentum dependent forces. In the latter case we will still obtain similar relations with the addition of a new ingredient, the effective mass, which we will define in the following section \ref{mdeos}.

In order to illustrate the importance of the symmetry energy and its relevance, for instance to understand neutron stars, we will assume that asymmetric nuclear matter undergoes a second order phase transition already at zero temperature. This is fulfilled by solving the equations:
\begin{equation}
\left\{
\begin{array}{l}
S(\rho_0) = a_a(\infty), \\
\frac{\partial P}{\partial \rho}\vert_{\rho=\rho_c}=0,\\
\frac{\partial^2 P}{\partial \rho^2} \vert_{\rho=\rho_c}=0.
\end{array}
\right. \label{spmck225}
\end{equation}
We fix $m_{\chi}=m_c$ (for a second order phase transition) close to $1$, then we solve for $c_1, c_2$ and $\rho_c$. Typical results are included in table \ref{c1c2}.

\begin{table}
\begin{center}
\begin{minipage}[t]{16.5 cm}
\caption{Values of the parameters assuming a second order phase transition. Cases (8) and (9) refer  respectively, to  equation (\ref{ck225}) and a free Fermi gas approximation. For cases (5) (QLG) and (6) (QGP), two different critical densities are obtained for the same concentration and symmetry energy. The compressibility is $K=225 MeV$ for all cases.}
\label{c1c2}
\end{minipage}
\begin{tabular}{|c|c|c|c|c|c|c|c|}
\hline
&&&&&&&\\[-2mm]
&$S(\rho_0)$ (MeV)  &  $\tilde \rho_c $  &  $m_c$  &     $c_1$   &   $c_2$     &  $L(\rho_0)$ (MeV)   &   $K_{sym}(\rho_0)$ (MeV) \\[2mm]
\hline
1&23.2      &     0.216256       &    0.92     &  -0.492449 &  -0.607513  &      16.0929          &    -189.028                     \\
\hline   
2&23.2      &     0.331363       &    0.92     &  -0.583811 &  -0.749632  &       6.49982          &    -227.401                      \\
\hline
3&32         &     0.188629       &    0.78     &  -0.730633 &  -0.847651  &       26.2835          &    -253.866                      \\
\hline
4&32         &     0.0715581     &    0.94     &  -0.164079 &  0.0336557  &       85.7718          &    -15.913                      \\
\hline
5&32         &     0.0638193     &    0.98     &  -0.0898674 &  0.149095  &       93.5639          &    15.2557                      \\
\hline
6&32         &     4.43276         &    0.98     &  -0.809792  &  -0.970788  &       17.9718          &    -287.113                      \\
\hline
7&32         &     18.4109         &   1.1     &  -0.701832  &  -0.80285        &       29.3076          &    -241.769                      \\
\hline
8&12.5      &     -                     &   -         &  0.0              &  0.0                &       25                   &    -25                           \\
\hline
9&12.5      &     -                     &   -         & $(A_1=B_1=0)$ 0.0    &  0.0                &       25                   &    -25                           \\
\hline
\end{tabular}
\end{center}
\end{table}     

Let us start from the easiest cases (8) and (9) in table \ref{c1c2}. Case (9) refers to a pure Fermi gas, while case (8) refers to equation (\ref{ck225}) when $c_1=c_2=0$. Those two cases have exactly
 the same values of the physical quantities defined in equations (\ref{s}, \ref{l}, \ref{ksym}), but differ for the ground state binding energy and pressure. All the other cases display a second order phase transition at low densities for $S(\rho_0)=23.2 MeV$ and at very high densities for $S(\rho_0)=32 MeV$. It is very surprising such a sensitivity of the NEOS by just changing the value of the
 symmetry energy obtained from the mass formula. This gives two completely different scenarios for our equation of state. For the lower symmetry energy value, we can  think of a quantum liquid-gas (QLG) phase transition (see next section)
  occurring already at zero temperature but for almost pure neutron matter. On the other hand, the values obtained for the larger symmetry energy could be associated to a phase transition at high densities, from neutron matter
  to the quark-gluon plasma (QGP). Case (7) gives a second order phase transition for $m_c=1.1$ which is unphysical and we used to mimic a cross-over to the QGP at high densities. At present there is no universal consensus on the
  values of $L$ and $K_{sym}$, if we use a `popular' value for $L=50\pm40 MeV$, we see that most values reported in the table could be accepted. The value for $K_{sym}$ is even more undetermined.
    
In figure \ref{slk} we plot the different physical quantities described above versus densities for cases (1-6) from table \ref{c1c2}. The critical densities are easily recognized and on the right panels we have indicated some `current' estimates of $S$, $L$ and $K_{sym}$. We stress that the speed of sound is always less than $c$ the speed of light for the cases reported in the figure, in particular it is zero at the phase transition densities, which is especially relevant for those NEOS exhibiting a 
second order phase transition at high densities. From these simple estimates we hope we have further highlighted the importance of determining the symmetry energy.

\subsection{\it Momentum dependent NEOS}\label{mdeos}
An important ingredient of the NEOS is its momentum dependence. Most experimental data require a non local potential  due to the fact that nucleons are not elementary particles \cite{bonn1, bonn2, parisnn1, parisnn2, argonnenn1, argonnenn2}. A large variety of momentum dependent interactions have been proposed, especially
to reproduce low excitation energy phenomena such as giant resonances. Most of those interactions are valid for relative momenta of the order of
the Fermi momenta. The phenomenology of high energy heavy ion collisions requires that the momentum dependence should not diverge
for relative momenta higher than the Fermi one. Several momentum dependent NEOS (MNEOS) have been proposed \cite{bertschrep88, baoanrep08, aldorep94, aichelinrep91, baranrep05, bertsch2, rizzo1, gogny1, gale1, das2, prakash1}. For instance,  following \cite{bertschrep88, bertsch2}, the potential energy density is 
\begin{equation}
V(\rho) = \frac{a\rho^2}{2\rho_0}+\frac{b\rho^{\sigma+1}}{(\sigma+1)\rho_0^\sigma}+c\frac{\rho}{\rho_0}\int \frac{f({\bf r, p})}{1+(\frac{{\bf p-\langle p \rangle}}{\Lambda})^2}d^3p,
\end{equation}
where $\Lambda$ is a constant, ${\bf \langle p \rangle}$ is the average momentum at position {\bf r}.  $f({\bf r, p})$ is the nucleon density in the phase space. When $T=0$, 
\begin{equation}
f({\bf r, p}) = \frac{g}{h^3} \Theta (p_F-p) \Theta (R-r),
\end{equation}
and
\begin{equation}
\frac{g}{h^3} = \frac{3}{4\pi} \frac{\rho}{p_F^3}, \quad p_F=(\frac{3}{4\pi}\frac{h^3}{g})^{1/3}\rho^{1/3} = d\rho^{1/3},
\end{equation}
where $d=(\frac{3}{4\pi}\frac{h^3}{g})^{1/3}$, $\Theta$ is the step function.

The corresponding one body potential $U(\rho, {\bf p})=\partial V/\partial \rho_p$, which will be used to calculate the effective mass, is
\begin{eqnarray}
U(\rho, {\bf p}) &=& a \frac{\rho}{\rho_0}+b(\frac{\rho}{\rho_0})^\sigma+c\frac{1}{\rho_0}\int \frac{f({\bf r, p})}{1+(\frac{{\bf p-\langle p \rangle}}{\Lambda})^2}d^3p+\frac{c\rho}{\rho_0}\frac{1}{1+(\frac{{\bf p-\bf \langle p \rangle}}{\Lambda})^2} \nonumber\\
&=&a \frac{\rho}{\rho_0}+b(\frac{\rho}{\rho_0})^\sigma+c\frac{\rho}{\rho_0} \langle \frac{1}{1+(\frac{{p-\bf \langle p \rangle}}{\Lambda})^2} \rangle +\frac{c\rho}{\rho_0}\frac{1}{1+(\frac{{\bf p-\bf \langle p \rangle}}{\Lambda})^2}, \label{onebu}
\end{eqnarray}
where $\langle \frac{1}{1+(\frac{{p-\bf \langle p \rangle }}{\Lambda})^2} \rangle =\int \frac{f({\bf p})}{1+(\frac{{\bf p-\langle p \rangle}}{\Lambda})^2}d^3p$. 

For static nuclear matter, ${\bf \langle p\rangle }= 0$, $f({\bf p}) = \frac{3}{4\pi p_F^3}\Theta (p_F-p)$. Therefore
\begin{eqnarray}
\langle \frac{1}{1+(\frac{{p-\bf \langle p \rangle}}{\Lambda})^2} \rangle&=&\int \frac{f({\bf p})}{1+(\frac{{\bf p- \langle p \rangle}}{\Lambda})^2}d^3p \nonumber\\
&=& \frac{3}{4\pi p_F^3}\int_0^{p_F} \frac{4\pi p^2}{1+(\frac{p}{\Lambda})^2}dp \nonumber\\
&=& 3 (\frac{\Lambda}{p_F})^3[\frac{p_F}{\Lambda}-\tan^{-1}(\frac{p_F}{\Lambda})].
\end{eqnarray}
The energy per particle is:
\begin{eqnarray}
\frac{E}{A} &=& \frac{3}{5}\frac{p_F^2}{2m}  + \frac{V(\rho)}{\rho}\nonumber\\
&=&\frac{3}{5}\frac{p_F^2}{2m} + \frac{a}{2} \frac{\rho}{\rho_0}+\frac{b}{\sigma+1}(\frac{\rho}{\rho_0})^\sigma+c\frac{\rho}{\rho_0}\langle \frac{1}{1+(\frac{{p-\bf \langle p \rangle}}{\Lambda})^2}\rangle \nonumber\\
&=& \frac{3}{5}\frac{p_F^2}{2m} + \frac{a}{2} \frac{\rho}{\rho_0}+\frac{b}{\sigma+1}(\frac{\rho}{\rho_0})^\sigma+c\frac{\rho}{\rho_0} 3 (\frac{\Lambda}{p_F})^3[\frac{p_F}{\Lambda}-\tan^{-1}(\frac{p_F}{\Lambda})]. \label{eamdi}
\end{eqnarray}
Thus the pressure is given by:
\begin{eqnarray}
 P &=& \rho^2 \frac{\partial (\frac{E}{A})}{\partial \rho} \nonumber\\
 %\frac{2}{5}\frac{p_F^2}{2m}\rho+\frac{a}{2} \frac{\rho^2}{\rho_0}+\frac{b\sigma}{1+\sigma}\frac{\rho^{\sigma+1}}{\rho_0^\sigma}+c\frac{\rho^2}{\rho_0}\frac{1}{1+(\frac{{\bf p-\bf \langle p\rangle}}{\Lambda})^2} \nonumber\\
 &=&\frac{2}{5}\frac{d^2}{2m}\rho^{5/3}+\frac{a}{2} \frac{\rho^2}{\rho_0}+\frac{b\sigma}{1+\sigma}\frac{\rho^{\sigma+1}}{\rho_0^\sigma}+c\frac{\rho^2}{\rho_0}\frac{1}{1+(\frac{{p_F}}{\Lambda})^2}\nonumber\\
 &=&\frac{2}{5}\frac{p_F^2}{2m}\rho+\frac{a}{2} \frac{\rho^2}{\rho_0}+\frac{b\sigma}{1+\sigma}\frac{\rho^{\sigma+1}}{\rho_0^\sigma}+c\frac{\rho^2}{\rho_0}\frac{1}{1+(\frac{{p_F}}{\Lambda})^2},
 \end{eqnarray}
 and the compressibility: 
 \begin{eqnarray}
 K &=& 9 \frac{\partial P}{\partial \rho} \nonumber\\
 &=&9[\frac{2}{3}\frac{d^2}{2m}\rho^{2/3}+a \frac{\rho}{\rho_0}+b\sigma\frac{\rho^{\sigma}}{\rho_0^\sigma}+2c\frac{\rho}{\rho_0}\frac{1}{1+(\frac{{p_F}}{\Lambda})^2} - c\frac{\rho^2}{\rho_0}(\frac{1}{1+(\frac{{p_F}}{\Lambda})^2})^2(2\frac{{p_F}}{\Lambda^2})(\frac{1}{3}\frac{p_F}{\rho})]  \nonumber\\
 &=&9[\frac{2}{3}\frac{p_F^2}{2m}+a \frac{\rho}{\rho_0}+b\sigma\frac{\rho^{\sigma}}{\rho_0^\sigma}+2c\frac{\rho}{\rho_0}\frac{1}{1+(\frac{{p_F}}{\Lambda})^2} - \frac{2}{3}c(\frac{{p_F}}{\Lambda})^2\frac{\rho}{\rho_0}(\frac{1}{1+(\frac{{p_F}}{\Lambda})^2})^2].
 \end{eqnarray}
 Using the same conditions, equation (\ref{ck225_eq2}), we can fix the parameters entering the MNEOS. However, the number of constraints is not enough,
 thus $\Lambda$ is a free parameter. It determines how fast the momentum dependent part becomes negligible, and should
 be larger than the Fermi momentum. In refs. \cite{bertschrep88, bertsch2}, $a=-144.9 MeV$, $b=203.3 MeV$, $c=-75 MeV$, $\sigma = \frac{7}{6}$ and $\Lambda=1.5 p_{F0}=1.5d\rho_0^{1/3}$ giving a compressibility $K=215 MeV$. Using this MNEOS, the collective flow observed in heavy ion collisions is well  reproduced, in particular a higher
 flow is observed as compared to a local NEOS.  In order to reproduce a similar flow, local NEOS must have a much larger compressibility $K=380 MeV$ \cite{bertschrep88, aichelinrep91}. Notice that the force acting on a particle now contains a term which is the gradient of the mean field respect to ${\bf p}$ \cite{bertschrep88, baoanrep08, aldorep94, aichelinrep91, cassingrep90}.

%\subsection{\it Effective mass}\label{effm}
The definition of the effective mass is: 
\begin{equation}
\frac{m^*}{m}=[1+\frac{m}{p}\frac{dU}{dp} ]^{-1}_{p=p_{F0}}.
\end{equation}
%The single particle energy is
%\begin{eqnarray}
%U&=&a \frac{\rho}{\rho_0}+b(\frac{\rho}{\rho_0})^\sigma+c\frac{\rho}{\rho_0}<\frac{1}{1+(\frac{{p-\bf <p'>}}{\Lambda})^2}>+\frac{c\rho}{\rho_0}\frac{1}{1+(\frac{{p-\bf <p'>}}{\Lambda})^2} \nonumber\\
%&=& a \frac{\rho}{\rho_0}+b(\frac{\rho}{\rho_0})^\sigma+c\frac{\rho}{\rho_0} 3 (\frac{\Lambda}{p_F})^3[\frac{p_F}{\Lambda}-\tan^{-1}(\frac{p_F}{\Lambda})]+\frac{c\rho}{\rho_0}\frac{1}{1+(\frac{p}{\Lambda})^2} 
%\end{eqnarray}
Using equation (\ref{onebu}) gives: 
\begin{equation}
\frac{dU}{dp} = -c\frac{\rho}{\rho_0} \frac{2\frac{p}{\Lambda^2}}{[1+(\frac{p}{\Lambda})^2]^2}.
\end{equation}
Giving an effective mass at ground state density:
\begin{eqnarray}
\frac{m^*}{m}&=& [1+\frac{m}{p}\frac{dU}{dp} ]^{-1}_{p=p_{F0}} \nonumber\\
&=& \{ 1-\frac{m}{p_{F0}}c\frac{2\frac{p_{F0}}{\Lambda^2}}{[1+(\frac{p_{F0}}{\Lambda})^2]^2} \}^{-1} \nonumber\\
&=& 0.7.
\end{eqnarray}
Using the values of $\Lambda$ and $c$ reported above.

 \subsection{\it Asymmetric nuclear matter EoS with momentum dependence}\label{anmmd}
% For symmetric nuclear matter EoS with momentum dependence, the energy per particle is
% \begin{equation}
 %\frac{E}{A} = \frac{3}{5}\frac{p_F^2}{2m} + \frac{a}{2} \frac{\rho}{\rho_0}+\frac{b}{\sigma+1}(\frac{\rho}{\rho_0})^\sigma+c\frac{\rho}{\rho_0} 3 (\frac{\Lambda}{p_F})^3[\frac{p_F}{\Lambda}-\tan^{-1}(\frac{p_F}{\Lambda})]
 %\end{equation}
 Using equation (\ref{eamdi}) we can define the MNEOS for asymmetric nuclear matter as:
 \begin{eqnarray}
 \frac{E}{A} &=&  \frac{E}{A}\Big\vert_p\times \frac{Z}{A}+ \frac{E}{A}\Big\vert_n \times \frac{N}{A} \nonumber\\
 &=&\frac{3}{5}\frac{p_{Fp}^2}{2m}\times \frac{Z}{A} + \frac{3}{5}\frac{p_{Fn}^2}{2m}\times \frac{N}{A} \nonumber\\
 && + \frac{a}{2}(1+c_1m_\chi^2) \frac{\rho}{\rho_0}+\frac{b}{\sigma+1}(1+c_2m_\chi^2)(\frac{\rho}{\rho_0})^\sigma \nonumber\\
 &&+c\frac{\rho}{\rho_0} 3 (\frac{\Lambda}{p_{Fp}})^3[\frac{p_{Fp}}{\Lambda}-\tan^{-1}(\frac{p_{Fp}}{\Lambda})] \times \frac{Z}{A}+ c\frac{\rho}{\rho_0} 3 (\frac{\Lambda}{p_{Fn}})^3[\frac{p_{Fn}}{\Lambda}-\tan^{-1}(\frac{p_{Fn}}{\Lambda})]\times \frac{N}{A}. \label{asmeos1}
 \end{eqnarray}
 Since 
 \begin{equation}
 p_{Fp} = d_p \rho_p^{1/3}, \quad p_{Fn} = d_n \rho_n^{1/3}, \quad p_F = d \rho^{1/3},
 \end{equation}
 and 
 \begin{equation}
 d_p = d_n = 2^{1/3} d,
 \end{equation}
 \begin{equation}
 \rho_p = \frac{1-m_\chi}{2}\rho, \quad \rho_n= \frac{1+m_\chi}{2}\rho,
 \end{equation}
 \begin{equation}
 p_{Fp}=p_F (1-m_\chi)^{1/3}, \quad p_{Fn}=p_F (1+m_\chi)^{1/3}.
 \end{equation}
 Thus
 \begin{eqnarray}
 \frac{3}{5}\frac{p_{Fp}^2}{2m}\times \frac{Z}{A} + \frac{3}{5}\frac{p_{Fn}^2}{2m}\times \frac{N}{A}&=&\frac{3}{5} \frac{p_F^2}{2m} [(1-m_\chi)^{2/3} \times \frac{1-m_\chi}{2}+(1+m_\chi)^{2/3}\times \frac{1+m_\chi}{2}] \nonumber\\
 &=&\frac{3}{5} \frac{p_F^2}{2m}\frac{1}{2} [(1-m_\chi)^{5/3}+(1+m_\chi)^{5/3}] \nonumber\\
 &\approx&\frac{3}{5} \frac{p_F^2}{2m}(1+\frac{5}{9} m_\chi^2),\label{asmeos2}
 \end{eqnarray}
\begin{eqnarray}
c\frac{\rho}{\rho_0} 3 (\frac{\Lambda}{p_{Fp}})^3[\frac{p_{Fp}}{\Lambda}-\tan^{-1}(\frac{p_{Fp}}{\Lambda})] \times \frac{Z}{A} &=& c \frac{\rho}{\rho_0} 3 (\frac{\Lambda}{p_{F}})^3\frac{1}{1-m_\chi}\nonumber\\
&&\times [\frac{p_{F}}{\Lambda} (1-m_\chi)^{1/3} -\tan^{-1}(\frac{p_{F}}{\Lambda} (1-m_\chi)^{1/3})] \times \frac{1-m_\chi}{2} \nonumber\\
&=& \frac{1}{2}c \frac{\rho}{\rho_0}3(\frac{\Lambda}{p_{F}})^3[\frac{p_{F}}{\Lambda} (1-m_\chi)^{1/3} -\tan^{-1}(\frac{p_{F}}{\Lambda} (1-m_\chi)^{1/3})], \label{asmeos3}
\end{eqnarray}
\begin{equation}
 c\frac{\rho}{\rho_0} 3 (\frac{\Lambda}{p_{Fn}})^3[\frac{p_{Fn}}{\Lambda}-\tan^{-1}(\frac{p_{Fn}}{\Lambda})]\times \frac{N}{A}  = \frac{1}{2}c \frac{\rho}{\rho_0}3(\frac{\Lambda}{p_{F}})^3[\frac{p_{F}}{\Lambda} (1+m_\chi)^{1/3} -\tan^{-1}(\frac{p_{F}}{\Lambda} (1+m_\chi)^{1/3})].\label{asmeos4}
\end{equation}
Substituting equations (\ref{asmeos2}, \ref{asmeos3}, \ref{asmeos4}) into equation (\ref{asmeos1}), one can obtain the energy per particle
\begin{eqnarray}
\frac{E}{A} &=&\frac{3}{5} \frac{p_F^2}{2m}(1+\frac{5}{9} m_\chi^2) + \frac{a}{2}(1+c_1m_\chi^2) \frac{\rho}{\rho_0}+\frac{b}{\sigma+1}(1+c_2m_\chi^2)(\frac{\rho}{\rho_0})^\sigma \nonumber\\
&&+\frac{1}{2}c \frac{\rho}{\rho_0}3(\frac{\Lambda}{p_{F}})^3[\frac{p_{F}}{\Lambda} (1-m_\chi)^{1/3} -\tan^{-1}(\frac{p_{F}}{\Lambda} (1-m_\chi)^{1/3})]\nonumber\\
&&+\frac{1}{2}c \frac{\rho}{\rho_0}3(\frac{\Lambda}{p_{F}})^3[\frac{p_{F}}{\Lambda} (1+m_\chi)^{1/3} -\tan^{-1}(\frac{p_{F}}{\Lambda} (1+m_\chi)^{1/3})].
\end{eqnarray}
Where we have followed the same philosophy of the local NEOS. The important difference is due to the momentum dependent interaction
which results in another contribution to the symmetry energy because of the difference of Fermi momenta of protons and neutrons when their densities
are different. 

Using equation (\ref{avEs}) the symmetry energy is:
\begin{eqnarray}
S(\rho) &=&\frac{1}{3} \frac{p_F^2}{2m}+\frac{a}{2}c_1 \frac{\rho}{\rho_0}+\frac{b}{1+\sigma}c_2 (\frac{\rho}{\rho_0})^\sigma\nonumber\\
&&+\frac{3}{2}c \frac{\rho}{\rho_0}(\frac{\Lambda}{p_{F}})^3[2^{1/3}\frac{p_{F}}{\Lambda} -\tan^{-1}(2^{1/3}\frac{p_{F}}{\Lambda})]-3c \frac{\rho}{\rho_0}(\frac{\Lambda}{p_{F}})^3[\frac{p_{F}}{\Lambda} -\tan^{-1}(\frac{p_{F}}{\Lambda})].
\end{eqnarray}
Thus 
\begin{eqnarray}
L(\rho) &=& 3\{\frac{2}{9}\frac{p_F^2}{2m}\frac{1}{\rho/\rho_0}+c_1\frac{a}{2}+c_2\frac{b\sigma}{1+\sigma}(\frac{\rho}{\rho_0})^{\sigma-1}\nonumber\\
&&+\frac{c}{2^{2/3}}(\frac{\Lambda}{p_{F}})^2[1-\frac{1}{1+2^{2/3}(\frac{p_{F}}{\Lambda})^2}]-c(\frac{\Lambda}{p_{F}})^2[1-\frac{1}{1+(\frac{p_{F}}{\Lambda})^2}]\},
\end{eqnarray}
\begin{eqnarray}
K_{sym}(\rho) &=& 9\{-\frac{2}{27}\frac{p_F^2}{2m}\frac{1}{(\rho/\rho_0)^2}+c_2\frac{b\sigma(\sigma-1)}{1+\sigma}(\frac{\rho}{\rho_0})^{\sigma-2}\nonumber\\
&&+\frac{2^{1/3}c}{3}\frac{1}{\rho/\rho_0}[-(\frac{\Lambda}{p_{F}})^2(1-\frac{1}{1+2^{2/3}(\frac{p_{F}}{\Lambda})^2})+\frac{2^{2/3}}{(1+2^{2/3}(\frac{p_{F}}{\Lambda})^2)^2}]\nonumber\\
&&-\frac{2c}{3}\frac{1}{\rho/\rho_0}[-(\frac{\Lambda}{p_{F}})^2(1-\frac{1}{1+(\frac{p_{F}}{\Lambda})^2})+\frac{1}{(1+(\frac{p_{F}}{\Lambda})^2)^2}]\}.
\end{eqnarray}

Similarly to equation (\ref{s0}), we obtain
\begin{eqnarray}
S(\rho_0)&=&\frac{1}{9}\frac{p_{F0}^2}{2m}-\frac{2}{27\sigma}\frac{p_{F0}^2}{2m}+\frac{L}{3}-\frac{K_{sym}}{9\sigma}\nonumber\\
&&+\frac{3}{2}c (\frac{\Lambda}{p_{F0}})^3[2^{1/3}\frac{p_{F0}}{\Lambda} -\tan^{-1}(2^{1/3}\frac{p_{F0}}{\Lambda})]-3c(\frac{\Lambda}{p_{F0}})^3[\frac{p_{F0}}{\Lambda} -\tan^{-1}(\frac{p_{F0}}{\Lambda})]\nonumber\\
&&-\frac{c}{2^{2/3}}(\frac{\Lambda}{p_{F0}})^2[1-\frac{1}{1+2^{2/3}(\frac{p_{F0}}{\Lambda})^2}]-c(\frac{\Lambda}{p_{F0}})^2[1-\frac{1}{1+(\frac{p_{F0}}{\Lambda})^2}]\nonumber\\
&&+\frac{1}{\sigma} \{\frac{2^{1/3}c}{3}[-(\frac{\Lambda}{p_{F0}})^2(1-\frac{1}{1+2^{2/3}(\frac{p_{F0}}{\Lambda})^2})+\frac{2^{2/3}}{(1+2^{2/3}(\frac{p_{F0}}{\Lambda})^2)^2}]\nonumber\\
&&-\frac{2c}{3}[-(\frac{\Lambda}{p_{F0}})^2(1-\frac{1}{1+(\frac{p_{F0}}{\Lambda})^2})+\frac{1}{(1+(\frac{p_{F0}}{\Lambda})^2)^2}]\}\nonumber\\
&=&\frac{5}{27}\bar\epsilon_f-\frac{10}{81}\frac{\bar\epsilon_f}{\sigma}+\frac{L}{3}-\frac{K_{sym}}{9\sigma}\nonumber\\
&&+\frac{3}{2}c (\frac{\Lambda}{p_{F0}})^3[2^{1/3}\frac{p_{F0}}{\Lambda} -\tan^{-1}(2^{1/3}\frac{p_{F0}}{\Lambda})]-3c(\frac{\Lambda}{p_{F0}})^3[\frac{p_{F0}}{\Lambda} -\tan^{-1}(\frac{p_{F0}}{\Lambda})]\nonumber\\
&&-\frac{c}{2^{2/3}}(\frac{\Lambda}{p_{F0}})^2[1-\frac{1}{1+2^{2/3}(\frac{p_{F0}}{\Lambda})^2}]-c(\frac{\Lambda}{p_{F0}})^2[1-\frac{1}{1+(\frac{p_{F0}}{\Lambda})^2}]\nonumber\\
&&+\frac{1}{\sigma} \{\frac{2^{1/3}c}{3}[-(\frac{\Lambda}{p_{F0}})^2(1-\frac{1}{1+2^{2/3}(\frac{p_{F0}}{\Lambda})^2})+\frac{2^{2/3}}{(1+2^{2/3}(\frac{p_{F0}}{\Lambda})^2)^2}]\nonumber\\
&&-\frac{2c}{3}[-(\frac{\Lambda}{p_{F0}})^2(1-\frac{1}{1+(\frac{p_{F0}}{\Lambda})^2})+\frac{1}{(1+(\frac{p_{F0}}{\Lambda})^2)^2}]\},
\end{eqnarray}
where the relation $\bar\epsilon_f=\frac{3}{5}\frac{p_{F0}^2}{2m}$ has been used. The latter equation shows the connections among the various terms of the NEOS including the momentum dependent part through the parameter $\Lambda$. The previous result, equation (\ref{s0}), can be easily recovered by taking the limit $\Lambda\rightarrow 0$. A simple estimate gives
 $K_{sym}=-19.6 MeV$ quite different from the estimate from  equation (\ref{s0}). Changing the symmetry energy of 1 MeV, changes the value of $K_{sym}=-30.1 MeV$, thus it is very sensitive
  to small changes.

Note that previously we have assumed that the same effective mass for both neutrons and protons. This might be not true and different options are discussed in the literature \cite{baoanrep08, baranrep05, rizzo1, zuoeffm, baoaneffm, liueffm, faesslerEM1, faesslerEM2, ad_zhang}. From the definition of effective mass, we can calculate it for the asymmetric part as well and we could  have different values of the $\Lambda$ parameters for $n$ and $p$ in the MNEOS discussed above. Detailed experimental data is needed to fix this point as well \cite{youngsthesis, couplandthesis}.

 \begin{figure}[tb]
\begin{center}
\begin{minipage}[t]{12 cm}
\includegraphics[width=1.0\columnwidth]{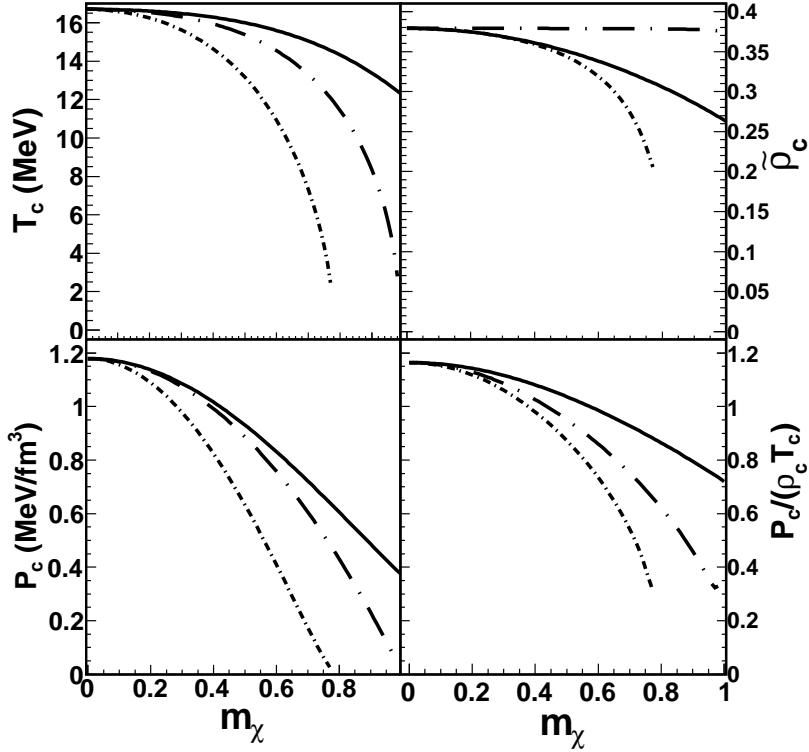}
\end{minipage}
\begin{minipage}[t]{16.5 cm}
\caption{$T_c$, $\tilde \rho_c$, $P_c$ and $\frac{P_c}{\rho_cT_c}$ versus $m_\chi$ for the modified CK225 NEOS using the low temperature Fermi gas approximation. 
Solid line for $(c_1=0, c_2=0)$ [case (8)], short dash-dotted line for  $(c_1=-0.730633, c_2=0.847651)$ [case (3)] and long dash-dotted line
 for  $(c_1=-0.501529, c_2=-0.66137)$.}\label{ck225fermit}
\end{minipage}
\end{center}
\end{figure}

\begin{figure}[tb]
\begin{center}
\begin{minipage}[t]{12 cm}
\includegraphics[width=1.0\columnwidth]{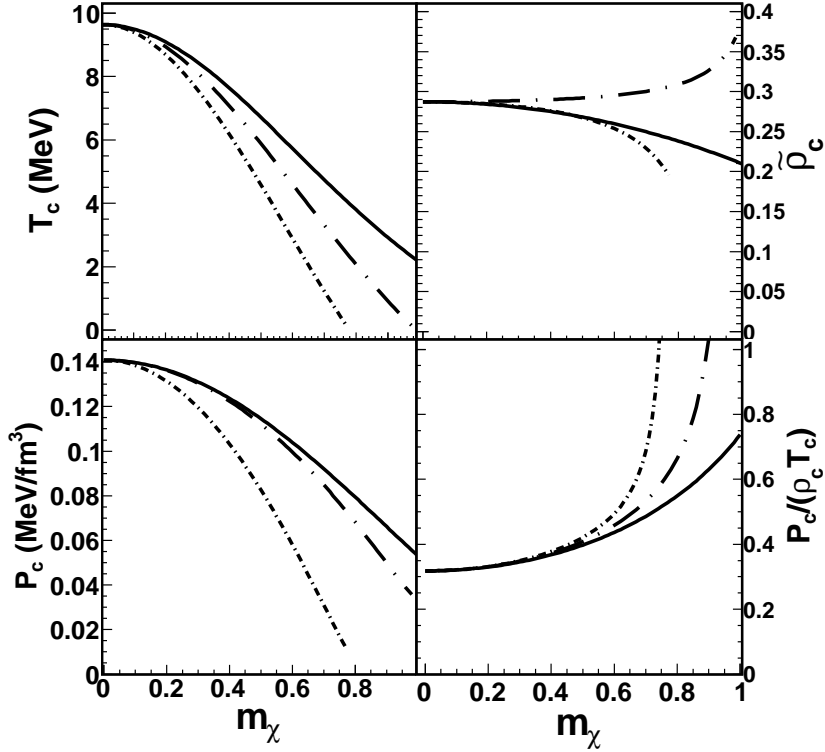}
\end{minipage}
\begin{minipage}[t]{16.5 cm}
\caption{Similar to figure \ref{ck225fermit} but for the classical approximation.}\label{ck225classt}
\end{minipage}
\end{center}
\end{figure}

\section{ The Nuclear Equation of State at Finite Temperatures } \label{eost}
At finite temperatures the NEOS can be simply obtained by modifying the kinetic part in equation (\ref{ck225}) for momentum independent interactions. The kinetic part can be obtained by solving the integral in equation (\ref{etfs}) and using a finite temperature Fermi-Dirac  distribution instead of a $\Theta$-function. For momentum dependent interactions, the potential energy must be obtained by folding the relevant integrals with the finite temperature 
 distributions. Various calculations can be found in the literature and we refer to those for details \cite{baoanrep08, dasrep05, hua1, su1}.

 It is instructive to derive the NEOS at finite temperatures in two limits. First let us assume that the ratio $\frac{T}{\epsilon_f}$ is much smaller than one and use the low temperature Fermi approximation. 
 The energy per particle can be written as:
 \begin{eqnarray}
\frac{E}{A}& =&(1+\frac{5}{9} m_{\chi}^2) \bar \epsilon_f \tilde \rho^{2/3} + (1+c_1 m_{\chi}^2) \frac{A_1}{2} \tilde \rho  + (1+c_2 m_{\chi}^2) \frac{B_1}{1+\sigma} \tilde \rho^\sigma + \frac{1}{1+\frac{5}{9} m_{\chi}^2} a_0T^2 \frac{1}{\tilde \rho^{2/3}},
\label{modifiedfermi}
\end{eqnarray}
where $a_0=1/13.3 MeV^{-1}$. 

For each value of $m_{\chi}$ the critical point can be calculated by finding the roots of the following equations, see also equation (\ref{spmck225})
\begin{equation}
\left\{
\begin{array}{l}
\frac{\partial P}{\partial \rho}\vert_{\rho=\rho_c}=0,\\
\frac{\partial^2 P}{\partial \rho^2} \vert_{\rho=\rho_c}=0.
\end{array}
\right.
\end{equation}

\begin{figure}[tb]
\begin{center}
\begin{minipage}[t]{12 cm}
 \includegraphics[width=1.0\columnwidth]{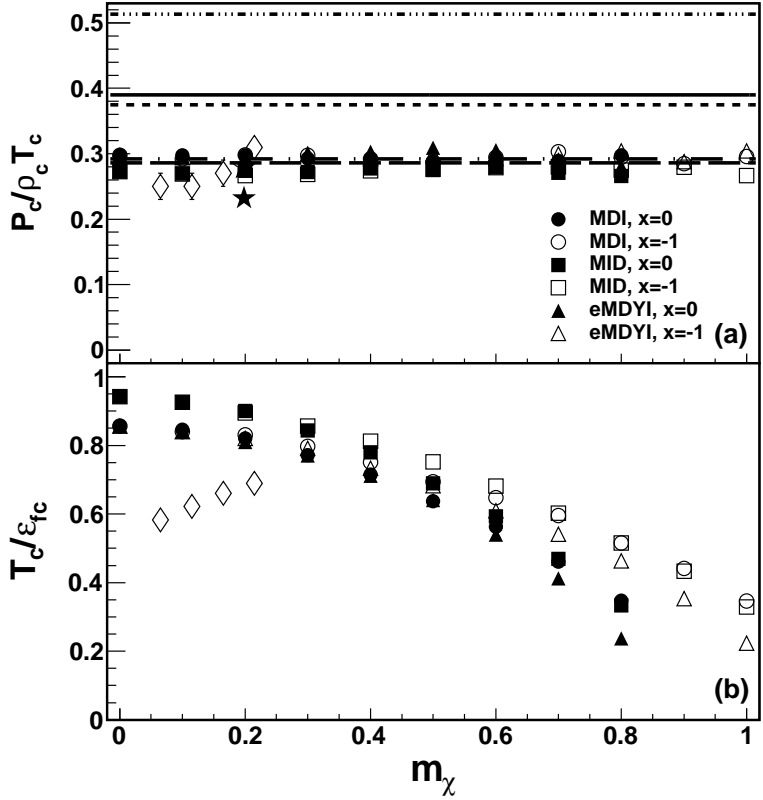}
\end{minipage}
\begin{minipage}[t]{16.5 cm}
\caption{The $\frac{P_c}{\rho_c T_c}$ (top) and $\frac{T_c}{\epsilon_{fc}}$ (bottom) versus $m_\chi$. The results for MDI, MID and eMDYI interactions are obtained from \cite{baoanrep08, junxu} which we refer for details. The dash-triple-dotted line is the result for an ideal Bose gas \cite{landau, khuang, pathria}, the solid line is the result from the Fisher model \cite{huang}, the dotted line is the result from a Van der Waals gas, the dash-dotted line is the result from Guggenheim \cite{gugg}, the dashed line is the result from the CMD model \cite{aldo2}, the diamond  from \cite{justin}, 
the open star from \cite{elliott1, elliott2} and the solid star from \cite{elliott3}. The $m_\chi$ for \cite{elliott1, elliott2, elliott3} is estimated from ${}^{197}Au$. \label{prhotc}}
\end{minipage}
\end{center}
\end{figure}

Those conditions, if fulfilled, give the critical temperature and the critical density, for fixed $m_{\chi}$, of the nuclear system and can be associated to a QLG phase transition. This is consistent with the description of the ground state of the nucleus as a quantum liquid drop, but we have to stress the fact that we have two liquid components: neutrons and protons. Using the low temperature approximation we get $T_c\approx 18MeV$ and $\rho_c\approx \frac{1}{3}\rho_0$ for symmetric nuclear matter. These values are consistent with those quoted in the literature which would suggest that our approximation is reasonable. However, when we look at the pressure and the ratio  $\frac{P_c}{\rho_c T_c}$, we find surprising values as illustrated in figure \ref{ck225fermit}. 
The value of the ratio for symmetric nuclear matter is larger than one and decreases for increasing asymmetries. Similar behavior for the other quantities plotted in the figure. In particular the NEOS
 in table \ref{c1c2}, case (3),  gives a critical temperature equal to zero at $m_{\chi}=0.78$. Experimental values of the critical ratio range somewhat around 0.28 for real gases \cite{gugg} to $0.4$ for the Van der Waals EOS \cite{landau, khuang, pathria}. This implies that our low temperature expansion is not yet convergent. If we include corrections to $T^4$ we obtain a shift to $T_c \approx 11MeV$ and values of the critical ratios less than one!
This implies that the low temperature approximations converge very slowly, a feature which should be kept in mind when dealing with quantities near the critical point for a QLG phase transition.

We can investigate a second limit which is the classical one. Then the modified CK225 EoS becomes
\begin{equation}
P  = \rho_0 [(1+\frac{5}{9} m_{\chi}^2) \bar \epsilon_f \frac{2}{3}\tilde \rho^{5/3} + (1+c_1 m_{\chi}^2) \frac{A_1}{2} \tilde \rho^2  + (1+c_2 m_{\chi}^2) \frac{B_1 \sigma}{1+\sigma} \tilde \rho^{\sigma+1}+(1+\frac{5}{9}m_{\chi}^2)^{3/2}\tilde \rho T ] . \label{modifiedclass}
\end{equation}
In equation (\ref{modifiedclass}) we have used the relation $\tilde \rho \rightarrow \tilde \rho(1+\frac{5}{9}m^2_\chi)^{3/2}$ suggested from the Fermi gas. This is for the purpose of illustration in order to include a concentration dependence in the temperature part of the NEOS. The critical values obtained in this extreme limit are reported in figure \ref{ck225classt}, now the behavior as function of $m_{\chi}$ is in contrast with the low temperature approximation. The critical ratio is close to 0.3 but increases for neutron rich nuclear matter. The critical $T_c\approx 9MeV$ for symmetric matter, is about a factor of two below the previous estimate. The message is that in these ranges of temperatures and densities it is dangerous to use either purely classical or low 
 $\frac{T}{\epsilon_f}$ approximations: even though the behavior might seem reasonable in a given region, it is not so in another.
 
 In figure \ref{prhotc} we display the results of  exact calculations for different NEOS \cite{baoanrep08, gugg, junxu,  justin, elliott1, elliott2, elliott3, aldo2, huang, mekjianasyeos}. Maybe not surprising, the critical ratio is constant as function of  $m_{\chi}$
  which suggests that the matter properties at the critical point are universal, i.e. independent of concentration. Furthermore, the calculated values are in agreement with real gases \cite{gugg}, and other values from the literature
  from heavy ion collision analysis that we will discuss later \cite{justin, elliott1, elliott2, elliott3}. Results from some theoretical models as well as the Van der Waals gas (which overestimates the ratio) are also displayed. The bottom part of the
  figure displays the behavior of $\frac{T_c}{\epsilon_{fc}}$ as function of concentration. Such a ratio becomes very small for increasing concentrations, which explains why the low temperature approximation improves
  for large concentrations, see figure \ref{ck225fermit}, while the opposite is true for the classical approximation reported in figure \ref{ck225classt}. 
  The values of the critical temperature and density are consistent with those estimated in the low temperature limit and decrease for increasing concentration, similar to figure \ref{ck225fermit}.
  
 It is important to stress that the features discussed above are valid in the mean field approximation. Such an approximation is questionable in the instability region and near the critical point. The values of the critical exponents are not correct \cite{landau, hua1}, for instance if we expand the compressibility $K(\rho, T)$ around the critical point:
\begin{eqnarray}
K(\rho, T) &=& K(\rho_c, T_c) \nonumber\\
&&+ K^{(1, 0)}(\rho_c, T_c) (\rho-\rho_c) + \frac{1}{2} K^{(2, 0)}(\rho-\rho_c)^2 + \frac{1}{6}K^{(3, 0)} (\rho-\rho_c)^3+ \frac{1}{24}K^{(4, 0)} (\rho-\rho_c)^4 \nonumber\\
&& +[K^{(0, 1)}+K^{(1, 1)}(\rho_c, T_c)(\rho-\rho_c) + \frac{1}{2}K^{(2, 1)}(\rho-\rho_c)^2 +\frac{1}{6}K^{(3, 1)}(\rho-\rho_c)^3+\frac{1}{24}K^{(4, 1)}(\rho-\rho_c)^4]\nonumber\\
&&\times (T-T_c) \nonumber\\
&=&0,
\end{eqnarray}
where $K^{(i, j)}$ are the $i$, $j$ derivatives respect to $\rho$ and $T$.  The terms
%\begin{equation}
$K^{(1, 1)}(\rho_c, T_c)$, $K^{(2, 1)}(\rho_c, T_c)$, $K^{(3, 1)}(\rho_c, T_c)$, $K^{(4, 1)}(\rho_c, T_c)$ can be neglected since they are of higher order in $(\rho-\rho_c)\times(T-T_c)$.
%\end{equation}
Using equation (\ref{spmck225}) we also have:
\begin{equation}
K(\rho_c, T_c) = K^{(1, 0)}(\rho_c, T_c)= 0.
\end{equation}
Thus \cite{hua1}:
\begin{equation}
\frac{1}{2} K^{(2, 0)}(\rho-\rho_c)^2+K^{(0, 1)}(T-T_c)=0,
\end{equation}
from which we recover $\beta=1/2$, i.e. one of the``classical" or mean field value for the critical exponents \cite{ landau, khuang, pathria, tokebeta05}.  
 As discussed by K. Huang \cite{khuang}, ``when you do not know what to do,  try the mean field approximation first". 
  We learn many lessons from the mean field, but we cannot stop there, and we should try to push forward. One possible path  is the use
  of molecular dynamics models which take into account quantum features such as the Pauli principle. It is clear that all of these are approximations and should be taken  ``cum grano salis".

\subsection{\it Finite Sizes \label{mf}}
We can study the properties of the NEOS at finite temperature by using heavy ion collisions at beam energies around the Fermi energy. Two major problems arise when doing that:
1) nuclei are finite;
2) Coulomb forces must be included and those are long range forces. Furthermore, it is meaningless to speak about an NEOS in presence of a long range force. However, in some approximations and some 
physical conditions (low densities, high temperatures), we can correct for Coulomb effects and constrain the NEOS. 
Critical behavior has been observed in finite size systems, for instance in percolation models which we will use as reference, as well as in experimental data such as \cite{aldo3, aldoworld}. There have been many attempts to correct for finite sizes \cite{elliott1, elliott2, aldo2, joe3, joe4} and we will discuss here the mean field approximation of ref. \cite{mekjian0}. The model is essentially based on the Hill-Wheeler approximation \cite{hill1} modified to take into account finite temperatures.

\begin{figure}[tb]
\begin{center}
\begin{minipage}[t]{12 cm}
 \includegraphics[width=1.0\columnwidth]{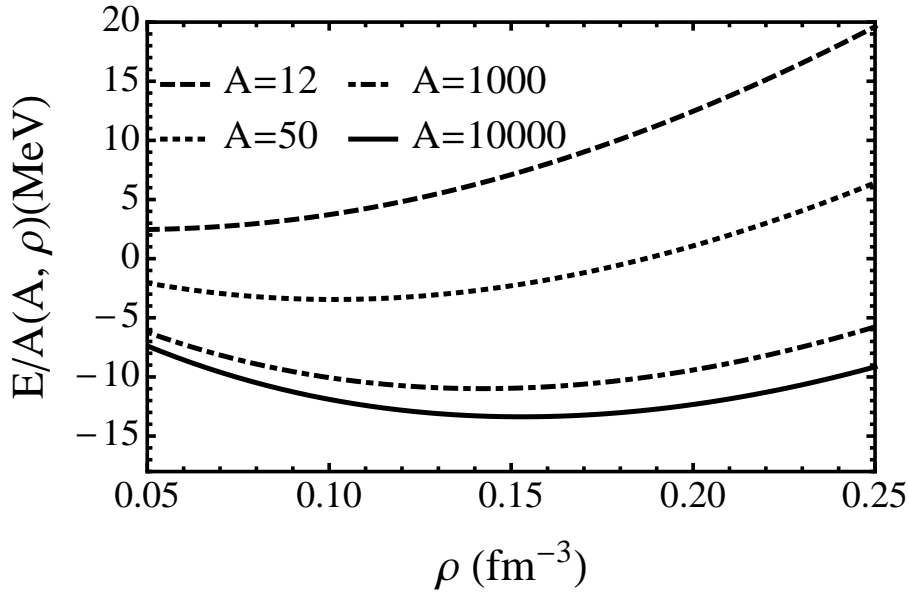}
\end{minipage}
\begin{minipage}[t]{16.5 cm}
\caption{Energy per particle versus density at zero temperature in the interacting Fermi gas model plus finite size corrections.}\label{eafinitesize}
\end{minipage}
\end{center}
\end{figure}

\begin{figure}[tb]
\begin{center}
\begin{minipage}[t]{12 cm}
 \includegraphics[width=1.1\columnwidth]{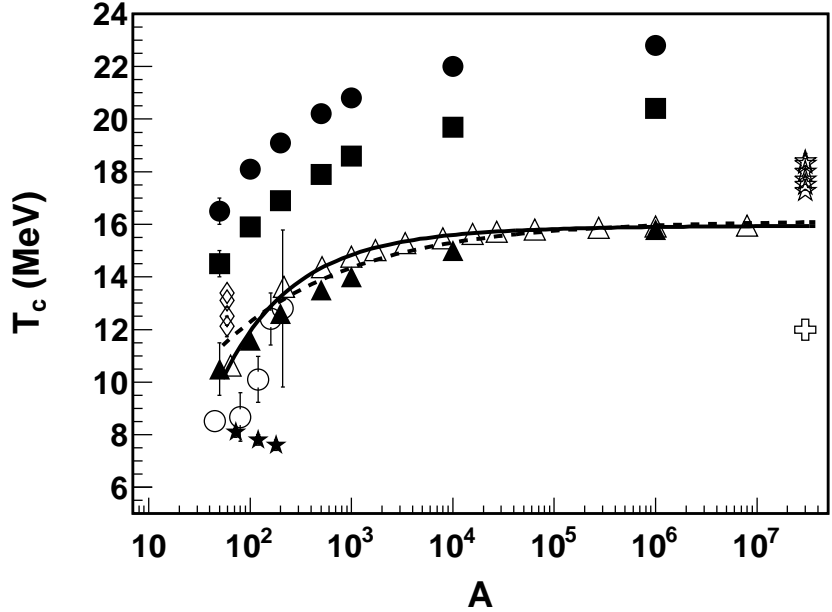}
\end{minipage}
\begin{minipage}[t]{16.5 cm}
\caption{The critical temperature $T_c$ versus the mass number $A$. Solid circles refer to ZR1, solid squares refer to ZR2, solid triangles refer to ZR3-NEOS \cite{mekjian0}; open triangles refer to the percolation data \cite{perc3}; open stars refer to the experimental results from Elliott \cite{elliott1, elliott2}, solid stars refer to the experimental results from Elliott \cite{elliott3}; open diamonds refer to Mabiala's experimental results \cite{justin}; open circles refer to Natowitz's results \cite{joe3, joe4}; open cross refers to Ono's AMD
 calculations \cite{onotc}. The percolation results are fitted with  $T_c(A)=15.949-\frac{4.6149}{A^{1/3}}-\frac{65.305}{A^{2/3}}$ (solid line) and
  with $T_c(A)=16.1346-\frac{17.8664}{A^{1/3}}$ (dashed line).}\label{finitetca}
\end{minipage}
\end{center}
\end{figure}

First we can use the proposed NEOS with compressibility $225 MeV$ and for symmetric matter, just by adding the finite size modifications to the simple Fermi gas model, equation (\ref{ck225}). The results are displayed in figure \ref{eafinitesize} which depicts the energy per particle as function of density at zero temperature for different finite nuclei. As it can be seen, not only the binding energy shifts towards $-8 MeV/A$, the value for finite nuclei, but also the equilibrium densities slightly decrease.  However, very small nuclei are unbound, thus the model is clearly just qualitative. Furthermore, adding the Coulomb force will decrease the binding energy further while pairing might go in the opposite direction. Of course finite size effects, i.e. the surface, will modify the potential term as well. However, the goal here is to show that finite size effects might modify the critical properties and the eventual phase transition of the nucleus. The calculated binding energies in the model are reasonable up to mass 50.

For finite temperatures, the starting point is the partition function
\begin{equation}
Q(\beta) = \frac{V e^{-\beta \epsilon_0}}{\lambda_T^3}(1-\frac{\lambda_T}{4}\frac{S}{V}+\frac{\lambda^2_T}{8}\frac{L}{V}),
\end{equation}
where $\lambda_T = (\frac{2\pi \hbar^2}{mkT})^{1/2}$ is the thermal wavelength of a nucleon \cite{mekjian0}, and the finite size corrections are given in equation (\ref{dnfs}). 
Standard thermodynamics techniques are used to calculate the NEOS for finite nuclei in ref. \cite{mekjian0} and the results for three different forms of the NEOS 
are reported in figure \ref{finitetca} where $T_c$ is plotted as function of the system size $A$. The three different NEOS give critical temperatures ranging from $14 MeV$ to $22 MeV$ for infinite systems. 
For systems of mass $A=50$ the critical temperature decreases as low as $10 MeV$.

We can estimate the behavior of the critical point as function of mass in a simple bond percolation model \cite{aldonc00, perc3, perc1, perc2, perc4}. 
If we assume that the critical temperature is proportional to the critical percolation bond probability \cite{perc4}:
\begin{equation}
T_{perc} \propto \frac{1}{p_c}. \label{tperc}
\end{equation}
We can normalize equation (\ref{tperc}) to any of the NEOS reported in the figure \ref{finitetca}. We can see that the behavior of the NEOS and the percolation model is surprisingly similar, giving a quick method to estimate the result of an infinite system, once the critical temperature for some masses are known. 
In the figure are also reported  some experimental results obtained from heavy ion collisions and different system sizes \cite{justin, elliott1, elliott2, elliott3, joe3, joe4}. The parametrization for the percolation model given in figure \ref{finitetca},
 inspired from the finite size Fermi gas results, could be used to derive $T_c(\infty)$. Of course, this discussion is for illustration only, since the critical temperature should depend on the $m_{\chi}$ of the emitting source as predicted by mean-field calculations, see figures (\ref{ck225fermit}, \ref{ck225classt}, \ref{prhotc}, \ref{eafinitesize}), and experimental results \cite{justin}. Notice, however, that the results of ref. \cite{justin} are obtained for fixed mass, varying neutron concentrations and are influenced by Coulomb effects.
 Refs. \cite{joe3, joe4} results are obtained by changing the mass size but no information is given on the values of 
$m_{\chi}$. Similarly for ref. \cite{elliott3} results,  solid stars, furthermore those authors have devised a method to correct for finite sizes \cite{elliott1, elliott2}. 
We will discuss the different methods more in detail below. 
Theoretical calculations have been performed in ref. \cite{onotc} using the Antisymmetrized Molecular Dynamics model (AMD) with periodic boundary conditions. The estimated critical temperature is about $12 MeV$. We stress
that using a similar NEOS but in a mean field approximation gives a critical temperature of about $18 MeV$ as discussed in the previous section. A similar decrease in temperature has been observed in Classical Molecular Dynamics (CMD) calculations which were compared
to the mean field approximation for the same interactions \cite{aldo2, schlagel1, belkacem1}. In the interesting work \cite{mekjian0} corrections due to Coulomb and concentrations are discussed as well. 
With these progresses
 and excellent experimental devices we should be able to pin down the NEOS at finite temperatures. 
 %As a quick reference, in figure \ref{finitetca}, we have given a parameterization of the percolation results. 
 %The parameter $T_c(\infty)$ can be determined through a fit of critical values for finite nuclei.

\section{ Neutron Stars }
The knowledge of the NEOS is necessary to explain observed celestial objects and events. We will discuss the relevance of the NEOS in the case of neutron stars (NS).
Those objects have been found so far with masses ranging from $1.4$ to about $2$ solar masses and a radius of the order of 10 km \cite{nsob1, nsob2, nsob3, nsob4, nsob5, nsob6, nsob7, nsob8, nsob9, nsob10, nsob11, nsob12}. These observations reveal that the
density of the neutron star is larger than the ground state density of a nucleus, and of course it decreases to zero at the surface.  Thus a neutron star is a big nucleus made mostly of neutrons.
Common understanding is that at the end of the evolution of a massive star, all the nuclear fuel, which contrasted the gravitational collapse, is used up and only heavy nuclei, around iron, remain. At this stage, the
 gravitational force continues to constraint the matter which collapses further. For some conditions, which depend on the NEOS of the system, it becomes energetically more convenient to transform protons into
 neutrons by capturing electrons and keep the system electrically neutral. Now the NEOS, which is strongly repulsive, as we have discussed when $m_\chi\rightarrow 1$, balances the Gravitational attraction.
 However, depending on the initial mass, dynamical equilibrium might be not reached such as in the observed Supernovae explosions \cite{bradly, suzuki1, supernova, add_nslattimer}. Explaining the observed masses and radii of neutron stars gives some constraints to
 the NEOS.  We will describe briefly in this section some of these constraints and refer to more in depth review for more considerations and observations \cite{bradly, add_nslattimer, nsob12, baoanrep08, stone3, skyrme3}.

 \begin{figure}[tb]
\begin{center}
\begin{minipage}[t]{12 cm}
 \includegraphics[width=1.1\columnwidth]{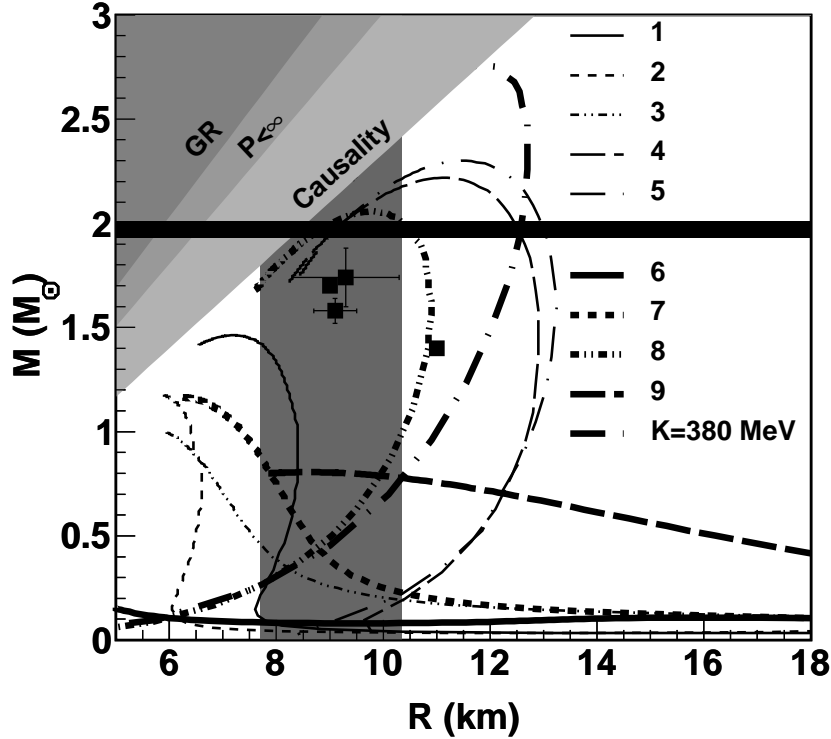}
\end{minipage}
\begin{minipage}[t]{16.5 cm}
\caption{Neutron star mass-radius relations. The data points are from \cite{nsob6, nsob8, nsob9}. The radius band is from \cite{nsob11}.  Different NEOS from table \ref{c1c2} are used to solve the TOV equations and are indicated in the figure.}\label{nsk225}
\end{minipage}
\end{center}
\end{figure}

 \begin{figure}[tb]
\begin{center}
\begin{minipage}[t]{12 cm}
\includegraphics[width=1.1\columnwidth]{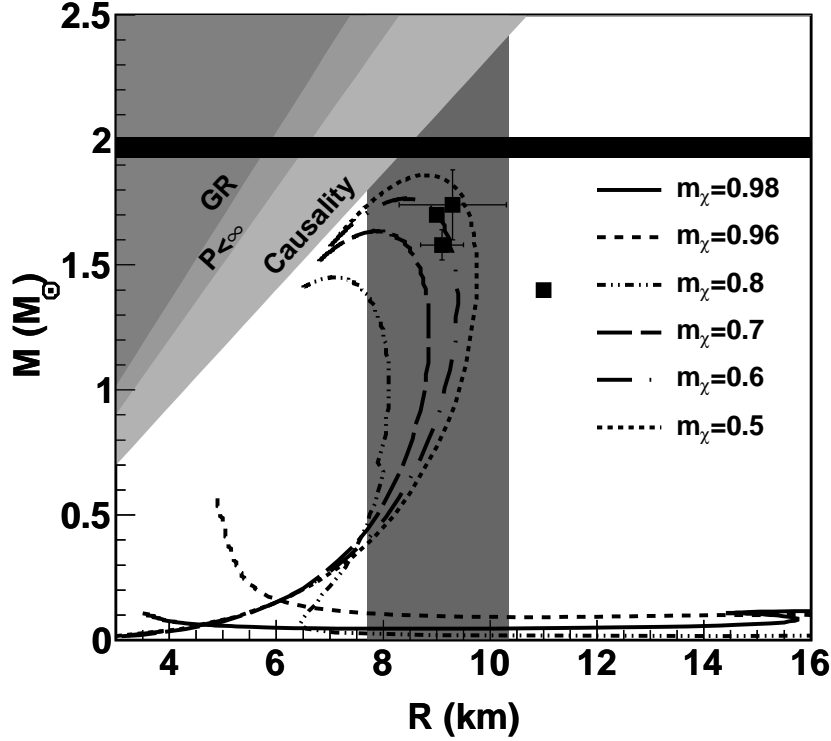}
\end{minipage}
\begin{minipage}[t]{16.5 cm}
\caption{Neutron star mass-radius relations for case (6) with different $m_\chi = 0.98, 0.96, 0.8, 0.7, 0.6, 0.5$. The data points are from \cite{nsob6, nsob8, nsob9}. The radius band is from \cite{nsob11}. }\label{nsk225case6}
\end{minipage}
\end{center}
\end{figure}
 
Taking into account corrections due to general relativity, the structure equations, which properly describe a neutron star, are given by the Tolman-Oppenheimer-Volkoff (TOV) equations \cite{tov1, tov2}:
\begin{equation}
\frac{dm(r)}{dr}=\frac{4\pi r^2\varepsilon(r)}{c^2},\label{tovm}
\end{equation}
\begin{equation}
\frac{dp(r)}{dr}=-\frac{G\varepsilon(r)m(r)}{c^2r^2}[1+\frac{p(r)}{\varepsilon(r)}][1+\frac{4\pi r^3 p(r)}{m(r)c^2}][1-\frac{2Gm(r)}{c^2r}]^{-1}. \label{tovp}
\end{equation}
$G$ is the gravitational constant and $c$ the speed of light, $m(r)$ is the mass inside the sphere of radius $r$,  $p(r)$ and $\varepsilon(r)$ are the pressure and energy density of the star at radius $r$ respectively. If the last term in equation (\ref{tovp}) becomes zero, the pressure diverges.  
This defines the Schwarzschild radius and the condition for the occurrence of a black hole \cite{bradly}.  The NEOS enters throughout the pressure $p(r)$ and the energy density $\varepsilon(r)$. The TOV equations are easily solved numerically \cite{svtov1, svtov2, svtov3, svtov4}.

\begin{figure}[tb]
\begin{center}
\begin{minipage}[t]{12 cm}
 \includegraphics[width=1.0\columnwidth]{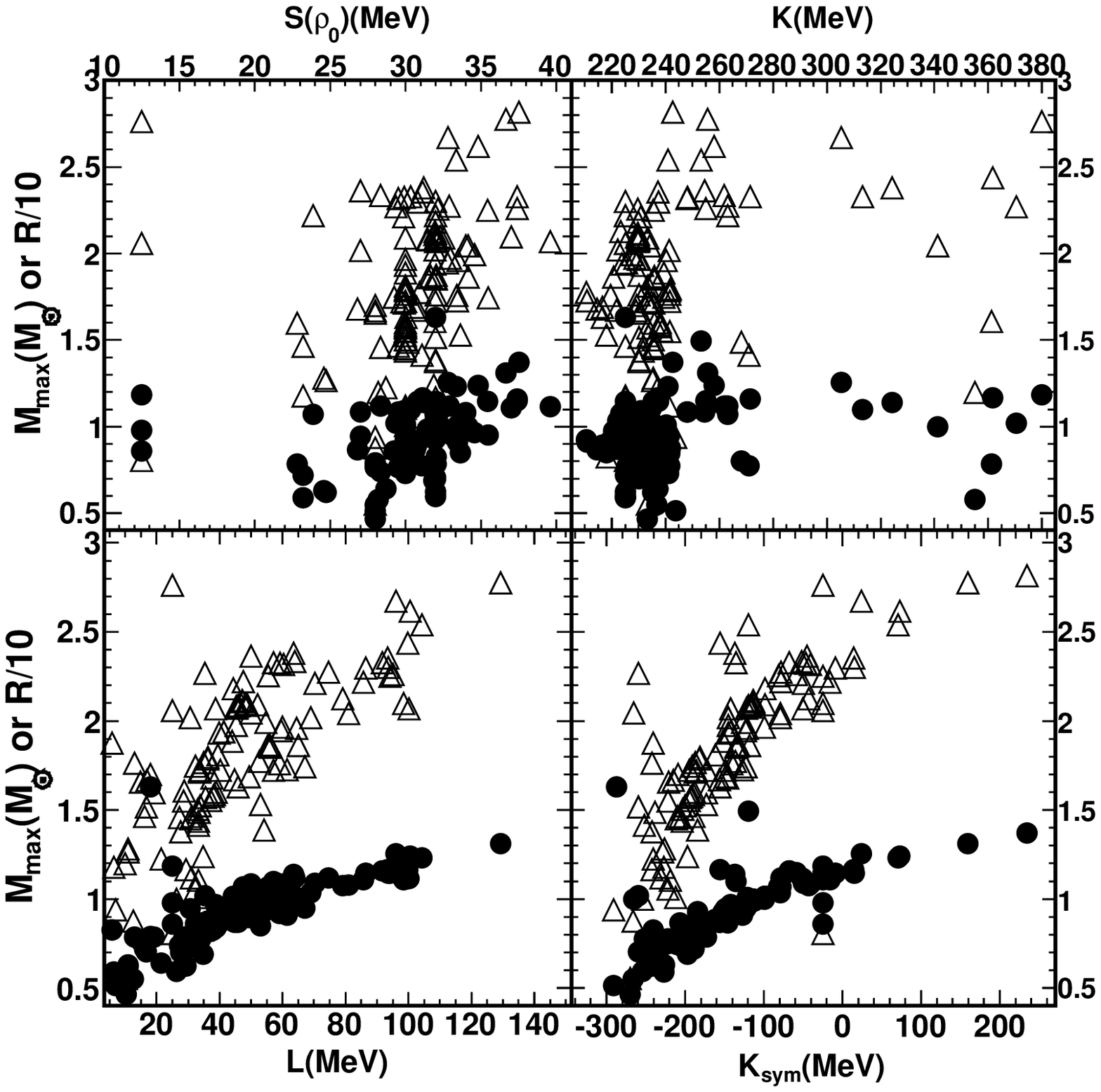}
\end{minipage}
\begin{minipage}[t]{16.5 cm}
\caption{The results for 159 Skyrme NEOS in table \ref{c1c2} and refs. \cite{shlomo2, stone2, stone3, shlomo0, shlomo1,  skyrme1_1, skyrme1_2, skyrme1_3, skyrme1_4, skyrme1_5, skyrme1_6, skyrme1_7, skyrme1_8, skyrme1_9, skyrme1_10, skyrme1_11, skyrme1_12, skyrme1_13,  skyrme1_14, skyrme1_15, skyrme1_16, skyrme2,  skyrme3,  skyrme4,   skyrme5,  skyrme6,   skyrme8,  skyrme9,  skyrme10,  skyrme11,  skyrme12,  skyrme13,  skyrme14,  skyrme15,  skyrme16, ad_skyrme1, ad_skyrme2, ad_skyrme3, ad_skyrme4, ad_skyrme5, ad_skyrme6, ad_skyrme7, ad_skyrme8, ad_skyrme9, ad_skyrme10, ad_skyrme11, ad_skyrme12, ad_skyrme13, ad_skyrme14, ad_skyrme15}. The open triangles are the maximum masses and the solid circles are the corresponding reduced radii (divided by $10$ km) of neutron stars.}\label{nssklksym}
\end{minipage}
\end{center}
\end{figure}

Two particular simple  but very instructive cases can be discussed first.  If we assume that the neutron is made of massless quarks, the NEOS is that of an ideal massless Fermi gas with $\frac{p(r)}{\varepsilon(r)}=\frac{1}{3}$, and
 $p\propto \rho^{4/3}$. This case gives an unphysical solution of the TOV equations, thus a simple non interacting, massless QGP can be excluded. On the other hand, assuming a non-interacting nucleon gas gives $\frac{p(r)}{\varepsilon(r)}=\frac{2}{3}$ in the non-relativistic case, 
and $p\propto \rho^{5/3}$. In the latter case the pressure increases faster with density and a solution to the TOV equations can be found. This solution is displayed in figure \ref{nsk225} as a thick-dashed line and corresponds to case (9) in table \ref{c1c2}.  In the figure the relation between the mass (in units of solar masses) versus its radius is given. The region in the top left corner is forbidden either by causality or constraints from the TOV equations as discussed above \cite{tov1, tov2}.
The thick horizontal line gives the maximum neutron star mass observed and, the shaded vertical region refers to the radii observed $\it {so-far}$. The simple Fermi gas NEOS is well below the observed values and can be 
safely excluded. Adding the interaction but without changing the compressibility ($c_1=c_2=0$, equation (\ref{ck225})), gives the dash-dotted line reported in the figure and corresponds to case (8) in table \ref{c1c2}. From the last two cases we understand 
that the actual value of the compressibility is not the only important ingredient but the density dependence of the pressure is. For $K=225$ MeV,   $p\propto \rho^{7/3}$, clearly if we increase the density dependence of the pressure even further, 
we can obtain  larger values of the neutron star mass. This is reported in the figure \ref{nsk225} for $K=380$ MeV, and given by the long dash-dotted line, in this case $p\propto \rho^{3}$ at high densities. Now we could have NS of the order
of 2.5 solar masses, but notice that this particular NEOS becomes acausal for radii below $12$ km. Thus also this form of NEOS can be excluded, and we had already excluded such a large compressibility from ISGMR studies \cite{shlomo2, youngblood1, youngblood2, youngblood3, piekarewicz1, chen1, lipparini1, cao1,  khan1, khan2, colo1, blaizotrep80}.

The situation becomes more complex when the interaction part of the symmetry energy is included, see figure \ref{nsk225}, corresponding to the cases of table \ref{c1c2}. Recall that all NEOS have $K=225$ MeV unless otherwise indicated. Most cases are excluded by the NS observations
and in particular the (6) and (7) NEOS which exhibit a phase transition at high densities. Notice the striking difference between case (5) and (6), derived from the same symmetry energy, but the first displaying a QLG and the second a QGP. These results together with the massless QGP result would suggest that the NS properties are mainly dependent on the EOS at a nuclear level. 
But this is of course not the entire story, we could for instance change the neutron concentration \cite{ad_pns1, ad_pns2, ad_pns3, ad_pns4}.

In figure \ref{nsk225case6} the NS mass-radius relation is now obtained for the NEOS, case (6), which undergoes a phase transition at $\rho=4.4\rho_0$. The concentration $m_\chi$ is now varied which results in larger NS masses
when including more and more protons. Eventually, the observations could be reproduced for a suitable choice of the concentration and its critical value, which, as we have seen, is also dependent on the symmetry energy. Qualitatively
we could expect that the high density region of the NS might be in the form of a QGP and, depending on the $m_\chi$ reached, the NS might become unstable. Of course other effects, such as strange matter \cite{prakash1}, a first order (or a cross-over) rather than a second order phase transition, can complicate 
the subject further, thus it is extremely important that the ingredients entering the NEOS for $m_\chi\ne 0$ be strongly constrained in a similar fashion that has been done through GMR studies.

In figure \ref{nssklksym} we recap different calculations using Skyrme forces found in the literature \cite{shlomo2, stone2, stone3, shlomo0, shlomo1, skyrme1_1, skyrme1_2, skyrme1_3, skyrme1_4, skyrme1_5, skyrme1_6, skyrme1_7, skyrme1_8, skyrme1_9, skyrme1_10, skyrme1_11, skyrme1_12, skyrme1_13,  skyrme1_14, skyrme1_15, skyrme1_16, skyrme2,  skyrme3,  skyrme4,   skyrme5,  skyrme6, skyrme8,  skyrme9,  skyrme10,  skyrme11,  skyrme12,  skyrme13,  skyrme14,  skyrme15,  skyrme16, ad_skyrme1, ad_skyrme2, ad_skyrme3, ad_skyrme4, ad_skyrme5, ad_skyrme6, ad_skyrme7, ad_skyrme8, ad_skyrme9, ad_skyrme10, ad_skyrme11, ad_skyrme12, ad_skyrme13, ad_skyrme14, ad_skyrme15}, together with the cases discussed in the previous figures. The NS radius is given in units of $10$ km which is
in the region of  the observed ones. No clear dependence on the compressibility and the symmetry energy is observed, while some linear relation is observed as function of $L$ and $K_{sym}$ \cite{add_nslattimer, nsob12, baoanrep08, ad_pns4}. In particular the NS
observed values seem to favor $L>30MeV$ and $K_{sym}>-150 MeV$. Such constraints are however not so strong, as we have seen above when changing some parameters in the NEOS (eg. concentration), which suggests
that the relevant quantities should be defined at higher densities. However, all these physical quantities can be constrained using heavy ion collisions varying the concentration and the  densities reached. These studies must be complemented 
with ground state studies of exotic nuclei.

\section{ A Reminder of  Some Microscopic Dynamical Models }
We have already referred to some models for heavy ion collisions in the previous sections. We will use model results in the following sections.
It is useful to recall some of their features and differences.  Review papers exist and we refer to those for details \cite{bertschrep88, baoanrep08, aldorep94, aldonc00, aichelinrep91, cassingrep90, onoprog04, stockerrep86, papajcp05,  fmd1, fmd7,  bondorf1, bauer1, gregoire1, gregoire2, pandharipande1, danielewicz1}. 

Ground state description of nuclei are well described within Shell model calculations \cite{MCshell92, brown1, brown2} or more involved microscopic Hartree-Fock (HF) \cite{dukelsky1, jeukenne1}. Correlations can be included at some 
level within the  Hartree-Fock-Bogoliobuv (HFB) method \cite{brown2, bombaci1, zuo1} and are necessary for the correct description of nuclei. For time dependent problems, i.e. heavy ion collisions at low beam energy,
Time Dependent HF (TDHF) has been widely used with a good reproduction, for instance of fusion cross sections \cite{ring, tdhffusion1, tdhffusion2}. At high bombarding energies, TDHF
becomes inadequate since two body correlations are relevant and should be included \cite{cassingrep90}. The Wigner transform of TDHF gives the Vlasov equation in the limit $\hbar\rightarrow 0$ \cite{aldotdhf}.
The Vlasov equation is easy to handle numerically and can be extended to include a two body collision term which takes into account the Pauli principle. The ground state of the nucleus in the Vlasov
equation is simply obtained starting from a Fermi gas model and including a mean-field, Coulomb term and surface corrections. The latter ingredients are similar to those used in TDHF, but at variance
with TDHF, there is not a real minimization procedure for the ground state. Since we are dealing with one body dynamics, it means that the Liouville theorem is satisfied already at the one-body level.
If the initial state is built in such a way to satisfy the Pauli principle, then the Liouville theorem ensures that it is never violated \cite{bertschrep88}.  This is true even after the inclusion of the collision term, since Pauli blocking
is explicitly taken into account after each nucleon-nucleon collision.

The method used to solve the Vlasov equation due to Wong \cite{wongtdhf1, wongtdhf2, wongtdhf3, wongtdhf4} is called the test particles method (tp). It consists in writing the one body distribution function $f({\bf r, p},t)$ as, in principle, an infinite sum of $\delta$ functions in coordinate and 
momentum space. The substitution of this ansatz in the Vlasov equation results in the classical Hamiltonian equation of motion of the tp moving under the influence of the mean field and the Coulomb potential.

Different (numerical) methods of solution of the Vlasov equation plus collision term have given rise to different names that can be found in the literature: Vlasov-Uheling-Uhelenbeck (VUU) \cite{cassingrep90, kruse1, kruse2, molitoris1},   Boltzmann-Uheling-Uhelenbeck
 (BUU) \cite{bertschrep88}, Boltzmann-Nordheim-Vlasov (BNV) \cite{aldobnv1, aldobnv2, aldobnv3, aldobnv4}. A particular solution of the Vlasov equation, dubbed Landau-Vlasov (LV), was proposed by C. Gregoire and collaborators using gaussians instead of $\delta$ functions \cite{gregoire1, gregoire2}. 
  
  Aichelin and St\"ocker proposed to use one test particle per nucleon and described the nucleon as gaussian distributions in phase space \cite{aichelinrep91, gregoire1, gregoire2, aichelin2, faesslerQMD1, faesslerQMD2}, this method was dubbed Quantum Molecular Dynamics (QMD).
  The name Quantum comes from the interpretation of the nucleon as a wave packet interacting through some suitable potential. Skyrme type potentials are sometimes used, which are often $\delta$  (contact) potential. When folding the 
  $\delta$ potential with gaussian distributions one obtains gaussian interactions. This method is exactly equivalent to describing the nucleons as $\delta$-functions (one per nucleon) interacting through a suitable gaussian 
  two body potential. This means that the system is completely classical, in fact the classical equations of motion are solved numerically. Thus quantum features are lost in this approach, but exact N-body correlations are included
  at the classical level. This means that the model can, for instance in fragmentation studies, form $d$, $\alpha$ and any other kind of clusters, and all possible symmetries are broken. This is at variance with mean-field type of approaches
  which describe well only average trajectories and fail if instabilities are present. Some authors have tried to correct for this by including fluctuations in the Vlasov dynamics \cite{chomaz1, colonna1}. Nevertheless,
  the problem of making light fragments remains in the (fluctuating) Vlasov equation and a possible way out is to stop the calculations at early times and use a coalescence approach in connection with an 
  `afterburner' which is a statistical model dealing with the decay of the hot source \cite{bondorf1, smm1, smm2, gemini,  gross1}. The possibility of correcting for this shortcoming in mean field dynamics makes molecular type approaches very appealing.
  
Many attempts to use Classical Molecular Dynamics (CMD) with a minimum of quantum requirements have been proposed \cite{molitoris1, bodmer1, lenk1}. In particular, including the Fermi motion results in
  very unstable  systems. In fact, if we give a Fermi motion to the particles, the classical time evolution solving the N-body dynamics brings classical correlations. Now the Liouville theorem is satisfied at the N-body level
  and the classical correlations can mimic a classical Boltzmann collision term. One can prove this rigorously by averaging over many ensembles the classical N-body evolution \cite{landaukin}. The initial `Fermi'
  momentum develops into a temperature $T$ and the particles get high momenta and are emitted from the system. The emission will stop when the remaining particles have small momenta. The real ground state of a classical system
  is a solid. The situation might improve if one introduces momentum dependent potentials. The parameters of the interaction can be chosen in such a way that in the ground state the particles have zero velocity but finite momenta.
  A particular solution was introduced in refs. \cite{dorso1, dorso2} using the so-called Pauli potential, which is a gaussian potential in phase space. The QMD model is very similar to these approaches with some important differences.
  For momentum independent potentials the Fermi motion is partly included through the widths of the gaussian used to describe a nucleon. In fact, folding the kinetic and potential energy terms with Gaussians gives rise to a term
  $A\frac{3\sigma^2_p}{2m}$ \cite{aichelinrep91, papa2}, where $\sigma_p$ is the width in momentum space. If such a term is of the order of $20 MeV/A$, then practically all the Fermi motion might be included in it. However, this term is a constant and
  it does not modify in any ways the equations of motion, which remain classical. This implies that the centroids of the gaussians are at rest, i.e. the ground state is a solid and the real binding energy is much higher.
  This is one of the ambiguities that we have when we try to solve quantum problems using classical equation of motion.  If we try to include a real Fermi motion in QMD, i.e. a kinetic energy is given to the centroids of the gaussians,
  the classical correlations make the system unstable.
  
  An elegant way to overcome this problem was first proposed by Feldmeier and it is dubbed Fermionic Molecular Dynamics (FMD) \cite{fmd1, fmd7, fmd2, fmd3, fmd4, fmd5, fmd6, fmd8}. He proposed to antisymmetrize the wave function to take into account
  the Pauli principle. This is done using gaussian wave functions as in QMD plus antisymmetrization. The equations of motion are obtained through a minimization procedure as usual. In FMD a realistic potential which
  includes the hard core is used, together with the possibility that the gaussian widths are time dependent as well. These most wished feature lead to large CPU times needed for calculations, thus reducing the 
  number of applications proposed so far \cite{fmd1, fmd7, fmd2, fmd3, fmd4, fmd5, fmd6, fmd8}. 
  
  \begin{figure}[tb]
\begin{center}
\begin{minipage}[t]{12 cm}
 \includegraphics[width=1.0\columnwidth]{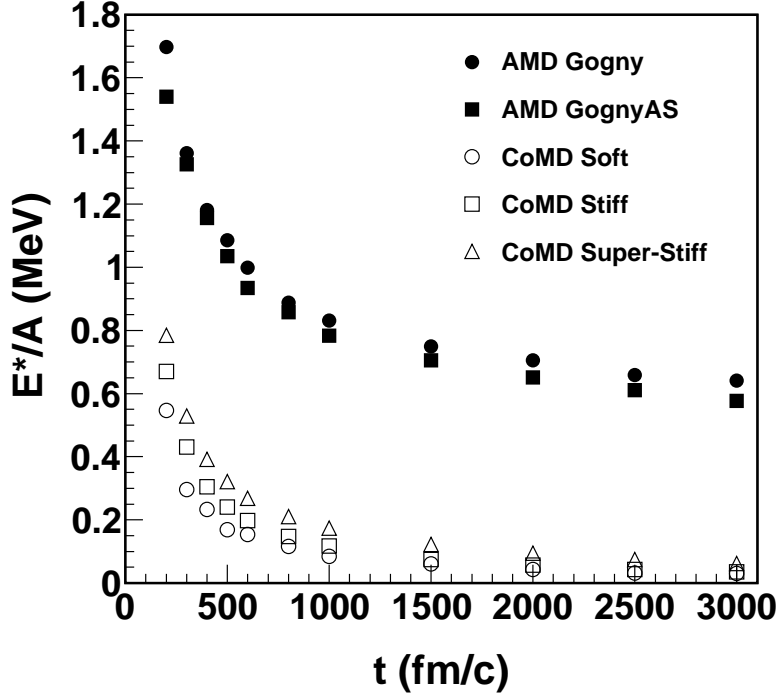}
\end{minipage}
\begin{minipage}[t]{16.5 cm}
\caption{The excitation energy per particle from AMD and CoMD calculations versus time for different NEOS for ${}^{64}Zn+{}^{64}Zn$ at 35 MeV/A\cite{andrew}.} \label{estart}
\end{minipage}
\end{center}
\end{figure}  

  A more practical way to include the Pauli principle has been proposed in ref. \cite{onoprog04}, dubbed Antisymmetrized Molecular Dynamics (AMD), and essentially it consists in fixing the width of the gaussians and
  including a collision term to mimic hard core collisions. The Pauli principle is enforced at all times. One further simplification was proposed in refs. \cite{papajcp05, papa2} where antisymmetrization is obtained through a constraint on the phase space occupation to be less than one at all times, dubbed Constrained Molecular Dynamics (CoMD). A collision term, similar to AMD, is also included . 
    
  FMD, AMD and CoMD are all essentially classical in nature plus a constraint to take into account the Pauli principle. To make an analogy with the Bohr model of the atoms, one solves the classical equation of motion and chooses only the trajectories constrained by $\hbar$ \cite{kimura1}. Being classical, the problem of what to do with the width of the gaussians when calculating the total energy remains. Even though this is becoming boring, we insist on this point since different choices are made by different authors on how to treat the gaussian's width. This implies that the models could give different results even though nominally they use the same interaction and solve the same equation of motion. To be more specific let us write down equation (9) from ref. \cite{onoprog04}, the total energy of the system $\mathcal{H}(Z)$:
  \begin{equation}
\mathcal{H} (Z)=\frac{\langle \Phi(Z)|H|\Phi(Z)\rangle}{\langle \Phi(Z)|\Phi(Z)\rangle}-\frac{3\hbar^2\nu}{2m}A+T_0(A-N_F(Z)).\label{ono9}
\end{equation}
Here, $H$ is the hamiltonian, $Z$ is the generalized coordinate of the wave packet $\Phi(Z)$, $\nu$ is its width in $fm^{-2}$, $m$ is the nucleon mass, $T_0$ is a free parameter and $N_F(Z)$ is the number of fragments. In CoMD $T_0=0$, the contribution on the width of the gaussian is subtracted as in equation (\ref{ono9}) and the total energy is a constant of motion, i.e. $\mathcal{H}$ is independent of the coordinate Z and thus the time. In AMD, the width of the gaussian is subtracted and a constant $T_0$ is added. The constant is fixed to reproduce the ground state binding energy of the nuclei and its value is $T_0=9.2 MeV$, while the contribution
$\frac{3\hbar^2\nu}{2m}=10MeV$. The difference is not just $0.8 MeV$ for the two terms but more importantly the number of fragments $N_F(Z)$ that the authors parametrize as a smooth function of the coordinate $Z$. For nuclear ground states of course $N_F(Z)=1$. This means that for a nucleon the correction is zero $MeV$, for $d$ it amounts to $\frac{9.2}{2} MeV/A$ and converges to $9.2 MeV/A$ for large nuclei.  With this ansatz, the authors are able to
reproduce the $BE$ of a large number of nuclei. However, the real binding energy of the system is the one with the $T_0$ term not included, since constant terms (in the ground state) do not give any contribution to the equations of motion.
This choice has important consequences in time dependent problems, such as fragmentation. In such a case $N_F(Z)$ changes when fragments are formed which results in a change of the total energy.
The `trick', as the authors define it, is to modify the kinetic energy of the particles which are emitted \cite{onoprog04} in order to conserve the initial total energy of the system. The extra energy is randomly distributed to the nucleons in some AMD versions or to the fragments in some other versions. The use of this constant has important consequences as we have seen in the calculated nuclear ground states, we might expect that a similar
effect will arise when calculating excitation energies. Results of a calculation for the system and the beam energy indicated are reported in figure \ref{estart} \cite{andrew}. The excitation energy per particle versus time is plotted for AMD and CoMD calculations using different NEOS. The two models give drastically different results as drastically different are the choices adopted in the models. In CoMD the Fermi motion
is given by the kinetic energy of the gaussians and $\it not$ by its width and the ground state of the nuclei is obtained for a given NEOS by fixing the width parameters and a surface interaction \cite{papajcp05, papa2}. None of those
ingredients are relevant for the calculation of the excitation energy which becomes negligible after a few hundred $fm/c$. In contrast AMD calculations display an almost constant value for very long times and systematically higher than CoMD due to the different assumptions and the inclusion of the parameter $T_0$, equation (\ref{ono9}). This feature must be taken into account also when using `hybrid' models, i.e. when AMD, CoMD or other models, are stopped at a certain time and an afterburner for the decay from excited states is coupled to them. Usually CoMD calculations
are followed for a long time (even up to 60000 fm/c for fission \cite{souliotis1} by choice of the authors). These differences should be kept in mind when trying to derive properties of the NEOS from a comparison
to experimental data. In the following we are going to rely heavily on the CoMD model which was proposed originally by one of us, thus the discussion above is biased, different point of view can be contrasted from the literature \cite{onoprog04, fmd1, fmd2}.

\section{Neutron Skin Thickness and Density Dependence of the Symmetry Energy}\label{neutronskin}
A very promising research line in the NEOS studies is represented by the
determination of the Neutron Skin Thickness (NST). This
quantity is defined as the difference between the root mean square radii of
neutrons and protons: 
\begin{equation}
\Delta r_{np}= \langle R^{2}_{n} \rangle^{\frac{1}{2}}-
\langle R^{2}_{p} \rangle^{\frac{1}{2}}. 
\end{equation}
Both relativistic and non-relativistic
mean field calculations for the $^{208}Pb$ nucleus have pointed out a linear relation between
$\Delta r_{np}$ and the slope of the symmetry energy $L$ \cite{Brown, Furns, Hor-Piek}. The neutron skin
formation has been explained as the consequence of two possible mechanisms: the
first one is related to the neutron concentration in the Fermi energy, the second refers to the less binding in neutron rich matter \cite{Brown}.

From the experimental point of view, the proton radii can be
determined with a precision of 0.02 fm or lower via electron scattering
\cite{Tab, qprofbrown03}, the determination of the neutrons radii represents an interesting
challenge whereas the choice of the probe is crucial. The goal of the Lead
Radius EXperiment (PREX) \cite{prex}, performed at the Jefferson Laboratory, is
to measure the parity violation in electron scattering. The advantage with
respect to hadron probes such as pions, protons and anti-protons is the
independence on strong-interaction uncertainties \cite{Had}. The current result
for the evidence of a neutron skin in $^{208}Pb$ provided by PREX is:
\begin{equation}
\Delta r_{np}(^{208}Pb)=0.33^{+0.16}_{-0.18} fm. 
\end{equation}
The Isovector Pygmy Dipole Resonance (IVPDR) collective mode in neutron-rich nuclei has
been interpreted as the mutual oscillation of the neutron skin with respect to a
symmetric nuclear core \cite{pdr1, pdr2, pdr3, pdr4, pdr5, pdr6, pdr7}, the NST exhibits a sensitivity to the fraction of the
Energy-Weighted Sum Rule exhausted by the IVPDR mode \cite{pdr8, pdr9, pdr10}.
The experimental data from the LAND
collaboration seem to confirm this evidence \cite{pdr_exp}. 

 \begin{figure}[tb]
\begin{center}
\begin{minipage}[t]{12 cm}
 \includegraphics[width=1.1\columnwidth]{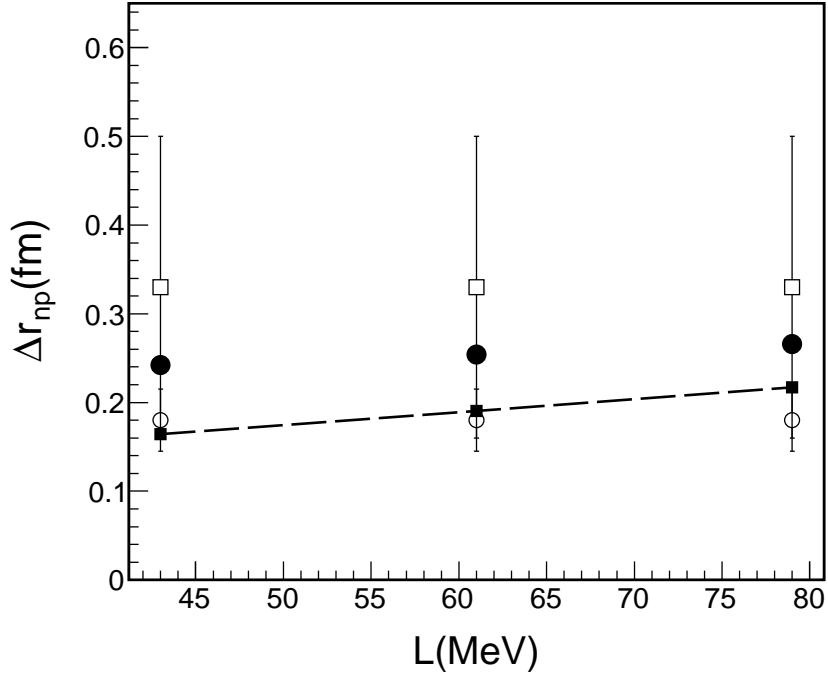}
\end{minipage}
\begin{minipage}[t]{16.5 cm}
\caption{NST of the $^{208}Pb$ nucleus as a function of the
slope of the symmetry energy $L$. Solid circles: CoMD model calculations, open
squares: data from the PREX experiment \cite{prex}, open circles: data from the LAND
collaboration experiment \cite{pdr_exp}. The dashed line connecting the solid squares represents the fit from \cite{lfit}.}\label{nstpb}
\end{minipage}
\end{center}
\end{figure}

\begin{figure}[tb]
\begin{center}
\begin{minipage}[t]{12 cm}
 \includegraphics[width=1.1\columnwidth]{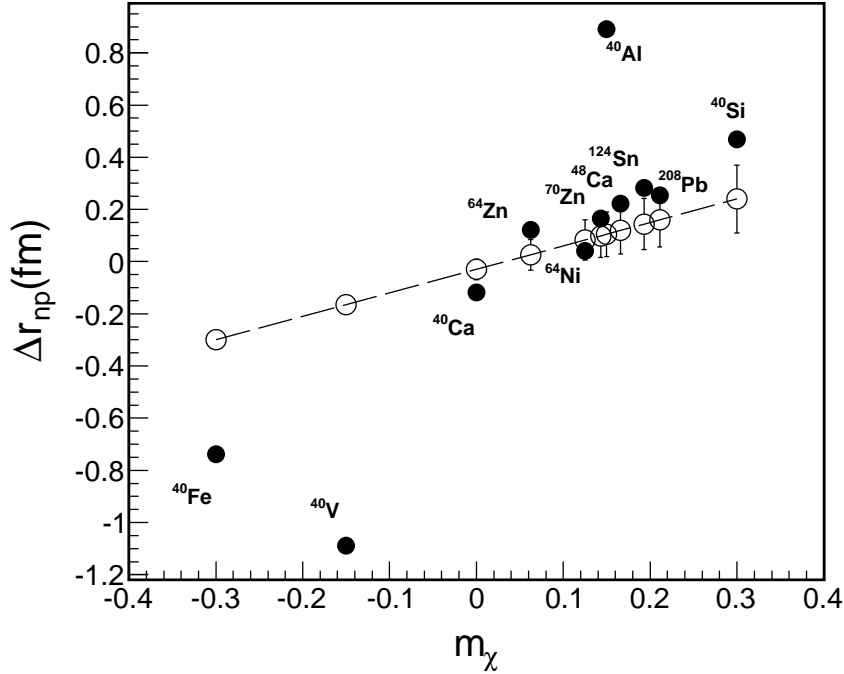}
\end{minipage}
\begin{minipage}[t]{16.5 cm}
\caption{NST $\Delta r_{np}$ versus neutron/proton
concentration $m_{\chi}$. Solid Circles: CoMD calculations. Open Circles:
$\Delta r_{np}$ from equation (\ref{nstmfit}); the line refers to the correlation between
$\Delta r_{np}$ and  $m_{\chi}$.}\label{nstm}
\end{minipage}
\end{center}
\end{figure}

\begin{figure}[tb]
\begin{center}
\begin{minipage}[t]{12 cm}
 \includegraphics[width=1.1\columnwidth]{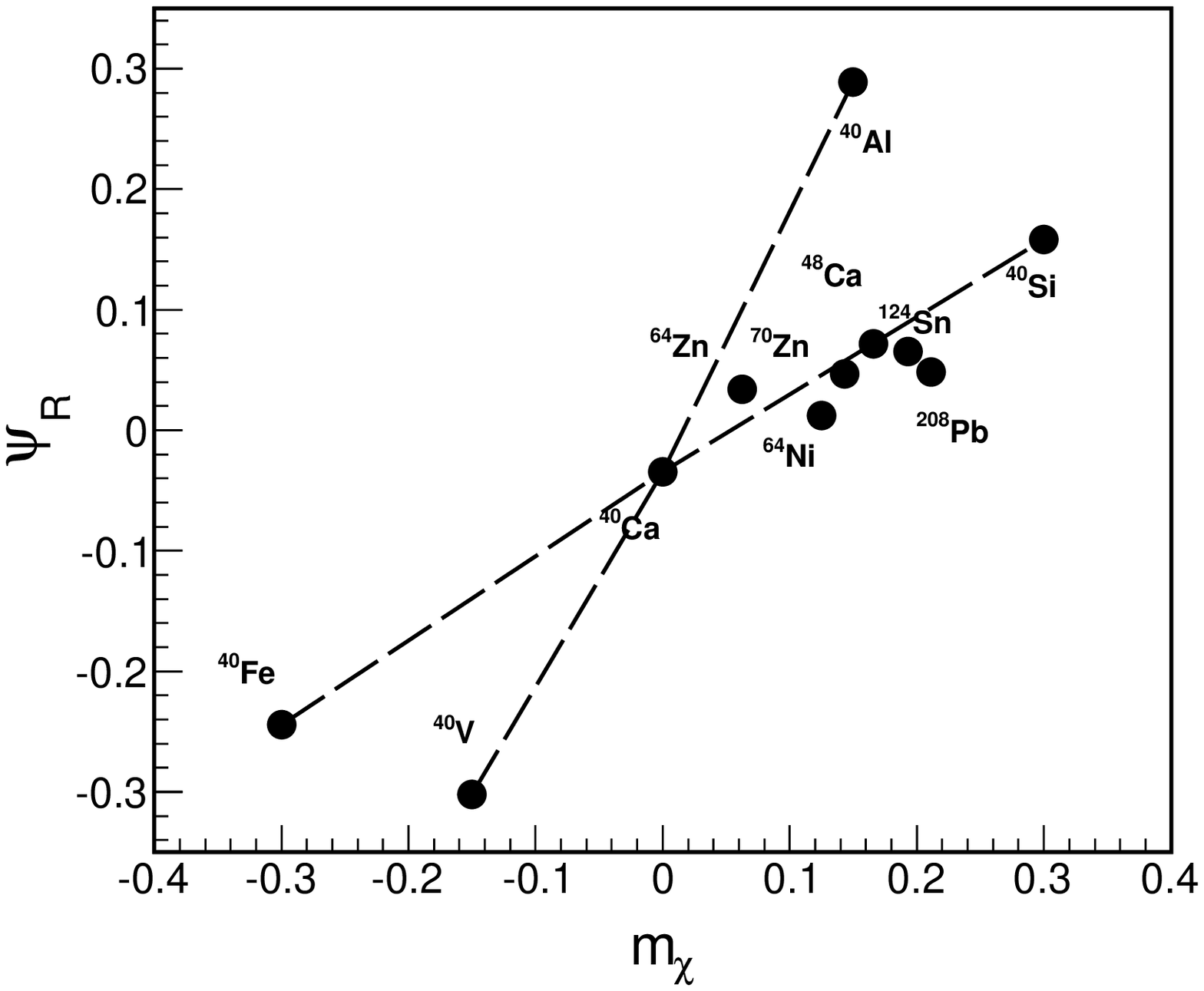}
\end{minipage}
\begin{minipage}[t]{16.5 cm}
\caption{$\Psi_{R}$ versus neutron/proton
concentration $m_{\chi}$. Solid Circles: CoMD calculations. The dashed lines
refer to the linear correlation for the systems having mass $A=40$ and
different neutron/proton concentrations.}\label{nstphi}
\end{minipage}
\end{center}
\end{figure}

An intriguing feature to pursue in the future for this research line could be
related to beyond mean-field calculations and the role of the formation of
clusters in the nuclear surface. Due to their peculiarities, molecular
dynamics models could be used for this purpose. For this reason we have
calculated $\Delta r_{np}$ for several nuclei by means of the CoMD
model; however, at this stage we are interested on the capability of the
model to produce reasonable results if compared with experimental data and
the $L$ dependence of NST provided by several theoretical models. Figure \ref{nstpb}
shows a comparison, for three $L$ values, between the CoMD calculations (solid
circles) of $\Delta r_{np}$ for the $^{208}Pb$ nucleus and the experimental
results (which are, of course, constant since the value of $L$ is not determined in the experiment) coming from both the PREX (open squares) and the LAND
collaboration (open circles) experiments. The line in the figure is
related to the correlation $\Delta r_{np}=0.101+0.00147 L$ fm which has been
obtained by fitting the results from several relativistic and non-relativistic
mean field models \cite{lfit}. As can be noted the CoMD points are inside the
PREX error bars but above the LAND experimental data. On the other hand the
reduced sensitivity on the slope of the symmetry energy as compared to the
points of the fit (solid squares) requires a further study.

The dependence of $\Delta r_{np}$ on the neutron/proton concentration
$m_{\chi}$ has been analyzed by collecting experimental data related to nuclei from $^{40}Ca$ to $^{238}U$  \cite{mchifit1, mchifit2, mchifit3}. The
data displayed a rough linear behavior that has been fitted  by the following
formula:
\begin{equation}
\Delta r_{np}=(0.9 \pm 0.15)m_{\chi}+(-0.03 \pm 0.02) fm. \label{nstmfit}
\end{equation}
The open circles in figure \ref{nstm} refer to the NST calculated
from equation (\ref{nstmfit}) for a group of eleven nuclei. CoMD calculations performed for $L=61$
MeV are close to the values obtained for the systems:
$^{64}Zn$, $^{64}Ni$, $^{70}Zn$, $^{48}Ca$, $^{124}Sn$, $^{208}Pb$. Not all the
nuclei that we have considered match with the ones used to obtain the
parameters of equation (\ref{nstmfit}), in particular, we have also taken under examination the
proton-riches $^{40}Fe$,$^{40}V$ and neutron-riches ${}^{40}Al, {}^{40}Si$. Since negative values
of $\Delta r_{np}$ for proton-rich systems could be obvious, we would like to
remark an odd-even staggering effect between $^{40}Fe$ (even-even) and
$^{40}V$ (odd-odd). The same trend, but in an opposite direction, can be
noticed between the neutron-rich systems $^{40}Al$ (odd-odd) and $^{40}Si$
(even-even). 

%\begin{figure}[tb]
%\begin{center}
%\begin{minipage}[t]{8 cm}
%\epsfig{file=psi_par.eps, scale=0.6} % 
%\end{minipage}
%\begin{minipage}[t]{16.5 cm}
%\caption{$\Psi_{R}^{'}$ versus $m_{\chi}$ for the $^{40}Fe$, $^{40}Ca$ and
%$^{40}Si$ systems. Solid circles: CoMD calculations. Solid Line: eq. (\ref{symexp}) fit.
%Dashed line: parabolic fit.}\label{psiprime}
%\end{minipage}
%\end{center}
%\end{figure}

In order to take phenomenologically into account the geometry of the
nuclear systems, that could play a role in the determination of the NST, we have defined the following quantity:
\begin{equation}
\Psi_{R}=\frac{ \langle R^{2}_{n} \rangle-
\langle R^{2}_{p} \rangle } {\langle R^{2}_{n} \rangle +
\langle R^{2}_{p} \rangle }. 
\end{equation}
Such a quantity becomes $\pm 1$ for pure proton or neutron systems. In figure \ref{nstphi} the $\Psi_{R}$ quantity is related to $m_{\chi}$ for the same
systems of figure \ref{nstm}.

An approximately linear relation between $\Psi_R$ and $m_\chi$ can be observed apart the very exotic and odd-odd nuclei, $^{40}Al$ and $^{40}V$. A certain linearity can be noticed for the
systems having $A=40$, moreover the slopes of the lines show a certain
sensitivity on whether the nuclei are even-even or odd-odd. By fixing the mass, 
and varying the neutron/proton concentration, is equivalent to fix the surface
and volume terms and vary the terms depending on the neutron and proton numbers
in the liquid drop formula: symmetry, Coulomb and pairing terms. By
considering one sequence of systems along the $^{40}Fe,{}^{40}Si$ even-even
or the $^{40}V, {}^{40}Al$ odd-odd nuclei lines, the pairing term contribution is
fixed. As a consequence the quantity $\Psi_{R}$ is directly dependent on the
symmetry coefficient $a_{a}$ (see equation (\ref{bind})); we notice that in these
calculations the Coulomb contribution is included. Therefore, the investigation
on the $\Psi_{R}$ quantity could be helpful in constraining the symmetry energy
on a ``nucleus-by-nucleus'' basis similar to the Isobaric Analog states
method discussed in section \ref{IAS}. 

The negative values for the $N=Z$ system $^{40}Ca$ of $\Delta r_{np}(^{40}Ca)
\approx -0.12 fm$ and $\Psi_{R}(^{40}Ca) \approx -0.035$, could be related to the
Coulomb repulsion pushing protons on the surface. We can consider such
an effect, which in principle is also present in the $N \neq Z$ nuclei, by
taking $\Psi_{R}(^{40}Ca)= \Psi_{0}$ as a reference point and define the
difference:
\begin{equation}
\Psi_{R}^{'}= |\Psi_{R} - \Psi_{0} |.
\end{equation}
This quantity should be dependent on $m_\chi$ and it could be expanded up to sixth order similar to the symmetry energy. However, the available data is scarce and it needs further studies.

We conclude this section by stressing how the NST represents a powerful
tool for future symmetry energy studies, especially with the new experiments
PREX-II and CREX (Calcium Radius Experiment) \cite{piek}. Moreover, further
improvements of the CoMD model could be helpful to approach a beyond mean-field
scenario. 

\section{ Giant Resonances} \label{giantR}
Important informations about the NEOS can be obtained from the study of giant resonances in nuclei. These are small oscillations of the system whose typical energy and width (or damping) depend
on the way the excitation energy is delivered to the system. For instance, if the nucleus is compressed,  then a breathing mode results commonly referred as Isoscalar Giant Monopole Resonances (ISGMR) \cite{shlomo2, youngblood1,  youngblood2, youngblood3, piekarewicz1, chen1, lipparini1, cao1, khan1, khan2, colo1, blaizotrep80}. Since the nucleus is compressed it follows that the compressibility $K$ can be constrained by such modes, equation (\ref{GMR}).
Another important example is the Isoscalar  Giant Quadrupole Resonance  (ISGQR) which can constrain the effective mass as well \cite{r2ad_gr1, r2ad_gr2, r2ad_gr3}. If we displace the protons from the neutrons, using energetic photons for instance, an oscillation results, commonly known as Isovector
Giant Dipole Resonances (IVGDR) \cite{ shlomo2, youngblood1, youngblood3, lipparini1, colo1,  trippa1, ring1, hamamoto1, yoshida1, shlomo3, suraud1}. In this case there is no real compression of the nucleus and intuitively we can think that the typical oscillation frequency must be connected somehow to 
the symmetry energy and Coulomb energy, see equation (\ref{bind}). Thus IVGDR could be a good study case to constrain some ingredients of the symmetry energy near the ground state density and at zero temperature.

The values of the centroid of the GDR resonances have been  measured for a number of ions, mostly stable, during the years. A suitable parametrization of the GDR resonance energy is \cite{lipparini1, aldorep94, woudebook, colonnappnp93}:
\begin{equation}
E_{GDR}=31.2 A^{-1/3}+20.6A^{-1/6}.\label{Egdr} 
\end{equation}
The latter is a smooth function of $A$ which might be surprising at first since we would expect to find a dependence on $m_{\chi}$ as well. Since the parametrization is obtained from stable nuclei, it might happen that some deviation from equation (\ref{Egdr}) can be found when exotic nuclei will be available for experimentation.

The IVGDR can be rather well described within a hydrodynamical model as proposed by Stringari and Lipparini \cite{lipparini1} and later on generalized in ref. \cite{moller1} displaying the dependence from the symmetry energy:
\begin{equation}
E_{-1}=\sqrt{\frac{6\hbar^2}{m\langle r^2 \rangle}g_A(\rho_0)(1+\kappa)}, \label{E-1}
\end{equation}
where $\kappa$ is a `well known' (whose nature is rather obscure) enhancement factor connected to the momentum dependent force (and maybe Coulomb and surface terms). Such a term
turns out to be of the order of $\kappa\approx 0.2$ \cite{lipparini1},
\begin{equation}
g_A(\rho) = \frac{S(\rho)}{1+\frac{5}{S(\rho)}[\rho \frac{dS}{d\rho}-\frac{\rho^2}{4}\frac{d^2S}{d\rho^2}]A^{-1/3}}.\label{grho}
\end{equation}
The latter equation contains all the terms entering the symmetry energy that we have discussed so far, i.e. $S$, $L$ and $K_{sym}$, see equations (\ref{s}, \ref{l}, \ref{ksym}),  and $\langle r^2 \rangle=\frac{3}{5}R^2$ as usual.
A simple inspection of equation (\ref{E-1}) gives a $A^{-1/3}$ dependence (from the radius $R=r_0A^{1/3}$ which is similar to the phenomenological formula
equation (\ref{Egdr})). One might also think that the effective mass is important if we define $m^*=\frac{m}{1+\kappa}$, indeed we expect the effective mass to play a role
in the exact determination of the IVGDR energy. Still other corrections due to Coulomb and finite sizes are possible and not included (at least not explicitly) in the formulas above.

\begin{figure}[tb]
\begin{center}
\begin{minipage}[t]{12 cm}
 \includegraphics[width=1.1\columnwidth]{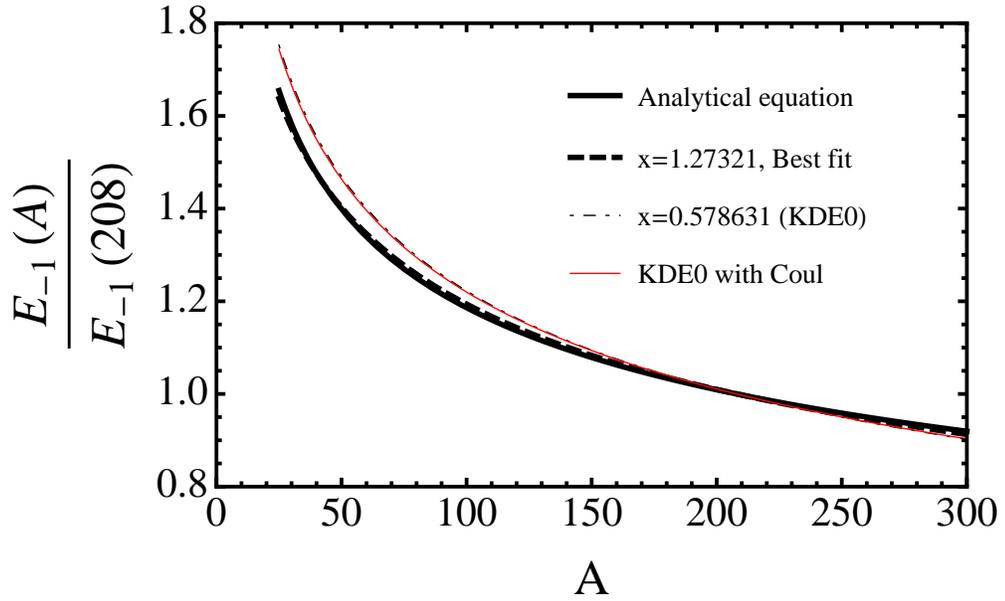}
\end{minipage}
\begin{minipage}[t]{16.5 cm}
\caption{(Color online) The fit of equation (\ref{ratiogdr}). The black thin dash-dotted line is obtained from the KDE0 parametrization \cite{shlomo2}, the red thin line includes (negligible) Coulomb corrections.} \label{egdrfit}
\end{minipage}
\end{center}
\end{figure}

\begin{figure}[tb]
\begin{center}
\begin{minipage}[t]{12 cm}
 \includegraphics[width=1.0\columnwidth]{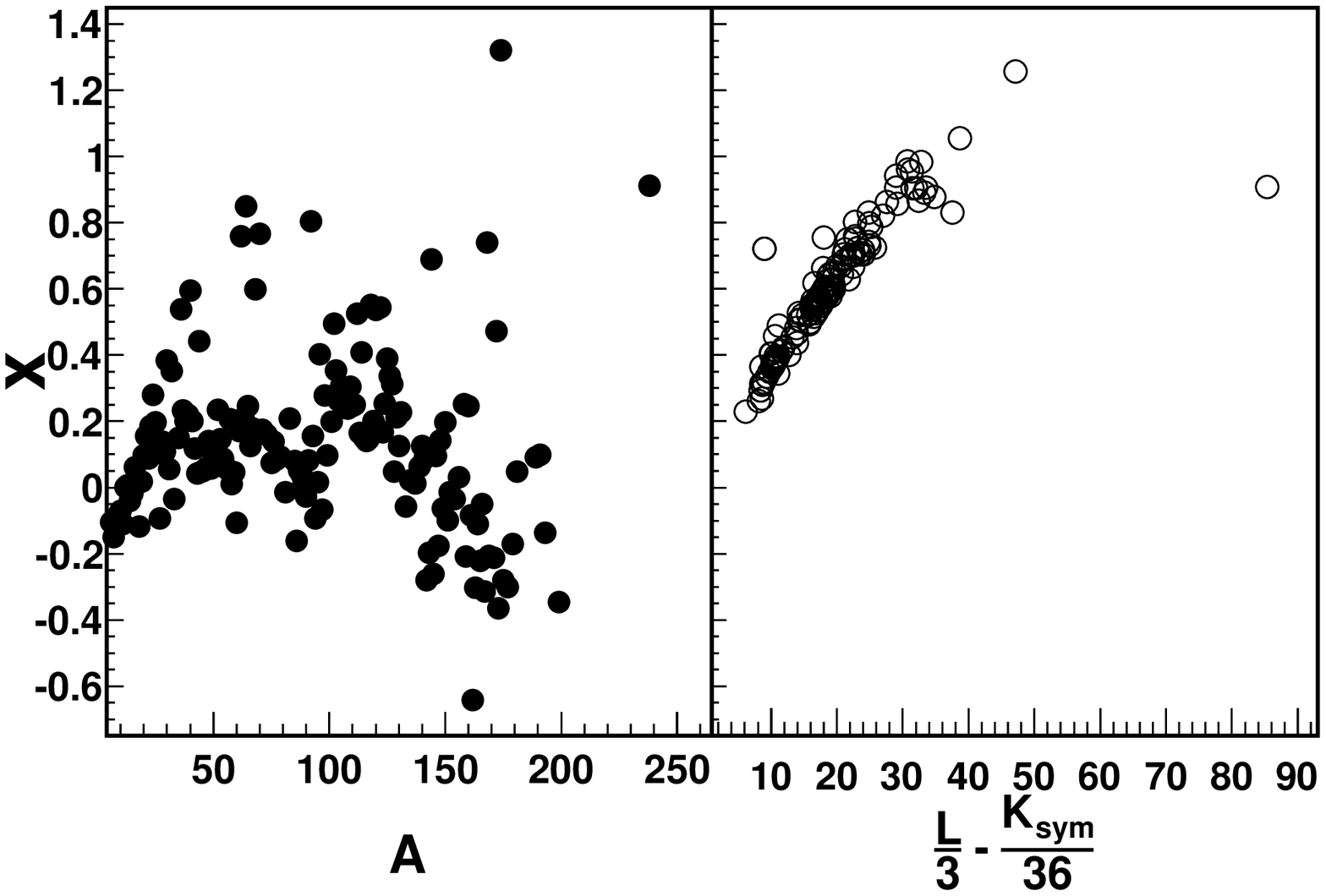}
\end{minipage}
\begin{minipage}[t]{16.5 cm}
\caption{(Left panel) $x$ versus the mass number $A$; (right panel) $x$ versus $\frac{L}{3}-\frac{K_{sym}}{36}$ for the NEOS in table \ref{c1c2} and refs. \cite{shlomo2, stone2, stone3, shlomo0, shlomo1,  skyrme1_1, skyrme1_2, skyrme1_3, skyrme1_4, skyrme1_5, skyrme1_6, skyrme1_7, skyrme1_8, skyrme1_9, skyrme1_10, skyrme1_11, skyrme1_12, skyrme1_13,  skyrme1_14, skyrme1_15, skyrme1_16, skyrme2,  skyrme3,  skyrme4,   skyrme5,  skyrme6, skyrme8,  skyrme9,  skyrme10,  skyrme11,  skyrme12,  skyrme13,  skyrme14,  skyrme15,  skyrme16, ad_skyrme1, ad_skyrme2, ad_skyrme3, ad_skyrme4, ad_skyrme5, ad_skyrme6, ad_skyrme7, ad_skyrme8, ad_skyrme9, ad_skyrme10, ad_skyrme11, ad_skyrme12, ad_skyrme13, ad_skyrme14, ad_skyrme15}.} \label{tryx}
\end{minipage}
\end{center}
\end{figure}

 In order to avoid the factor $\kappa$ and put some constraints on the
ingredients of the symmetry energy we define the ratio:

\begin{eqnarray}
\frac{E_{GDR}(A)}{E_{GDR}(208)} &=&\frac{E_{-1}(A)}{E_{-1}(208)}\nonumber\\
&=& \frac{\sqrt{\frac{6\hbar^2}{m\langle r^2 \rangle_A}g_A(\rho_0)(1+\kappa)}}{\sqrt{\frac{6\hbar^2}{m\langle r^2 \rangle_{208}}g_{208}(\rho_0)(1+\kappa)}} \nonumber\\
%&=& \frac{\sqrt{\frac{1}{\langle r^2 \rangle_A}g_A(\rho_0)}}{\sqrt{\frac{1}{\langle r^2 \rangle_{208}}g_{208}(\rho_0)}} \nonumber\\
%&=& \frac{\sqrt{\frac{1}{A^{2/3}}g_A(\rho_0)}}{\sqrt{\frac{1}{208^{2/3}}g_{208}(\rho_0)}} \nonumber\\
&=&\frac{\sqrt{\frac{1}{A^{2/3}} \frac{S(\rho_0)}{1+\frac{5}{S(\rho_0)}[\frac{L}{3}-\frac{K_{sym}}{36}]A^{-1/3}} }}{\sqrt{\frac{1}{208^{2/3}} \frac{S(\rho_0)}{1+\frac{5}{S(\rho_0)}[\frac{L}{3}-\frac{K_{sym}}{36}]208^{-1/3}} }} \nonumber\\
&=&\frac{\sqrt{\frac{1}{A^{2/3}} \frac{S(A)}{1+5xA^{-1/3}} }}{\sqrt{\frac{1}{208^{2/3}} \frac{S(208)}{1+5x 208^{-1/3}}}}. \label{ratiogdr}
\end{eqnarray}
where the left hand side can be obtained from the experimental values, equation (\ref{Egdr}), and $x =\frac{1}{S}(\frac{L}{3}-\frac{K_{sym}}{36})$ can be fitted to the data. $S(A)$ is the symmetry energy which could depend on the mass as for instance in equations (\ref{iasaafgia1}, \ref{iasaafgia2}). If we assume that the symmetry energy is mass independent, we can simplify equation (\ref{ratiogdr}), under such an assumption, it is reasonable to assume that $x$ is mass independent as well. The result of the fit is given in figure \ref{egdrfit}. From the fit we find $x=1.27321$ which
could be used as a constraint to the symmetry energy parameters. But..., we can repeat the same procedure for a model where, in principle, we know all the ingredients. For instance, using the
NEOS of \cite{shlomo2, shlomo0, shlomo1}, we can calculate analytically the ratio given in equation (\ref{ratiogdr}), which gives $x=0.578631$, more than a factor two below the fit to the experimental data.
Here comes the problem, we know from refs. \cite{shlomo2, shlomo4} that the IVGDR energies of $Ca$ and $Pb$ are quite well reproduced by the RPA calculations using the KDE0-NEOS. This implies that
in the formula above we are missing some ingredients. One of these ingredients is the surface which is quite difficult to parametrize or predict its effect on the resonance. The other effect
is Coulomb which we have elegantly ignored so far. Notice that if we use the mass formula, equation (\ref{bind}) we easily obtain:
\begin{eqnarray}
\bar E_{sym}^{corr}=\frac{\frac{E}{A}(1)+\frac{E}{A}(-1)}{2}-\frac{E}{A}(0)=a_a+\frac{a_c}{4}A^{2/3}+pairing-corrections=a'_a. \label{symcou}
\end{eqnarray}

\begin{figure}[tb]
\begin{center}
\begin{minipage}[t]{12 cm}
 \includegraphics[width=1.0\columnwidth]{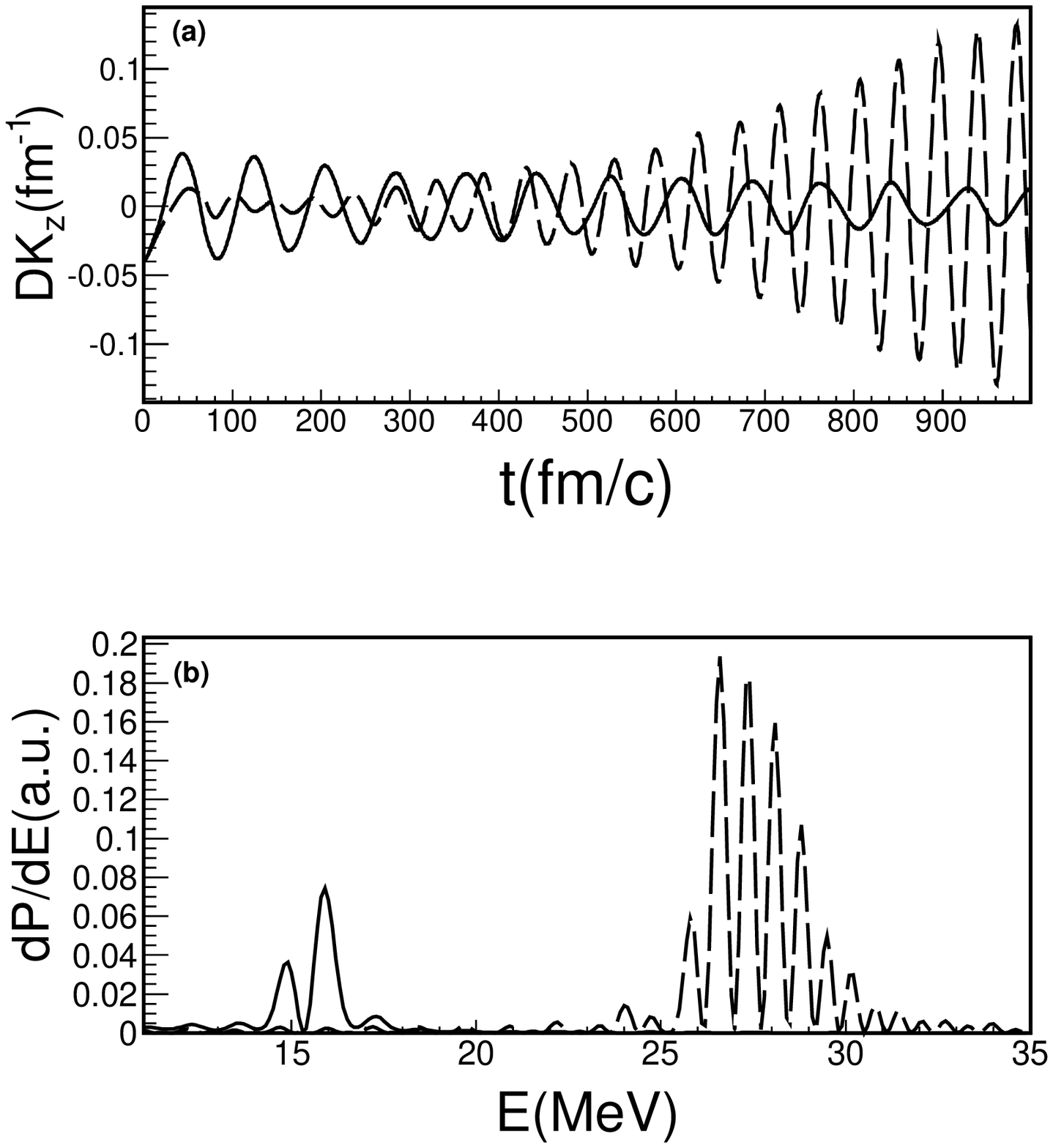}
\end{minipage}
\begin{minipage}[t]{16.5 cm}
\caption{BNV calculations for $^{40}Ca$ nuclei, with (dashed line) and without (solid line) momentum dependent interactions. Panel (a): dipolar signals as a function of time; Panel (b):  related oscillation energy distributions.}\label{bnvmdi}
% (a): Solid Line: BNV calculations without momentum depedent interaction. Dashed line: BNV calculations with momentum depedent interaction; (b): oscillation energy for the  momentum indepedent interaction case.
\end{minipage}
\end{center}
\end{figure}

\begin{figure}[tb]
\begin{center}
\begin{minipage}[t]{12 cm}
 \includegraphics[width=1.0\columnwidth]{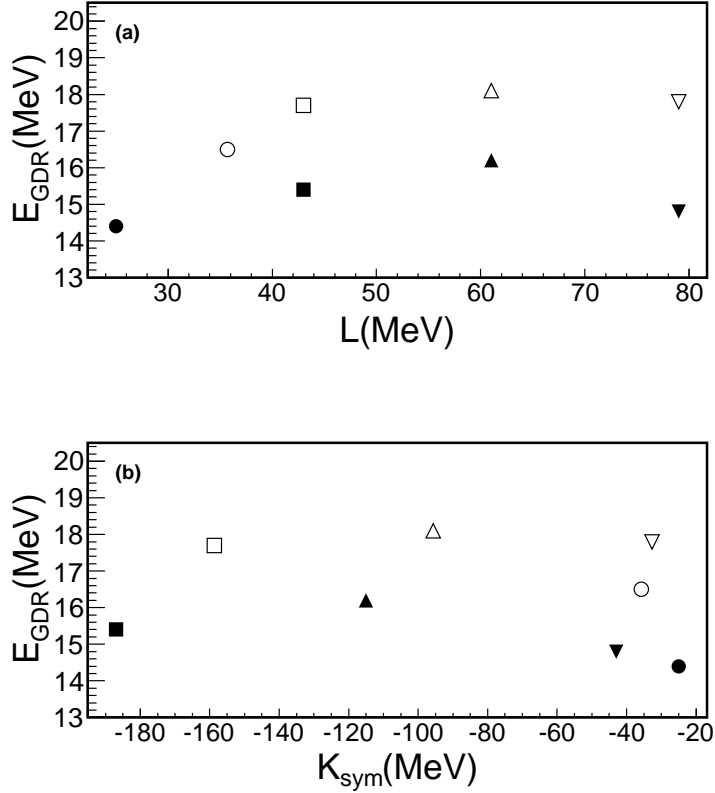}
\end{minipage}
\begin{minipage}[t]{16.5 cm}
\caption{$E_{GDR}$ versus $L$ and $K_{sym}$ from BNV calculations with (open symbols) and without (solid symbols) momentum dependent interactions. The different symbols refer to different $L$ and $K_{sym}$ couples, for instance the open circle refers to a momentum dependent calculation with $L=36 MeV$ and $K_{sym}=-36 MeV$. No signal smoothing of the Fourier transform has been performed.}\label{bnvlksym}
\end{minipage}
\end{center}
\end{figure}

\begin{figure}[tb]
\begin{center}
\begin{minipage}[t]{12 cm}
 \includegraphics[width=1.0\columnwidth]{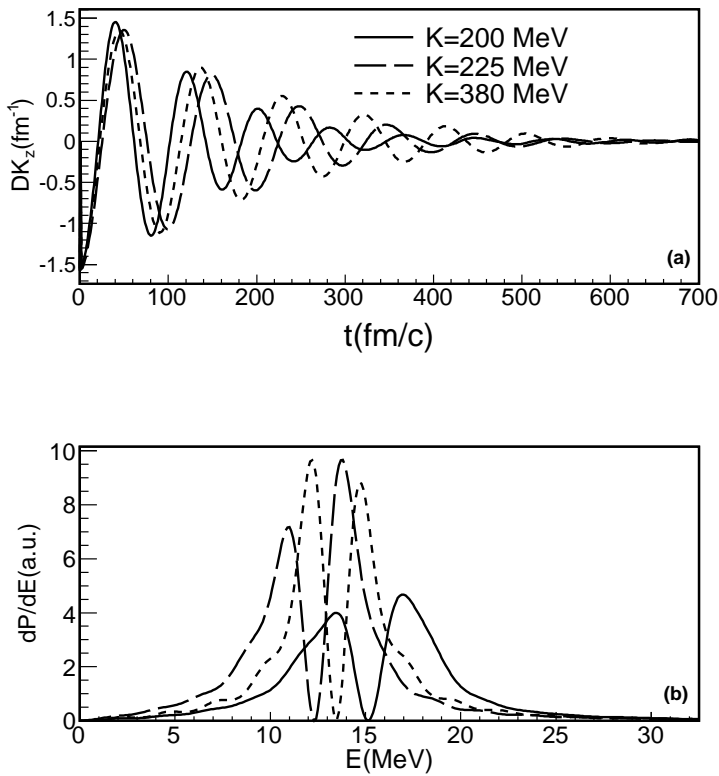}
\end{minipage}
\begin{minipage}[t]{16.5 cm}
\caption{CoMD calculations for $^{40}Ca$ nuclei. Panel (a): for three NEOS corresponding to the compressibilities $K=200, 225, 380 MeV$, we show the behavior of the dipolar signals as a function of time; Panel (b) the corresponding oscillation energy distributions. Solid line for $K=200$ MeV,  long dashed line for $K=225$ MeV and short dashed line for $K=380$ MeV.} \label{comdegdr}
\end{minipage}
\end{center}
\end{figure}

\begin{figure}[tb]
\begin{center}
\begin{minipage}[t]{12 cm}
\includegraphics[width=1.1\columnwidth]{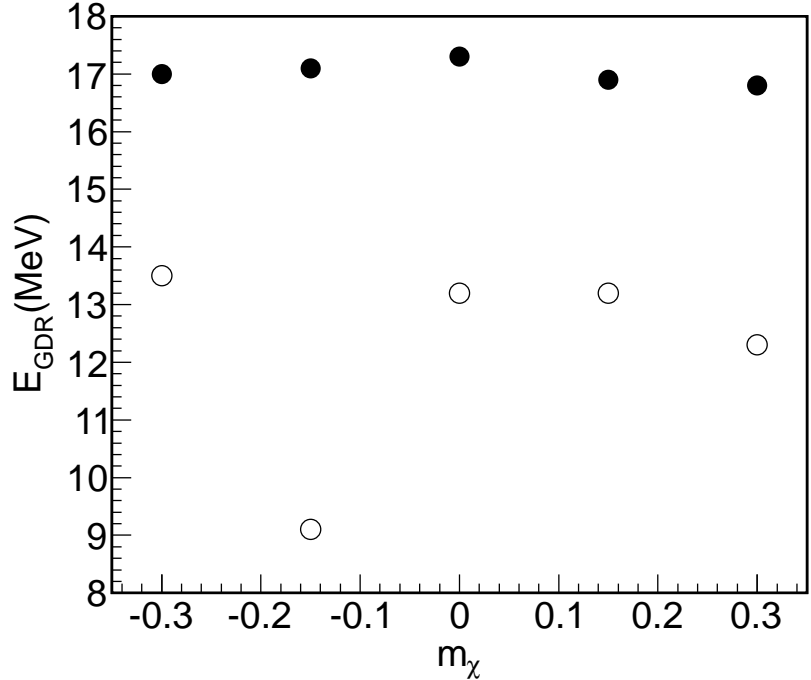}
\end{minipage}
\begin{minipage}[t]{16.5 cm}
\caption{$E_{GDR}$ versus $m_\chi$ from CoMD calculations. Solid circles refer to high energy peaks and open circles refer to low energy peaks.} \label{egdrm}
\end{minipage}
\end{center}
\end{figure}

\begin{figure}[tb]
\begin{center}
\begin{minipage}[t]{12 cm}
\includegraphics[width=1.1\columnwidth]{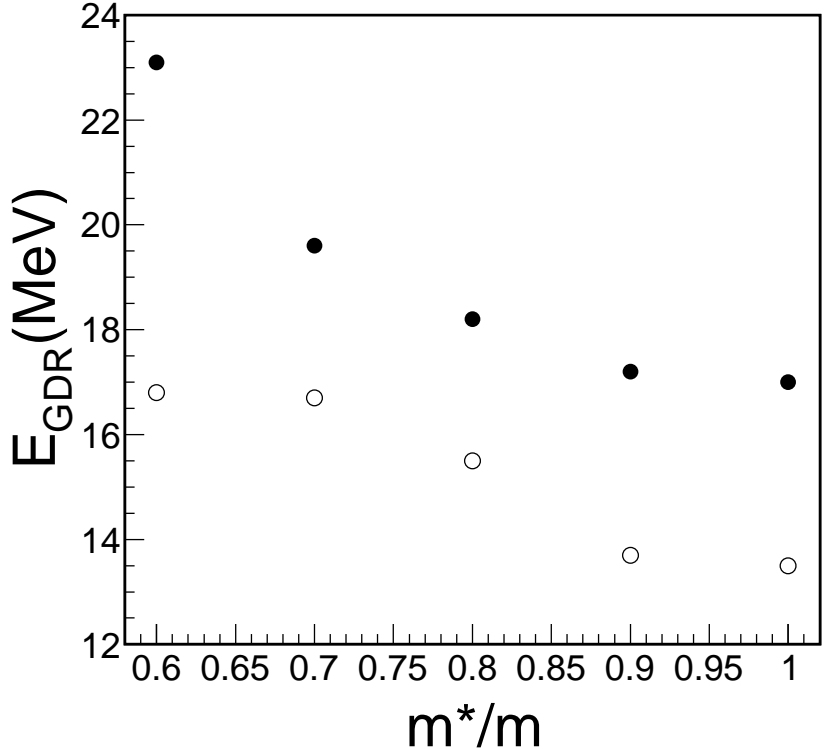}
\end{minipage}
\begin{minipage}[t]{16.5 cm}
\caption{$E_{GDR}$ versus $\frac{m^*}{m}$ from CoMD calculations for $^{40}Ca$ nuclei. Solid circles refer to the high energy peaks and open circles refer to the low energy peaks.}\label{egdrmstar}
\end{minipage}
\end{center}
\end{figure}

\begin{figure}[tb]
\begin{center}
\begin{minipage}[t]{12 cm}
\includegraphics[width=1.1\columnwidth]{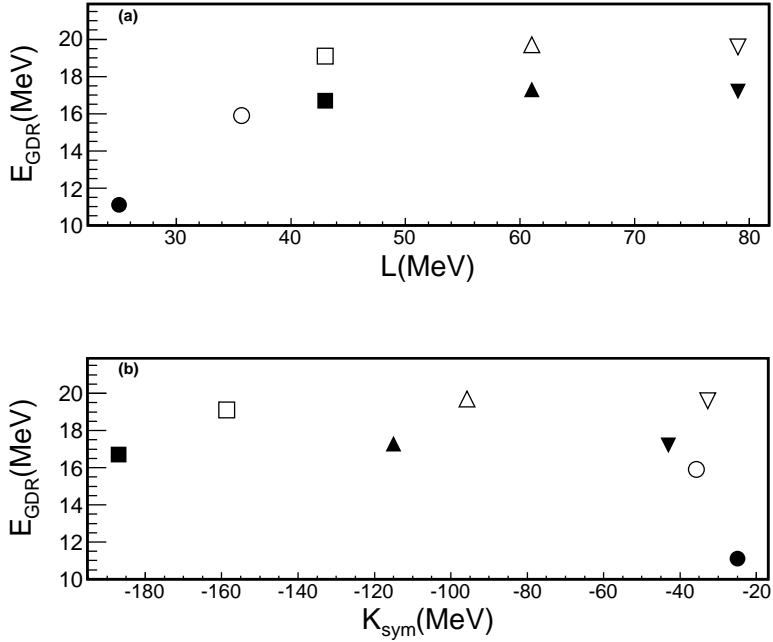}
\end{minipage}
\begin{minipage}[t]{16.5 cm}
\caption{$E_{GDR}$ versus $L$ and $K_{sym}$ from CoMD calculations with (open symbols) and without (solid symbols) momentum dependent interactions for $^{40}Ca$ nuclei. The different symbols refer to different $L$ and $K_{sym}$ couples.} \label{egdrmstarcomd}
%$E_{GDR}$ versus $L$ and $K_{sym}$ with and without effective mass from CoMD.
\end{minipage}
\end{center}
\end{figure}

From the mass formula, assuming the nucleus is a uniform sphere of density $\rho$ we can derive the Coulomb contribution to the symmetry energy for finite nuclei :
\begin{equation}
S_c(\rho) =\frac{a_c}{4}\tilde \rho^{1/3} A^{2/3},\label{sc}
\end{equation}
\begin{equation}
L_{c} = 3\rho_0 \left. \frac{d S_{c}(\rho)}{d \rho}\right\vert_{\rho=\rho_0} = 3(\frac{1}{3}\frac{a_c}{4}A^{2/3}),\label{lc}
\end{equation}
\begin{equation}
K_{symc}=9\rho_0^2 \left. \frac{d^2 S_{c}(\rho)}{d \rho^2}\right\vert_{\rho=\rho_0} = 9(-\frac{2}{9}\frac{a_c}{4}A^{2/3}).\label{ksymc}
\end{equation}
These corrections can be added to the definition of the parameter $x$ for the KDE0-NEOS calculations. The reproduction of the ratio from experimental
data improves as shown in figure \ref{egdrfit} but overall is negligible. This prediction can be tested in RPA calculations by turning off the Coulomb field.

We can try to introduce the surface effects using the values obtained from the IAS analysis \cite{Dan1, Dan2, Dan3}. In particular in equation (\ref{ratiogdr}) we can insert the experimental values of $S(A)=a_a(A)$ but we still have the problem of finite sizes corrections on $L$ and $K_{sym}$ through the quantity $x$.  One simple assumption is to parametrize $a_a=a_\infty f(A)$ as in equations (\ref{iasaafgia1}, \ref{iasaafgia2}) and do the same for the other quantities : $L=L_\infty f_L(A)$ and $K_{sym}=K_{sym\infty} f_{K_{sym}}(A)$. If we further assume $f(A)\approx f_L(A)\approx f_{K_{sym}}(A)$, then the mass dependence is approximately canceled in the quantity $x$ defined above. Notice that such an approximation is true if  Coulomb would be dominating, equations (\ref{sc}, \ref{lc}, \ref{ksymc}). With such an assumption, we can invert equation (\ref{ratiogdr}) and find $x$ for each mass number.  The results are plotted in figure \ref{tryx} where the quantity $x$ is plotted versus $A$ (left panel) and it shows some scattering consequent to the IAS analysis and in turn to nuclear effects not properly corrected for. In the same figure (right panel) we have included, for reference, the values of $x$ versus $\frac{L}{3}-\frac{K_{sym}}{36}$ from some NEOS reported in the literature \cite{shlomo2, stone2, stone3, shlomo0, shlomo1,  skyrme1_1, skyrme1_2, skyrme1_3, skyrme1_4, skyrme1_5, skyrme1_6, skyrme1_7, skyrme1_8, skyrme1_9, skyrme1_10, skyrme1_11, skyrme1_12, skyrme1_13,  skyrme1_14, skyrme1_15, skyrme1_16, skyrme2,  skyrme3,  skyrme4,   skyrme5,  skyrme6, skyrme8,  skyrme9,  skyrme10,  skyrme11,  skyrme12,  skyrme13,  skyrme14,  skyrme15,  skyrme16, ad_skyrme1, ad_skyrme2, ad_skyrme3, ad_skyrme4, ad_skyrme5, ad_skyrme6, ad_skyrme7, ad_skyrme8, ad_skyrme9, ad_skyrme10, ad_skyrme11, ad_skyrme12, ad_skyrme13, ad_skyrme14, ad_skyrme15} and the NEOS in table \ref{c1c2}, open symbols. Of course this analysis is strongly dependent on the assumptions made but it shows how important is the surface term in the IVGDR. Microscopic calculations of course include a surface term and it is usually chosen in order to reproduce ground state properties of nuclei, such as the radii and binding energies. From the figure \ref{tryx} we see that there is an overlap of the $x$ values used in the literature with the result obtained from the ratio, equation (\ref{ratiogdr}). However, the average value of $x$ obtained from the fit gives $x\approx 0.3$, while most NEOS plotted start from about $x=0.4$. The lesson to learn here is that the IVGDR depends on the symmetry part of the NEOS but crucially on surface effects as well. This means that models calculations might reproduce the IVGDR as a consequence of a cancellation among different terms.  

The IVGDR resonances, as well as other resonances, have been explored in RPA calculations \cite{shlomo2, lipparini1}. The calculations have been used to constraint the NEOS, both for the symmetric and the asymmetric part. Unfortunately, some difficulties have been  encountered because of finite sizes, which means
 surface and curvature terms, plus Coulomb.  These calculations and the used mean fields are obtained in the small amplitude limit. However, to get informations
 on the NEOS at densities different from the ground state and high excitation energies, we need suitable models such as TDHF, VUU or AMD. To validate those models, tests should be performed to estimate GR energies and widths. This has been done by some authors \cite{ baranrep05, suraud1, smerzi1} especially for mean field calculations. We have repeated
 some of those calculations for illustration, using a BNV model and checked that the results do not change when increasing the  number of tp. The nuclei were prepared in their
 ground state, choosing a surface term which reproduces fusion cross-sections and the binding energy of nuclei \cite{kondratyev1}. Protons and neutrons are displaced at time zero in momentum space of the quantity:
 \begin{equation}
\Delta P_{z}=\frac{E_{GDR}}{A \hbar},
\end{equation}
\begin{equation}
P^{n}_{z} \rightarrow P^{n}_{z}-(\frac{Z}{A} \Delta P_{z}),  
\end{equation}
\begin{equation}
P^{p}_{z} \rightarrow P^{p}_{z}+(\frac{N}{A} \Delta P_{z}).
\end{equation}

During time evolution of the system we calculate the Dipole moment of the
system in the coordinate and in the momentum spaces:
\begin{equation}
\textbf{DR}(t)=\frac{NZ}{A}[\textbf{R}_p-\textbf{R}_{n}],
\end{equation}
\begin{equation}
\textbf{DK}(t)=\frac{NZ}{A}[\textbf{K}_{p}-\textbf{K}_{n}].
\end{equation}
In figure \ref{bnvmdi} results of BNV calculations with and without momentum dependent interactions are displaced as function of time for $^{40}Ca$. The differences are striking, the calculations underestimate the experimental value $E_{GDR}=20.3 MeV$ when momentum independent
interactions are used. We tried to change the form and strength of the symmetry energy (within reasonable values) but we were not able to reproduce the resonance energy. Also from a search in the literature it seemed that it is a quite general result that local interactions cannot reproduce the IVGDR, thus one could question
the meaning of using those interactions in calculations. When switching to momentum dependent resonances the improvement is dramatic and a good reproduction
of $E_{GDR}$ can be obtained using the NEOS interaction discussed in section \ref{mdeos} with an effective mass $\frac{m*}{m}=0.7$ and a suitable
symmetry energy. The Fourier transforms of the dipole operator, for 
some typical cases, are displayed in figure \ref{bnvlksym}, where we plot the $E_{GDR}$ versus $L$ and $K_{sym}$ for
both momentum dependent and not NEOS.

As we have seen, it is very difficult to reproduce the data with momentum independent interactions which points to some severe constraints to the NEOS.
However, the mean field model could be at fault as well, for instance we know that some nuclei (${}^{12}C$, ${}^{40}Ca$ etc.) display some degree of clusterization
in $\alpha$ particles and this might influence the GR. Thus we performed some calculations using the CoMD model, which has been applied already to study the pre-equilibrium GDR $\gamma$-ray emission in nuclear reactions \cite{papagdr1, papagdr2, papagdr3}, and we believe 
that the GR are a good benchmark for models. The results of the Fourier transform of the oscillations without momentum dependent interactions are displayed in figure \ref{comdegdr}.
Similar features observed for the mean field calculations are obtained in CoMD, with the striking difference that the model displays two peaks whose magnitude
depends on the NEOS-used. While we are expecting the width of the resonance to be overestimated because of the semiclassical nature of the model and
the Monte Carlo method used to fulfill the Pauli principle \cite{papajcp05, papa2} the double peak is a surprise. The best explanation that we have to offer at the moment is
that the model has some degree of clusterization into deuterons and $\alpha$ particles which give rise to anharmonicites of the oscillations. Notice however, in figure \ref{bnvmdi}, that some BNV calculations displayed two peaks as well. Thus anharmonicities might be at play, and those, by definition are not present in RPA calculations. Notice that RPA calculations show peaks in the low energy region that have been interpreted as Pygmy resonances for $m_\chi\ne 0$ nuclei. However, we found now some better
agreement to data even for no momentum dependent interactions. In the calculations, different compressibilities are used and some sensitivity of the resonance is shown. This is a consequence of equations (\ref{l0}, \ref{ksym0}, \ref{s0}) where we showed that the symmetric and asymmetric part of the NEOS cannot be completely disentagled.

We can also explore some questions we asked at the beginning: why the IVGDR parametrization,
equation (\ref{Egdr}) is independent on $m_\chi$?  We performed some calculations for exotic nuclei with fixed mass $A=40$, such as ${}^{40}Fe$, ${}^{40}V$, ${}^{40}Al$, ${}^{40}Si$. The result of the IVGDR frequencies are displayed in figure \ref{egdrm}. The double peaks are present in all cases and we tentatively associate them to the Pygmy resonances \cite{pdr9, pygmy1, pygmy2, pygmy3, pygmy4, pygmy6, pygmy7, pygmy8, pygmy9, pygmy10} even though they are too much
enhanced in the model. The dependence on $m_{\chi}$ is rather weak as it can be seen in the figure.

There is no momentum dependent built into current versions of  the CoMD model available to us.
 Thus we simply modify the model to include a momentum dependent potential $U_p({\bf p}_{i,j})=\alpha  {\bf p}_{i,j}^2$
where ${\bf p}_{i,j}$ is the relative momentum of two nucleons. This gives rise to an effective mass, and should be a valid 
prescription at low excitation energy such as in GR studies. This assumption is similar to the NEOS used in many RPA calculations.
New ground state configurations where obtained for different values
of the effective mass, and the NEOS was adjusted to fix the compressibility $K=200 MeV$ and different values of $L$ and $K_{sym}$. The results of the energy centroid
are plotted in figure \ref{egdrmstar} as function of the effective mass. As for the BNV case, the resonance is very sensitive to the effective mass, the two peaks are always present as reported in figure \ref{egdrmstar}. The dependence from the $L$ and $K_{sym}$ parameters of the symmetry energy is reported in figure \ref{egdrmstarcomd}, the results are in agreement with the BNV
results and those found in the literature \cite{shlomo2}. The interesting new features are that a molecular dynamics model gives reasonable
resonance energies and the appearance of a second resonance which is most probably a model artifact to be refined with detailed studies or anharmonicities also due to some partial $\alpha$ clusterizations in the model. Those resonances could 
have the same origin of the Pygmy ones \cite{pdr9, pygmy1, pygmy2, pygmy3, pygmy4, pygmy6, pygmy7, pygmy8, pygmy9, pygmy10}.

\section{Constraints to the NEOS from heavy ion collisions}
In order to get valuable information on the NEOS we use heavy ion collisions (HIC).  Depending on the beam energy, the system is compressed and excited, expands and fragments if the excitation energy is large enough to bring the system into an instability region \cite{bertschfrag83, latoradyn94}.
At lower excitation energy, fusion or deep inelastic reactions might occur. The nuclei now have a smaller excitation energy and they might emit
small fragments, typically, $p$, $n$ and $\alpha$. Typically, as function of excitation energy, we observe a U-shape mass distribution at small energy
and an exponentially decreasing one at high excitation energy. The mass distributions display a very high yield of $\alpha$ particles with respect to
nearby mass fragments. This is due to the large $\alpha$ binding energy and it could also be a signature of Bose condensation. The  average mass 
distribution behavior as function of excitation energy is typical of a system undergoing a QLG phase transition and it is actually what we expect
from the NEOS at finite temperature discussed in section \ref{eost}. The relevant question is if, in the collision, thermal equilibrium is reached, or if the system is chaotic
enough for us to be able to define a temperature. Furthermore, the system is expanding, thus the density is a function of time \cite{friedman92, tokeexpansion05}. In some statistical models \cite{bondorf1, gross1, aldostat05} a `freeze out density' is defined where the fragments are formed at temperature $T$ and then it expands because of the thermal motion
and Coulomb repulsion. In some models, it is assumed that the fragments formed in the freeze out region are at ground state density and hot, thus a possible
evaporation mainly of $p$, $n$ and $\alpha$ is still possible. The hot fragments are dubbed as primary while the cold ones that reach the detectors are secondary or final fragments. Of course the concept of a system at low density and high $T$ which breaks into pieces at ground state density for later emitting other fragments might be a
little bit over-simplified. Probably, because of fluctuations and N-body correlations, from the low density region different fragments start to form, maybe
with exotic shapes and excitation energies.  The following time evolution cools down the fragments, shape oscillation might produce $\gamma$'s on the way to
the ground state. According to Dorso {\it et al.} \cite{dorso1, dorso2} in dynamical models, such as CMD, one can determine the fate of the system, i.e. predict the yield distribution and
kinetic energy of the final fragments, already in the early stages of the reaction, as short as $10 fm/c$. In the calculations, we know the positions and momenta of each
nucleon, through a minimization procedure in phase space, we can calculate the final distributions \cite{dorso1, dorso2}. This view is in contrast with the assumptions of statistical models, but still might admit the possibility of thermalization since the way the fragments are determined is similar to maximize the entropy.

From the discussion above we infer that the collision is clearly a dynamical process and the challenge is to define quantities as close as possible to temperature, density
 and pressure in order to put constraints on the NEOS. To make this task even harder, we have the Coulomb force which distorts the informations coming from the
 charged fragments, and the neutrons, which are difficult to measure, especially on an event by event basis.
 
\subsection{\it Density and temperature determination in the classical limit}
 Classically, the yield distribution can be calculated in the grand canonical ensemble, for a system in equilibrium \cite{landau, khuang, pathria}. 
 The well known Saha equation gives the ratio of the density of two different fragments from the ratio of their yields \cite{albergot}:
 \begin{equation}
\frac{Y_1}{Y_2}=\frac{\rho(A_1, Z_1)}{\rho(A_2, Z_2)}=(\frac{A_1}{A_2})^\frac{3}{2}(\frac{\lambda^3_{T, \mathcal{N}}}{2})^{A_1-A_2}\frac{2s_1+1}{2s_2+1}\rho_p^{Z_1-Z_2}\rho_n^{N_1-N_2}\exp[\frac{B_1-B_2}{T}],\label{albergo_eq1}
\end{equation}
where $\lambda_{T, \mathcal{N}}=\frac{h}{\sqrt{2\pi m_0T}}$ is the thermal wavelength, $\lambda^3_{T, \mathcal{N}}=4.206\times 10^3 T^{-\frac{3}{2}} fm^3$, $s_i$ are the spins and $B_i$ are the binding
energies of the $i$-fragment. The ratio above depends  on the unknown densities of $p$ and $n$, as well as the temperature. We can write a similar ratio for other fragments, for instance: 
\begin{equation}
\frac{Y_3}{Y_4}=\frac{\rho(A_3, Z_3)}{\rho(A_4, Z_4)}=(\frac{A_3}{A_4})^\frac{3}{2}(\frac{\lambda^3_{T, \mathcal{N}}}{2})^{A_3-A_4}\frac{2s_3+1}{2s_4+1}\rho_p^{Z_3-Z_4}\rho_n^{N_3-N_4}\exp[\frac{B_3-B_4}{T}].\label{albergo_eq2}
\end{equation}
Now we have two equations but still three unknowns. A particular method to obtain the temperature was devised by Rubbino and collaborators \cite{albergot}, and consists in taking
 the ratio of equation (\ref{albergo_eq1}) and equation (\ref{albergo_eq2}):
\begin{equation}
\rho_p^{(Z_1+Z_4)-(Z_2+Z_3)}\rho_n^{(N_1+N_4)-(N_2+N_3)}=\frac{\frac{Y_1Y_4}{Y_2Y_3}}{(\frac{A_1A_4}{A_2A_3})^\frac{3}{2}(\frac{\lambda^3_{T, \mathcal{N}}}{2})^{(A_1+A_4)-(A_2+A_3)}\frac{(2s_1+1)(2s_4+1)}{(2s_2+1)(2s_3+1)}\exp[\frac{(B_1+B_4)-(B_2+B_3)}{T}]}.\label{albergo_eq3}
\end{equation}
By imposing $(Z_1+Z_4)-(Z_2+Z_3)=0$ and $(N_1+N_4)-(N_2+N_3)=0$ we can eliminate the densities from equation (\ref{albergo_eq3}). The equation can be inverted to obtain
$T$, since the binding energies of the fragments are well known. This is a very elegant method and let us obtain the temperature once the fragments yields are known for a given excitation energy and a source size (mass and charge). However, different particles ratios might be taken and it is not guaranteed that for given source condition, they will provide the same temperature. Actually, different fragments might form during the time evolution at different densities \cite{joe2, wada1} or temperatures, which makes
the freeze out assumption questionable. From another point of view, assuming a freeze out, the hot fragments have different excitation energy, thus the
final yields are distorted by secondary evaporation which results in different temperatures for different fragment double ratios. Of course, another natural reason why
different ratios result in different temperatures, is because the system is quantal and not classical, furthermore particles with different quantum statistics, i.e. Bosons or Fermions,
might be mixed in the double ratio.

Within the same classical approximation we can derive the density of protons and neutrons as well, and hence of all particles. For instance let us consider the double ratio
formed with $p$, $n$, $t$ and $^3He$ which are all Fermions, but still using classical statistics.
\begin{equation}
\frac{\rho_p}{\rho_n}=\frac{Y(p)}{Y(n)}, \label{pnth1}
\end{equation}
\begin{equation}
\frac{\rho_p}{\rho_n}=\exp[\frac{0.765}{T}]\frac{Y({}^3He)}{Y(t)}, \label{pnth2}
\end{equation}
\begin{equation}
\rho_p\rho_n=4.35\times 10^{-8}T^3\exp[-\frac{7.716}{T}]\frac{Y({}^3He)}{Y(p)}, \label{pnth3}
\end{equation}
\begin{equation}
\rho_p\rho_n=4.35\times 10^{-8}T^3\exp[-\frac{8.481}{T}]\frac{Y(t)}{Y(n)}, \label{pnth4}
\end{equation}
\begin{equation}
T=\frac{0.765}{\ln \left[\frac{Y(p)Y(t)}{Y(n)Y({}^3He)}\right]}. \label{pnth5}
\end{equation}

From the set of equations (\ref{pnth1}, \ref{pnth2}, \ref{pnth3}, \ref{pnth4}, \ref{pnth5}), one can easily obtain the $p$ and $n$ densities and the temperature $T$. Similarly, it can be done for other particle double ratios \cite{albergot}. Notice that in experiments, usually the neutrons are not measured and they are inferred by assuming that the ratio of $p$ to $n$ is equal to the ratio $^3He$ to $t$. This is not strictly correct since the binding energies of $^3He$ and $t$ are not exactly the same.

We can test this classical method using the CoMD model. A large number of $^{40}Ca+^{40}Ca$ events were generated at $b=1 fm$ impact parameter
and for beam energies ranging from $4 MeV/A$ to $100 MeV/A$. From the beam energy it is possible to define a thermal energy $E^*$ looking at the kinetic energy
distribution of the particles in the transverse direction to the beam \cite{ hua6, hua3, hua4, hua5, hua7}. Looking for physical quantities in the transverse direction gives
a better chance of getting `equilibrated' events.
\begin{figure}[tb]
\begin{center}
\begin{minipage}[t]{14 cm}
\includegraphics[width=1.1\columnwidth]{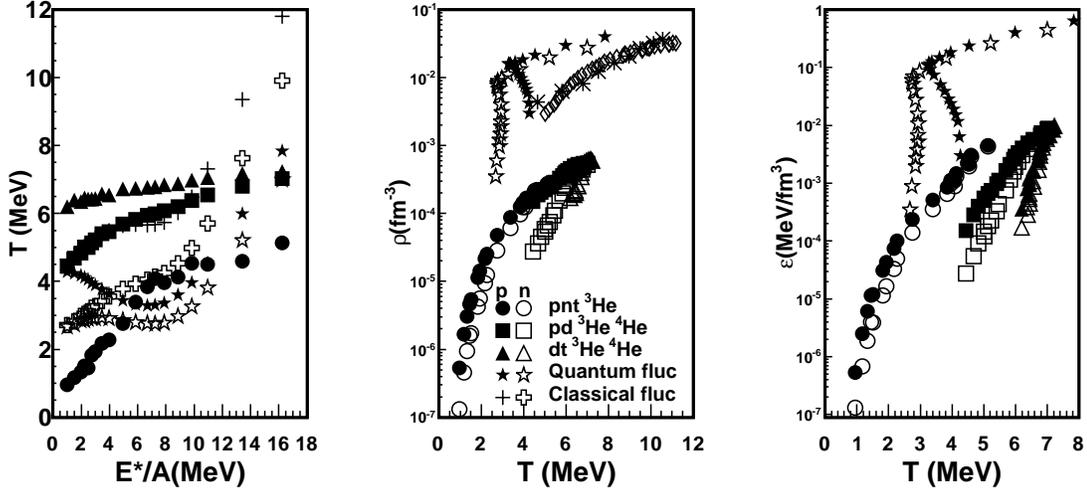}
\end{minipage}
\begin{minipage}[t]{16.5 cm}
\caption{(Left panel) the temperatures from different thermometers versus excitation energy per particle; (Middle panel) densities versus temperatures, the asterisks \cite{justin} and the diamonds refer to experimental results \cite{joe2, wada1, joe1}; (Right panel) energy densities versus temperatures. The calculations are performed with CoMD for $^{40}Ca+^{40}Ca$ collisions.}\label{drcomd}
\end{minipage}
\end{center}
\end{figure}
In figure \ref{drcomd} we plot the $T$ versus $E^*/A$ (left panel), $\rho_{p,n}$ (middle panel) and energy density $\varepsilon_{n,p}$ (right panel) versus $T$ for different
particle double ratios: (1) $pnt{}^3He$, (2) $pd{}^3He{}^4He$, (3) $dt{}^3He{}^4He$. The three cases give quite different temperatures when plotted versus the excitation energy. However, when density and energy density are plotted versus $T$, it seems almost as one case is the continuation of another. The collapse onto a curve is not perfect,
 which could be the consequence of model calculations stopped at $1000 fm/c$. However, in all cases the neutron densities are much smaller than the proton ones at the same $T$, similarly for energy densities. This is clearly a Coulomb effect which results on the proton densities being larger than the neutron ones. As we will show this effect is common to other methods discussed later and it is a consequence of the difference between proton and neutron radii discussed in section \ref{neutronskin}. In the figure we have plotted results from the model and experiments using different methods. The striking feature to be noticed is that the different methods gives similar ranges of $T$ while the densities differ some order of magnitude. This is a feature similar to the one discussed for the NEOS in the classical limit. Some quantities are reasonable while others differ from the exact calculations depending how close we are to the classical limit. As we stressed before,
 the classical limit is never recovered in the NEOS studies and the same effect can be observed in the density results.
 
\subsection{\it Coalescence Model}
 The failure of the classical thermal model \cite{ropke2} suggests that there might be another mechanism for cluster formation at play at higher densities. In heavy ion collisions
 the excitation energy or temperature might be high and the system expands quickly. Cluster formation might occur during the expansion in presence of a third body. For instance,
 the mechanism of deuterium formation might be $p+n\rightarrow d+\gamma$, but such a mechanism is too slow as compared to the expansion time of the nucleus \cite{csernairep86, mekjian1, mekjian2}.
 More phase space is available, thus larger reaction rates, in presence of a third body, for instance  $p+n+N\rightarrow d+N$, in this process the extra energy and momentum in the fusion process are taken
 by a third particle \cite{aldorep94, danielewicz1}. This is the basis of the coalescence model and essentially we assume that if two particles are within a sphere of radius $P_0$ in momentum space,
 they can coalesce to form a new species. If, in the experiments, we can measure precisely the momentum or energy distributions of fragments of mass $A$ and charge $Z$, plus the distributions of protons and neutrons,
 we can derive the value of $P_0$ \cite{mekjian1, mekjian2}. Neutrons are usually not measured, thus one uses the proton distribution and a correction for Coulomb \cite{awes1}. With all those assumptions
 and simplifications we can write a relation from which $P_0$ can be derived from the energy distributions of the fragments:
 \begin{equation}
\frac{d^2N(Z,N,E_A)}{dE_Ad\Omega}= R^N_{np}\frac{1}{N!Z!A}(\frac{4\pi P_0^3}{3[2m^3(E-E_c)]^{0.5}})^{A-1}(\frac{d^2N(1,0,E)}{dEd\Omega})^A.\label{coal}
\end{equation}
 $R^N_{np}$ is the ratio of neutrons to protons of the source, $E_c$ is the Coulomb correction. Thus for each fragment type, $d$, $t$, etc.,  a value of $P_0$ can be obtained. Notice that the coalescence model
 takes into account the presence of other bodies when fragments are formed. Pauli blocking is one of those effects which could be taken into account indirectly through the value of $P_0$ \cite{ropke2}.
 Compared to the classical thermal model, coalescence occurs at relatively high densities and temperatures. It does not occur, for instance, during the big-bang expansion, since densities are too low 
 at the time when nuclear processes are dominant \cite{bigbang1, bigbang2, bigbang3, bigbang4, mekjian1, mekjian2}. It might occur in relativistic heavy ion collisions when the quarks and gluons coalesce to form hadrons \cite{greco1, starantim}. 
 
 A further variation of the coalescence model, was proposed by Natowitz and collaborators \cite{ joe2, wada1, joe1, hagel3}, which consists in deriving the parameter $P_0$ as function of the velocity 
 of the particles in the reference frame of the emitting source after correcting for Coulomb. This is like following the time evolution of the system, in fact to higher velocities correspond shorter times and higher temperatures.
 The question now is how to derive the density of the system from the knowledge of $P_0$. A non-equilibrium model was proposed in ref. \cite{sato1} which assumes the knowledge of the fragment wave function, say the deuteron
 in the source, and connects $P_0$ to its volume in coordinate space. A less general approach, but more suitable to our goals was proposed by Mekjian \cite{mekjian1, mekjian2} and assumes thermal and chemical equilibrium:
 \begin{equation}
V=\frac{3(2\pi\hbar)^3}{4\pi P_0^3}[\frac{Z!N!A^3}{2^A}(2s+1)e^{\frac{B_A}{T}}]^\frac{1}{A-1}.\label{VP0}
\end{equation}
$B_A$ and $s$ are the ground state binding energy and the spin of the fragment. The temperature $T$ can be determined using other methods such as the double ratio method discussed in the previous section \cite{ joe2, wada1, joe1, hagel3, tsang3, tsang4, tsang5}. Experimentally derived values of $P_0$ as function of the surface velocity 
are within $90 MeV/c$ and $160 MeV/c$ for $^{40}Ar+^{112}Sn$ at $47 MeV/A$ \cite{wada1} similar to early results \cite{csernairep86, mekjian1, mekjian2}.
 The temperature can be derived from the double ratio method using $d,\alpha,t,^3He$ measured yields and the volume from equation (\ref{VP0}). In particular a source radius can be defined as:
 \begin{equation}
V=\frac{4\pi}{3}\bar R^3.\label{rsource}
\end{equation}

From the experiment we know the multiplicities of the particles as well, thus the density can be obtained for each source velocity together with the temperature from the double ratio method \cite{ joe2, wada1, joe1, hagel3}.
The results are plotted in figure \ref{drcomd} (middle panel) and they are much higher than the CoMD calculations in the classical statistical limit discussed in the previous section \cite{ropke2}, while they are in closer agreement
with model calculations and experimental data \cite{justin} using fluctuations to derive $\rho$ and $T$ as discussed below.

 \subsection{\it Density and temperature from quantum and classical fluctuations}
 The spirit of the Thomas-Fermi (TF) approximation \cite{ring} is to consider the density locally constant and derive the Fermi momentum for each density.  This approximation can also be generalized at finite temperatures. Now we can try to `invert' this procedure, i.e. starting from some
 physical observables, average multiplicities, kinetic energies etc., can we derive the (local) density and temperature of the system?  This is what we have done in the previous sections  assuming classical distributions. In the TF approximation we assume that the distribution is given by a finite temperature Fermi-Dirac. In nuclei we have Fermions as well as Bosons, $d$, $\alpha$ etc., thus we can generalize the approach
 to include Bosons. In general the distribution function for elementary particles at temperature $T$ is given by:
  \begin{equation}
f(p,\rho)=\frac{1}{e^{(\varepsilon-\mu)/T}\pm 1},\label{FB}
\end{equation}
where the chemical potential $\mu$ is connected to the density. The equation above refers to elementary particles and we can consider $p$ and $n$ as elementary particles at the excitation energies of interest in this work. Other particles are composites, i.e. made of $p$ and $n$, thus for instance an $\alpha$ particle is a Boson made of Fermions, thus the Pauli principle will play a role in all cases \cite{hua4}. We will not discuss this problem further in this work and we will consider them as elementary particles for illustration. For further discussion see \cite{hua6, hua3, hua5, hua7}.

Equation (\ref{FB}), contains two unknowns, $T$ and $\mu$, thus in order to fix them we need two observables. Since we are dealing with heavy-ion collisions, where non-equilibrium effects are important, we should choose observables which can give the closest approximation to a `temperature'. In the spirit of the fluctuation-dissipation theorem \cite{landau, khuang, pathria}, looking at fluctuations  gives the largest chaoticity, thus the closest approximation to an ergodic system. Of course chaoticity is not enough to ensure that the system is in thermal (and chemical) equilibrium, but it is probably  the closest we can get. Once the most chaotic 
`events' have been determined (largest fluctuations) we can also look at average quantities and check how close are the derived $T$ and $\rho$ for different observables. For instance we can use fluctuations and also the transverse kinetic energy to
determine $T$, the closest the results are, the closest to thermal equilibrium the system is. To think that we can get completely equilibrated events is a dream and we should settle for the closest we can get and, helped by models, correct for finite sizes,
Coulomb and dynamical effects.

In \cite{wuenschel1} it was proposed to look at quadrupole fluctuations in momentum space to determine the classical temperature. The quadrupole in the transverse direction was defined in order to minimize non-equilibrium effects:
\begin{equation}
Q_{xy} = p_x^2-p_y^2.
\end{equation}
In the classical case, i.e. Maxwell distribution, the temperature and density are decoupled, thus the quadrupole fluctuations properly normalized are function of $T$ only:
\begin{equation}
\langle \sigma^2_{xy} \rangle =(2mT)^2.
\end{equation}
If we use the correct quantum distribution for Fermions equation (\ref{FB}) to calculate the quadrupole fluctuations, our results will depend on the chemical potential $\mu$ as well. We need another condition which we choose to be
the multiplicity fluctuations. This  choice is dictated by the fact that multiplicity fluctuations are equal to one for a classical ideal system, while for Fermions (Bosons) are smaller (larger, near and above the critical point for Bose condensation)
than one \cite{esteve1, sanner1, muller1, westbrook1}. For Fermions we get \cite{hua4}:
\begin{equation}
\langle \sigma_{xy}^2 \rangle = (2mT)^2[0.2(\frac{T}{\varepsilon_f})^{-1.71}+1],
\end{equation}
\begin{equation}
\frac{T}{\varepsilon_f}=-0.442+\frac{0.442}{(1-\frac{\langle(\Delta
N)^2\rangle}{\bar N})^{0.656}}+0.345\frac{\langle(\Delta N)^2\rangle}{\bar N}-0.12(\frac{\langle(\Delta N)^2\rangle}{\bar N})^2\label{fitfm}.
\end{equation}
For Bosons the results are a little bit more complicated but more interesting, since they might undergo Bose-Einstein condensation (BEC) \cite{ohkubo1, ohkubo2, kokalova1, tohsaki1, ropke1, sogo1, funaki1, funaki2, raduta1, schuck1, pibec81}. If BEC could be somehow confirmed in HIC, it would open an interesting field of research since we have a system where Fermions and Bosons are somewhat mixed \cite{paolabec}. We have some debate on the possibility of BEC started from
the observation of Hoyle states \cite{hoyle1, chernykh1, jensen1, epelbaum1}, $\alpha$ decay, large $\alpha$ yields, densities in HIC \cite{joe2, hagel1, hagel2} and so on. 

To explore this problem even more, we have to deal with some effects that might distort
the results, the first one is Coulomb. The ideal would be to measure neutron distributions and multiplicities as function of the excitation energy, mass and charge of the source, which is complicated from an experimental point of view,
but maybe not impossible. A comparison of $p$ and $n$ will point to Coulomb corrections. The dynamics of the nucleons inside the nuclei are of course affected by Coulomb and there is nothing we can do about it. But a charged particle which
leaves an excited system will experience a Coulomb acceleration.  Thus we expect that the quadrupole fluctuations will be distorted by Coulomb, since a quadrupole distribution which changes in time, or in different
events, will result to different accelerations to the charged fragments which leave the surface of the system. In refs. \cite{hua6, hua7} it was proposed to take into account the Coulomb distortions and correct for them when possible. A method
borrowed from electron scattering was adopted and applied to classical as well as quantum systems. In the model calculations, it was observed that when taking into account those effects, the $T$ of $p$ and $n$ (as well as composite Fermions in the classical case)
are very similar, while the densities are not affected by the corrections. This is simple to understand, a proton at the surface of the system can be emitted because of the extra Coulomb push as compared to a neutron. Thus neutrons might explore lower densities \cite{joe2}, the physics is similar to the $\Delta r_{np}$ we discussed in section \ref{neutronskin}. 

\begin{figure}[tb]
\begin{center}
\begin{minipage}[t]{12 cm}
\includegraphics[width=1.0\columnwidth]{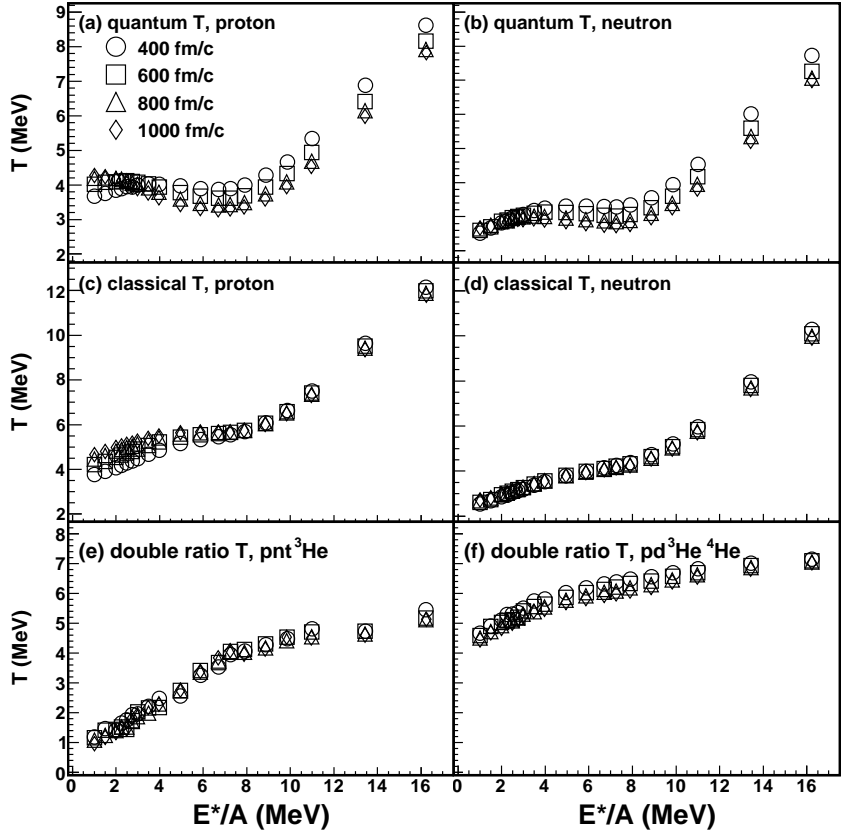}
\end{minipage}
\begin{minipage}[t]{16.5 cm}
\caption{Temperature time evolution for different thermometers from CoMD calculations.} \label{temptevo}
\end{minipage}
\end{center}
\end{figure}

The Coulomb correction is more interesting for the Boson cases, since we now introduce a repulsive force in the system. As it is well known, for an ideal Bose gas, the compressibility below the critical point
is zero, which results in diverging fluctuations and unphysical results. This can be corrected in presence of repulsive forces as it is well known from the literature \cite{khuang}. In our case we are using the Coulomb term which is a long
range force, thus we are in a situation very different from previous studies.

In all these discussions the problem remains on the role of interactions. We have assumed that density and temperature can be obtained using the quantum distributions equation (\ref{FB}), plus Coulomb corrections, and now
we can take into account interactions in the determination of the excitation energy. In fact, we can determine, both experimentally (using $4\pi$ detectors) or theoretically, the fragment distributions event by event. From their kinetic energies and 
difference of Q-value (which includes the interactions) of the fragments and the initial source, it is easy to determine the excitation energy. The same assumption were of course made in the previous section for the classical
and coalescence models, where there is no information about the interactions among fragments and the source.  In figure \ref{drcomd} some results are reported from CoMD calculations and experiments \cite{justin, joe2, wada1, hua3, hua4, joe1}. The densities obtained are much higher than the classical thermal model, but they are very close to the coalescence approach, which can be explained from the fact that the $P_0$ derived from the data
contains many-body effects such as Pauli blocking which is an essential ingredient of the fluctuations approach.

The use of multiplicity and quadrupole fluctuations is not the only possibility to obtain density and temperature in HIC. Borderie {\it et al.} \cite{borderie1, borderie2, borderie3} have derived the temperature from quantum and classical limits
as above and the density from other methods, such as interferometry. In particular, through selections of the data, either the volume and/or the pressure have been constrained, and caloric curves
have been derived. When constant pressure is imposed, a back bending in the caloric curve results which suggests a first order phase transition as predicted long ago in micro canonical statistical model calculations 
by D. Gross \cite{gross1, borderie3, pochodzalla1}. 

It is instructive to study the `time evolution' of the temperature using the different methods discussed. In figure \ref{temptevo} the temperature time evolution is displayed starting from 400 fm/c, a time when fragments are reasonably recognized in coordinate space. The fluctuation method has been applied for $p$ (without Coulomb corrections) and $n$. In the quantum case we notice that the temperature is saturating around 800 fm/c and the difference with earlier times is however small. It is interesting to notice a change in the time behavior of $T$ for protons. For low excitation energies the temperature increases with time until saturation, the opposite occurs at high excitation energies. The neutron temperature is saturated already at earlier times for the smallest excitation energies. Furthermore, for the classical case, this is true for $n$ and all excitation energies, while for $p$, it takes sometime until saturation is reached at small excitation energies. This is an effect of the Coulomb potential which distorts the fluctuations slightly. A similar behavior is observed when calculating $T$ from double ratios, even though the actual values of $T$ are different for each cases. The model seem to indicate an early saturation of the temperature similar to ref. \cite{aldonc00}. Probably, the time variation observed in figure \ref{temptevo} are comparable with the experimental error bars using the different methods \cite{justin, joe3, wada1, ropke2, joe1, wuenschel1, borderie1, borderie2, borderie3, pochodzalla1}.

The fluctuations method has many similarities with $pp$ correlation functions as first proposed by Koonin  \cite{koonin1, pratt1, gong1, korolija1, handzy1} for heavy ion collisions. Indeed both methods are based on the quantum nature of the 
nucleons and composite fragments, thus the informations obtained from both methods should be complementary. It would be very interesting to apply all these methods to data coming from the same experimental
scenario, using high resolution $4\pi$ detectors.

 \subsection{\it The Fisher model}
 The fragmentation observed in heavy ion collisions might be a signature of the QLG phase transition discussed in section \ref{eost}. Looking at typical
 mass distributions, we observe a U-shape for moderate or low excitation energy. In the QLG framework, this signals a liquid surrounded by small drops.
 For increasing excitation energies, the size of the liquid becomes smaller and the amount of drops increases while bigger sizes are observed at smaller
 probabilities. At very high excitation energies the liquid disappears, only small fragments are observed, and the mass distribution becomes exponentially decreasing.
 In between these two limiting cases, we can get a particular excitation energy (or temperature if the system is in equilibrium), 
 where the mass distribution is a power law with a typical exponent $\tau\approx 2.3$ \cite{cpma95, cpma99, cpmaprl99, cpmaepja99, cpma05, huangfisher10}.
 When this occurs the system is near a second order phase transition and $\tau$ is one critical exponent \cite{landau, khuang, pathria, aldonc00}. This feature is observed in macroscopic systems, as well
 as some geometrical models such as percolation \cite{perc1}. Other critical exponents can be defined and we have already discussed $\beta$.  The values of the critical exponents can be
  defined in classes such as the liquid-gas has $\beta=0.33$, $\gamma=1.23$, $\nu=0.63$, $\sigma=0.64$, while 3D percolation has different values: $\beta=0.41$, $\gamma=1.8$, $\nu=0.88$, $\sigma=0.45$. Apart these differences, the qualitative behavior of
  systems undergoing a second order phase  transition is very similar, displaying power laws, for instance the mass distribution or the specific heat, near the critical point.

  The mass distribution is usually discussed in the 
  framework of the Fisher model.  In general, for an equilibrated system, the probability of finding a fragment of mass $A$ can be calculated from the free energy. If the system is at constant 
  pressure, the Gibbs free energy is the suitable quantity to look at, while at constant volume (as assumed in many models) the Helmholtz free energy is the relevant one. One or the other prescription gives some differences especially above the critical point \cite{sobotka1}. Assuming a free energy $F$, we can write the mass yield as:
  \begin{equation}
Y(A,Z)=Y_0A^{-\tau}e^{-\frac{F(A,Z)}{T}}.\label{fisher}
\end{equation}
 $Y_0$ is a constant which can be obtained normalizing  to the total mass of the source: $A_s=\sum AY(A,Z)$. If  charges only are detected, then the normalization must 
 be performed on the charge. The power law term was first introduce by Fisher \cite{fisher} and it is due to the increase in entropy when a new fragment with circumference $2\pi R$
 is created. The free energy might be written as: $F=E-TS$, the internal energy $E$ and the entropy $S$; thus the Fisher entropy can be included in the last term as well.
  The Purdue group was the first to utilize the Fisher model and reproduce experimental data on $p$ induced fragmentation \cite{purdue1, purdue2, purdue3}. They assumed that the internal energy
  had similar terms to the mass formula for $T<Tc$, i.e. a volume, surface, symmetry, Coulomb and pairing terms, the difference now that all those terms depend on the 
  T and $\rho$ of the source, i.e. on the NEOS. Since the density is predicted to be lower than the ground state density, the Coulomb term will be reduced. Near the critical point, the surface term depends on the critical temperature as:
  \begin{equation}
a_s(T,\rho)=a_{s0}(T,\rho)(T_c-T)^{2\nu}.\label{critical}
\end{equation}
Which defines the critical exponents $\nu=0.65$ \cite{landau} for a liquid-gas phase transition and $\nu=0.5$ in mean field theory \cite{landau,khuang, pathria}.
The term $a_{s0}(T,\rho)$ might still contain a smooth temperature dependence as well density dependence, but at zero temperature should coincide with the surface
term in the mass formula \cite{dasrep05, purdue2, purdue3, goodmanfish84, belkacemfish95}. The volume and the symmetry terms are essentially given by the finite temperature NEOS discussed
in section \ref{eost}, thus they might be constrained by precise data. The density and the temperature explored by the systems might be determined using
 the methods discussed in the previous sections.
 Above the critical point, $T>T_c$, the surface term disappears and only the volume term (including the symmetry energy and entropy) remains, plus Coulomb. The fate of
 the pairing term is not well known, we might speculate that it will disappear at some temperature, not necessarily the same temperature of the QLG phase transition. While at $T<T_c$,
 the mass yield displays a U shape because of the competition of the surface and volume term, for $T>T_c$ the mass distribution becomes exponentially decreasing
 due to the volume term (and Coulomb). At the critical temperature the mass distribution is a power law and the volume and surface term cancel each other. Of course
 the power law is distorted by Coulomb if this term is strong enough, thus we have always an exponentially decreasing term due to Coulomb which naturally
 does not favor large charges. However, since the density is low and fragments might be deformed, the Coulomb term becomes smaller. We would like to stress that in finite systems, we might not be able to reach and go above the critical temperature. In fact, increasing the excitation energy results in an increase of collective flow, since the system cannot be confined long enough to develop large fluctuations. This might be the reason for the limiting temperature discussed in the literature \cite{joe4} and it has been shown in dynamical models from the saturation of the Lyapunov exponents \cite{aldo3}.
 
  \begin{figure}[tb]
\begin{center}
\begin{minipage}[t]{12 cm}
\includegraphics[width=1.1\columnwidth]{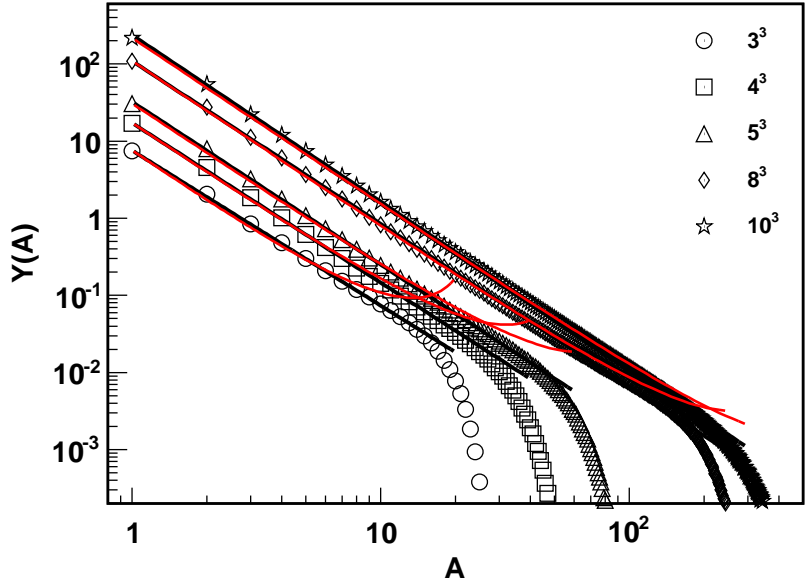}
\end{minipage}
\begin{minipage}[t]{16.5 cm}
\caption{(Color online) The mass distribution of the percolation model near the critical point for different sizes.  The black lines are fitted with $cA^{-\tau}$, the red lines are fitted with $c [\frac{(A_s-A)A}{A_s}]^{-\tau}$.} \label{percolationya}
\end{minipage}
\end{center}
\end{figure}
 
 Another feature which could be confused with the Coulomb suppression of large masses is the finite system sizes. Indeed, we are mimicking the QLG phase transition of finite nuclei
 with an infinite repetition of ensembles, but clearly we cannot take the limit of a fragment mass going to infinity: at most it can be the size of the initial source. A similar problem can be studied in percolation models which exhibit a second order phase transition at the critical control parameter $p_c$ which plays the role of the critical temperature in QLG, see equation (\ref{tperc}). In the model,
 different system sizes can be studied and a change of the critical point and a modification of the critical exponents are found \cite{perc1} as we have already discussed in figure \ref{finitetca}. In refs. \cite{elliott1, elliott2, elliott3}, modifications to the Fisher model have been proposed to take into account the finite size of the system. A first modification regards the Fisher term which
 becomes:
\begin{equation}
A^{-\tau}\rightarrow [\frac{(A_s-A)A}{A_s}]^{-\tau}. \label{elliot}
\end{equation}
 Taking the limit of infinite mass source, we recover the usual Fisher term, and it has been tested in the Ising model for finite sizes. The same test can be performed 
 in the percolation model and the results are displayed in figure \ref{percolationya}. As we see the modification increases for $A\approx A_s$. This increase mimics a liquid part which is not present at these $p_c$ and thus it is unphysical.  From the figure we can conclude that the error done by using the original power law fit is smaller than that of the proposed correction, equation (\ref{elliot}).

 On similar grounds the surface term has been modified to take into account finite sizes as:
 \begin{equation}
a_{s0}(T,\rho)(T_c-T)^{2\nu}A^{2/3}\rightarrow a_{s0}(T,\rho)(T_c-T)[(A_s-A)^{2/3}+A^{2/3}-A_s^{2/3}].\label{alliot2}
\end{equation}
Notice that, because of the assumptions made to derive equation (\ref{alliot2}), the value of the critical exponent is $\nu=0.5$, which is the mean field in contrast with the experimental value $\nu=0.6-0.7$ \cite{landau,khuang, pathria}. The coefficient $a_{s0}=18 MeV$ is  constant in ref. \cite{elliott2}. Other authors have used a slightly different form for the surface term:
\begin{equation}
a_s=18(1+\frac{3}{2}\frac{T}{T_c})(1-\frac{T}{T_c})^{3/2},
\end{equation}
which corresponds to a critical exponent $\nu=0.75$ \cite{purdue1, goodmanfish84, belkacemfish95}. The analysis of ref. \cite{elliott2} is performed in a temperature range $\frac{T}{T_c} \approx 1/3$ where the two approximations differ most. In their experiment, the authors \cite{elliott1, elliott2, elliott3} have measured the excitation energy of the fragments and obtained the temperature from the parametrization:
% \begin{equation}
% E^*=\frac{1}{k}A T^2,\label{elliot3}
 %\end{equation}
 \begin{equation}
 E^*_s=\frac{1}{8}\frac{1}{1+\frac{A_sE_s^*}{E_s^{bind}}}T^2,
 \end{equation}
where $E_s^{bind}\approx 8A_s MeV$ is the binding energy of the source. The values of $T$ obtained with such a parametrization are in very good agreement with the values obtained from quantum fluctuations \cite{justin} and
 double ratio \cite{joe3} reported in figure \ref{drcomd}. A careful fit of the experimental mass distribution gives $T_c=17.9\pm 0.4 MeV$, $\rho_c=0.06\pm 0.01 nucleons/fm^3$
 and it is plotted in figure \ref{finitetca} and previously discussed in section \ref{mf}. Notice that the other quoted experimental results have not been corrected for finite sizes and Coulomb,
 but they seem to indicate a slightly lower temperature if the percolation model \cite{perc1} or the ZR1 NEOS results are used as a guide \cite{mekjian0, goodmanfish84}, see section \ref{mf}. The reason for the discrepancy could be the same which gives a different critical exponent $\nu$ and surface term, and could be investigated. Furthermore, as we have discussed different neutron concentrations might change the values of $T_c$ and $\rho_c$, and no information on the concentration of the source is given \cite{elliott1, elliott2, elliott3}. The small modifications discussed here do not diminish the elegance and power of the method envisaged by J. Elliott {\it et al.} \cite{elliott1, elliott2, elliott3} to derive the NEOS near the critical point after correcting for finite sizes.
 
 Within the Fisher model, we can derive the pressure, thus constrain the NEOS \cite{elliott1, elliott2, elliott3, aldo2, huang, fisher,  kubo1, campi1, campi2}:
 \begin{equation}
 \frac{P}{\rho T}=\frac{M_0}{M_1}.\label{pfish}
 \end{equation}
 $M_k$ are the mass moment distributions given by \cite{aldonc00, campi1, campi2}:
 \begin{equation}
 M_k=\sum A^kY(A,Z).\label{Mfish}
 \end{equation}
 Knowing the density and temperature, for instance from fluctuations or from the coalescence model, discussed in the previous sections, we can constrain the pressure and
 thus the NEOS in the QLG region. We could estimate the pressure using the quantum fluctuation method applied to $p$ and $n$, however in such approach interactions
 are neglected. On the other hand, in the Fisher model, interactions are included through the fragment distributions. This is similar to the estimate of the excitation 
 energy of the system where all the fragments kinetic energies and their Q-values are included. At the critical point, the Fisher model gives:
 \begin{equation}
 \frac{P_c}{\rho_c T_c}=\frac{M_{0c}}{M_{1c}}=0.39,\label{pfishc}
 \end{equation}
 which is the Van Der Waals value in disagreement with experimental values plotted in figure \ref{prhotc}. This is a shortcoming of the model, which could partially be
 resolved by determining density and $T$ experimentally as in section \ref{eost}. This approach has been adopted by different authors \cite{justin, elliott1, elliott2, elliott3, aldo2, huang}, and the results reported in figure \ref{prhotc}, are in good agreement with real gases \cite{gugg}.

 \subsection{\it Free energy}
\begin{figure}[tb]
\begin{center}
\begin{minipage}[t]{12 cm}
\includegraphics[width=1.1\columnwidth]{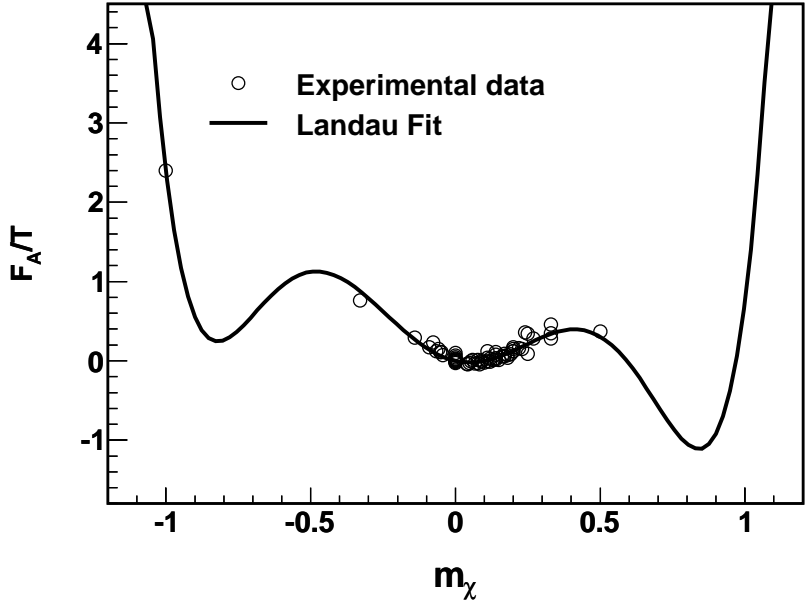}
\end{minipage}
\begin{minipage}[t]{16.5 cm}
\caption{Free energy versus the order parameter $m_\chi$. The data have been corrected for pairing \cite{justin1}. The full line is a fit using Landau's theory $O(m_\chi^6)$.} \label{isofreeEm}
\end{minipage}
\end{center}
\end{figure}

 As reported in equation (\ref{fisher}), the free energy determines the mass distributions. Knowing from experiments or models the distributions for each temperature and density,
 we can actually derive the free energy by inverting equation (\ref{fisher}). In particular we can define the ratio:
 \begin{equation}
R=\frac{Y(A,Z)A^\tau}{Y(^{12}C)12^\tau},\label{ratio}
 \end{equation}
where we have normalized to the yield of $^{12}C$ to get rid of the normalization constant. The free energy per particle $F_A$ is simply given by:
\begin{equation}
 \frac{F_A}{T}=-\frac{\ln R}{A}.\label{free}
 \end{equation}
As we discussed above, the free energy should contain the same terms included in the mass formula plus corrections due to the entropy. Note that in the definition
 equation (\ref{ratio}) we have taken care of the Fisher entropy. Thus if we plot the ratio versus the nuclear ground state binding energies we should get scaling. The experimental data \cite{huang, aldo5} does not show such a scaling, not even when the ratio is plotted as function of each term in the mass formula, i.e. Coulomb, surface or volume.
  It scales nicely when we plot as function of the symmetry energy.  This suggests that in the QLG region, the symmetry energy becomes dominant, thus $F_A\rightarrow F_A(m_\chi)$ and
  $m_\chi$ is an order parameter. A plot of the free energy as function of the order parameter is displayed in figure \ref{isofreeEm}. The data was obtained in the reactions ${}^{64}Zn+{}^{64}Zn$, ${}^{64}Ni+{}^{64}Ni$ and ${}^{70}Zn+{}^{70}Zn$ at 35 MeV/A. The experiments have been performed at the Cyclotron Institute of Texas $A\&M$ University \cite{huang, justin1}. The free energy was corrected for pairing, similar to the mass formula, and a good scaling is seen in the figure \cite{justin1, tripathi1, marini1}.
  
  Since the free energy depends on essentially one order parameter, we can write it in terms of Landau's theory of phase transitions:
  \begin{equation}
\frac{F_A}{T}= \frac{1}{2}a m_\chi^2+\frac{1}{4}b m_\chi^4+\frac{1}{6}c m_\chi^6-\frac{H}{T}m_\chi.\label{O6}
 \end{equation}
Notice the similarity to equation (\ref{symexp}) for the ground state symmetry energy. $\frac{H}{T}$ is the `external' field which enters in Landau's theory to take into account,
for instance, of the difference of protons and neutrons of the source \cite{landau, khuang, pathria}. It could be an external magnetic field when dealing with a spin model \cite{khuang}. Landau's theory is an expansion in terms of the order parameter and in equation (\ref{O6}) we have truncated it to sixth order. Fits to different data show that lower order
expansions, in particular second order, does not reproduce the data well. A fourth order expansion, which would indicate a second order phase transition, does not reproduce the
data as well \cite{huang, aldo5, justin1, tripathi1, marini1}. The expansion in equation (\ref{O6}) gives a good reproduction of the experimental data and in particular suggests a first order
phase transition \cite{ad_nsc1, ad_nsc2} and the possibility of a tri-critical point, similar to a ${}^3He-{}^4He$ liquid mixture \cite{khuang}. 

It is surprising that scaling occurs as function of isospin only, in fact we would expect other terms in the mass formula to be at least as important. A careful analysis of the same
experimental data, including all terms in the mass formula plus entropy corrections, gives indeed a good reproduction of the data, but again the dominant term is the symmetry
part as it is clear from the scaling plotted in figure \ref{isofreeEm} \cite{huangfisher10}. 

\begin{figure}[tb]
\begin{center}
\begin{minipage}[t]{12 cm}
\includegraphics[width=1.1\columnwidth]{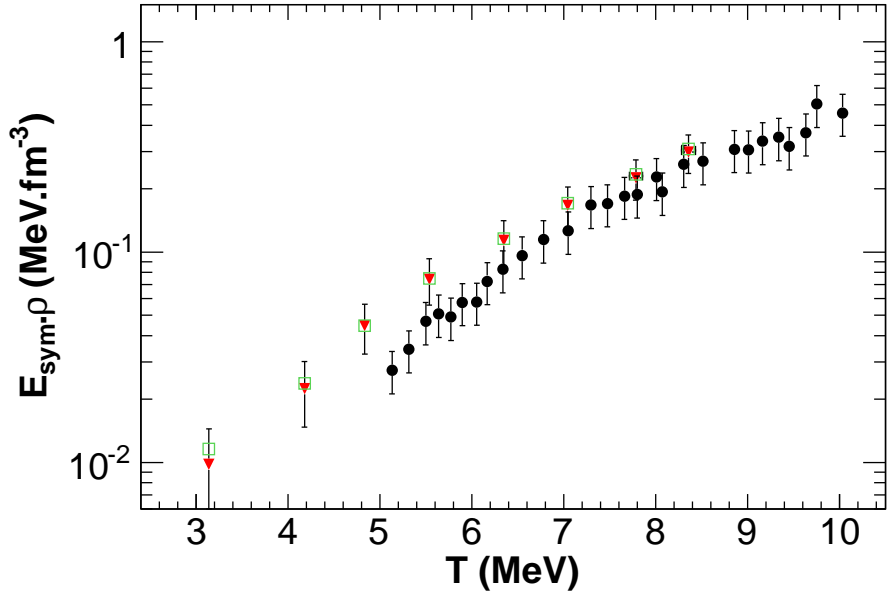}
\end{minipage}
\begin{minipage}[t]{16.5 cm}
\caption{(Color online) Symmetry energy density obtained from the Landau's theory $O(m_\chi^6)$ and quantum fluctuations, full triangles. Full circles are obtained from the coalescence model and isoscaling \cite{wada1}. The open squares are obtained from the excitation energy per particle times density, obtained from the fragments \cite{justin, justin1, justinpri}.}\label{seneden}
\end{minipage}
\end{center}
\end{figure}

There are some important consequences to draw from this result:\\
1) we discussed in section \ref{eost0}, that a particular choice of the coefficients $c_1,c_2$ in equation (\ref{ck225}) might give a phase transition. In particular we discussed the case of a
second order phase transition in order to fix the values of the parameters. A first order phase transition is of course possible as well, and the good reproduction of the
data within Landau's $O(m_\chi^6)$ confirms this. If the phase transition is driven by the symmetry term, it explains why the other terms contained in the mass formula 
become negligible.\\
2) A phase transition indicates a breaking of a particular symmetry, in this case $n-p$ symmetry, and the excitation of a Goldstone boson \cite{khuang}. This symmetry breaking might be accompanied
by the disappearance of a particular mode which we might associate to the disappearance of the IVGDR at high temperature. Coincidence measurements between fragments and 
photons in the IVGDR region as function of $T$ could confirm this. Such results might be associated to the occurrence of ``nuclear critical opalescence". \\
3) From Landau's free energy we can obtain the total symmetry energy similar to the zero temperature symmetry energy in equation (\ref{symexp}):
\begin{equation}
\frac{\bar F_A}{T}=\frac{\frac{F_A}{T}(m_\chi=1)+\frac{F_A}{T}(m_\chi=-1)}{2}-\frac{F_A}{T}(m_\chi=0)=\frac{1}{2}a +\frac{1}{4}b +\frac{1}{6}c. \label{FO6}
\end{equation}
This quantity is of course function of density and temperature. On similar grounds we can calculate the average symmetry energy
\begin{eqnarray}
\bar E_s &=& \frac{E_A(m_\chi=1)+E_A(m_\chi=-1)}{2}-E_A(m_\chi=0)\nonumber\\
&=& \bar F_A(m_\chi=1)- \bar F_A(m_\chi=0)+T[S_A(m_\chi=1)-S_A(m_\chi=0)],\label{EO6}
\end{eqnarray}
where the mixing entropy per particle is given by:
\begin{equation}
S_A(m_\chi)=-\Big[\frac{(1-m_\chi)}{2}\ln[\frac{(1-m_\chi)}{2}]+\frac{(1+m_\chi)}{2}\ln[\frac{(1+m_\chi)}{2}]\Big].\label{Smix}
\end{equation}
From experiments we can determine $T$ and $\rho$ using the methods outlined in the previous sections and therefore the symmetry energy. Because of the observed scaling of the free energy
as function of the order parameter $m_\chi$, we expect that the values of the symmetry energy are equal to the excitation energy corresponding to the $T$ and $\rho$ . In figure \ref{seneden} we plot the energy density versus $T$ from \cite{justin1}. The symmetry energy given by the full triangles is obtained by a careful fitting of the data using Landau's theory, equation (\ref{O6}) and multiplied by the 
density in order to obtain the symmetry energy density. $T$ and $\rho$  were obtained from proton quantum fluctuations and corrected for Coulomb \cite{hua6, hua7}. The full circles were obtained in ref. \cite{wada1} using the coalescence model and isoscaling as discussed in the next section.  The open squares are obtained from the excitation energy times the density obtained from the detected fragments and their Q-value.
It is amazing that such different methods agree within the error bars. The fact that all the excitation energy is transformed into symmetry energy is a further proof that it is the dominant factor in the QLG region we are 
analyzing and puts some further constraints on the NEOS.  We notice that plots of the symmetry energy as function of $T$ $(\rho)$, might be misleading since none of these can be obtained by keeping $\rho$ $(T)$ fixed. The energy density as function of temperature makes use of all the quantities that can be experimentally determined. Another possibility would be the symmetry energy divided by $T$ as function of $\rho$. A comparison of theoretical models to these quantities should be more constraining.

  \subsection{\it Isoscaling}
 We have discussed a large amount of data which suggests that thermal and possibly chemical equilibrium might be reached in the collisions. This explains the large success 
 of statistical models \cite{dasrep05, bondorf1, gross1, botvina1, botvina2} in the interpretation of a large amount of data with some simple assumptions and a few free parameters. For instance within a grand canonical model
 we can write the  production probability of a fragment of mass A as \cite{tsang6}:
 \begin{equation}
 Y(N,Z)\propto \exp(\frac{Z\mu_p+N\mu_n+B_{N,Z}}{T}),\label{gc}
 \end{equation}
 where $\mu_i$ are the chemical potential of protons and neutrons, and $B$ is the binding energy of the fragment. Notice that while the chemical potentials depend on the number of neutrons and protons
 of the source, all the other quantities, including the proportionality constant depend on the fragment. Tsang and collaborators \cite{tsangsym1, tsangsym2, tsangsym3} proposed to take the ratio of the yield of the same fragment from two different reactions,
 thus:
 \begin{equation}
 R_{21}=C \exp(\alpha N+\beta Z),\label{iso}
 \end{equation}
 with:
 \begin{equation}
 \alpha=\frac{\mu_{n,2}-\mu_{n,1}}{T},\quad \beta=\frac{\mu_{p,2}-\mu_{p,1}}{T},\label{alpha}
 \end{equation}
 which can be easily obtained from equation (\ref{gc}). Thus a plot of $R_{21}$ versus $N$ for fixed $Z$ and vice versa, should be exponentially increasing with coefficients given by equation (\ref{alpha}).
 This approach was dubbed as $\it {isoscaling}$ and the main assumption at this stage is that the two reactions reach the same temperature $T$. If one further imposes that the two reactions produce hot equilibrated
 sources at temperature $T$ and density $\rho$, have the same mass number $A_s$ but different $m_{\chi s}$ we get \cite{tsangsym1, tsangsym2, tsangsym3, toke3}:
 \begin{equation}
 \alpha\approx 4C_{sym}[(Z_{s1}/A_s)^2-(Z_{s2}/A_s)^2]/T.\label{alpha2}
 \end{equation}
 Equation (\ref{alpha2}) is intuitively easy to understand since we have restricted the mass number of the source to be the same and changed the concentration only. In such a limit we would expect that
 most contributions to the chemical potentials will cancel out apart the symmetry free energy and Coulomb. In effect we have seen in the previous section that Landau's free energy seem to be dependent on
 the order parameter $m_\chi$ only and Coulomb is negligible. Thus, within isoscaling, correcting for entropy, we can derive the symmetry energy \cite{wada1} and one result of such analysis is plotted
 in figure \ref{seneden} where the symmetry energy density is plotted and compared to the Landau's method discussed in the previous section \cite{justinpri}. In order to get the symmetry energy coefficient
it is necessary to determine the source concentration entering equation (\ref{alpha2}). Initially this quantity, indicated usually as $\Delta$, was estimated from the concentration of the two reactions at the same beam energy, 
 used to perform the isoscaling analysis. It was realized theoretically by Ono {\it et al.}, that the use of the initial concentrations of the ions used for the collisions was not the correct one but rather one should use
 the average concentration of the biggest fragment \cite{onoiso03}. Later on, and we will discuss this in the next section, it was shown that one could use $H/T$ obtained from the Landau's approach, the two latter methods
 gave similar results \cite{marini1}.
 
 \subsection{\it $m_\chi$ scaling}
 Within Landau's theory we can derive $R_{21}$ as well and we simply obtain:
 \begin{equation}
 R_{21}=C \exp(\frac{\Delta H}{T} m_{\chi}A),\label{isol}
 \end{equation}
 where $\frac{\Delta H}{T}=\frac{H_2-H_1}{T}$. Comparing equation (\ref{isol}) and equation (\ref{iso}) gives:
\begin{equation}
\frac{\Delta H}{T}m_{\chi}A=\alpha N+\beta Z\rightarrow \alpha=-\beta=\frac{\Delta H}{T}. \label{Halpha}
 \end{equation}
 Thus isoscaling and Landau's approach are equivalent especially if Coulomb effects are negligible \cite{huangiso10}, this is also confirmed 
 by a large amount of investigations which suggest $\alpha\approx -\beta$ \cite{ tripathi1, marini1, huangiso10, tripathi2, iacoll05, maisoscaling}.
 
 From a comparison of isoscaling and $m_\chi$-scaling, it is clear that the difference in the external fields is proportional to the differences between chemical potentials of neutrons and protons in the two systems.
 Using a grand canonical approach, a relation has been found between the differences of chemical potentials and the symmetry energy, equation (\ref{alpha2}), which involves the different concentrations of the sources.
 It is not clear how the concentrations should be determined, if it is the initial concentration of the two different reactions, or other quantities \cite{onoiso03}. From Landau's theory we know that,
 for a given $T$ and density, the corresponding order parameter can be found by minimizing the Free-energy \cite{khuang}. In presence of an external field such a minimum is shifted \cite{huang, aldo5, tripathi1, tripathi2}
 for instance as:
 \begin{equation}
 \epsilon_0=\frac{H/T}{a},\label{tri1}
 \end{equation}
where $\epsilon_0$ is dependent on the initial concentration of the emitting source. Tripathi and collaborators \cite{tripathi1, tripathi2} demonstrated that this quantity can be derived from the experimental concentrations
of light fragments, i.e. excluding $p$ and $n$, thus improving on the initial assumption of isoscaling to use the concentration of the reacting systems. Furthermore, the same authors showed that it was not necessary to use two different reactions 
to do the isoscaling or $m_\chi$-scaling analyses but it suffices to select sources from the same experiment with different initial concentrations. A further improvement proposed in \cite{tripathi1, tripathi2} was to use the Landau's free energy
approach to estimate the ratios of mirror nuclei and from these the quantity $\epsilon_0$ was estimated. In ref. \cite{marini1}, a comparison among the different methods was performed, including a Fisher-type analysis \cite{wada1} and it was
concluded that an estimate of $\epsilon_0$ or $\Delta$ in isoscaling from the concentration of small fragments observed in the experiments \cite{tripathi1, tripathi2} or the concentration of the biggest fragment \cite{onoiso03} gave similar results.

The analysis above showed that the external field is connected to the $a$ parameter of the Landau's free energy and to the concentration of the source. Of course if one is able to derive directly the free energy for one system, there is no need
for taking ratios since all the quantities are already present in the Landau's approach. Furthermore, most data is usually selected at the same excitation energy, while isoscaling and $m_\chi$-scaling require the same temperature and density, thus
distortions might be coming from these approaches.

\section{Conclusion}
 Various aspects of the NEOS have been discussed in this work and some conclusions can be drawn. We have a good understanding of symmetric nuclear matter, the compressibility is fixed by the ISGMR and collective flow observed in heavy ion collisions to a value $K=250\pm 25 MeV$ \cite{bertschrep88, shlomo2,  youngblood1, youngblood2, youngblood3, piekarewicz1,  chen1, lipparini1, cao1, khan1, khan2, colo1, blaizotrep80, bertsch2}. The momentum dependence of the NEOS and the effective mass (equal to about $m^*/m=0.7$ near the gs density)  \cite{ r2ad_gr1, bertschrep88, shlomo2, baoanrep08, aldorep94, baranrep05, stone3, bertsch2, gogny1, gale1, prakash1, r2ad_gr2, r2ad_gr3} is crucial to understand these phenomena. For IVGDR the symmetry energy plays an important role but the effective mass and surface effects are  crucial to reproduce the experimental results. Microscopic models using local potentials, can only give qualitative features. In order to describe the available data quantitatively, the interaction must be tested to reproduce ground state properties of nuclei and Giant Resonances which might be a little bit optimistic at this stage. The analysis of binding energies, isobaric analog states and giant resonances give a symmetry energy $a_a(\infty)=30\pm5MeV$, the error on such an estimate is too large and should be reduced including further constraints.  The values of the parameters $L$ and $K_{sym}$ are
 much more unconstrained and as our personal feeling $L=50 \pm 40MeV$ and $K_{sym}>-150MeV$, however, we did not really find any physical observable which is sensitive to one of them, but rather to their combination, including the symmetry energy and nuclear compressibility. We have seen that fixing the symmetry energy to 32 MeV, zero temperature and the order parameter $m_\chi=0.98$, we can get a second order phase transition. In particular for those conditions we obtain two sets of parameters, $c_1,c_2$, one gives a phase transition at low densities, the quantum liquid gas phase transition. The second set of parameters gives a transition at high densities which we have associated to the occurrence of the quark-gluon plasma. The parameters obtained give reasonable values of $a_a(\infty)$, $L$ and $K_{sym}$, thus they cannot be dismissed so easily. Neutron star masses and radii might exclude some of these NEOS, but we can change a little bit the composition of the star and be able to reproduce the observation available so far. Naturally, other conditions might give a first order or a cross-over but what we wanted to stress is the extreme sensitivity of the NEOS to the symmetry part.
 
 Astrophysical observations alone cannot constrain the NEOS, we need to do this in terrestrial laboratories as well especially producing and studying exotic nuclei, both neutron or proton rich.
 We have discussed some methods to derive from experimental observations of heavy ion collisions, the relevant quantities to the NEOS, i.e. $T$, $\rho$ and pressure.
 Those methods have given some constraints on the NEOS and more will give in the future. An important experimental quantity to determine is the neutron multiplicity and 
its energy distribution. This is still lacking because of experimental difficulties. Quantum and classical methods have been proposed to derive these quantities and we
 are starting to understand their validity and realm of application.
 
 Ground states and excited states of nuclei, including fragmentation, indicate  that the nucleus can be discussed in terms of strongly interacting Fermions, protons and neutrons, but also as a mixture of Bosons ($\alpha$, d..) and Fermions. This is due to the strong correlations which result in the high binding energy of $\alpha$ particles. To highlight these possible effects, the methods used to analyze experimental data must contain the relevant quantum statistics. Classical methods are clearly inadequate for nuclear systems, even though they might give some guidance at high entropies. However, to insist on the validity of classical approaches and include modifications to try to reproduce some data, might be misleading. When possible, one should start from quantum models or at least the correct quantum statistics and if some features are classical in nature, they will also
 be reproduced by the quantum approach. But a Bose condensate or Fermion quenching cannot be described by classical approaches, thus a large amount of interesting
 physics might be missed when using those models.
 
 We have only dealt with `near-equilibrium' features, but there is a large variety of data for pre equilibrium particle production and collective flow among others. Those phenomena constraint the NEOS as well and detailed modeling and experimental investigations are needed. 
 
 \section{Acknowledgement}
We are grateful for discussions with M.R. Anders, V. Baran, B. Borderie, A. Chbihi, M. Colonna, P. Danielewicz, C. Dorso, F. Gulminelli, K. Hagel, Z. Kohley, X. Liu, Y.W. Lui, J. Mabiala, P. Marini, T. Maruyama, L. May, A. McIntosh, J. Natowitz, M. Papa, A. Raphelt, G. R\"opke, U. Schroeder, S. Shlomo,  G.Souliotis, I. Toke, G. Verde, R. Wada,  T. Werke, S. Wuenschel, J. Xun,  S. Yennello,   D.H. Youngblood.

\end{document}